\documentclass[11pt]{article}
\usepackage{amssymb}
\usepackage[sectionbib]{natbib}
\usepackage{array,epsfig,rotating}
\usepackage{amsmath}
\usepackage{bbm}
\usepackage{amsfonts}
\usepackage{multirow}
\usepackage{amsthm}
\usepackage{array}
\usepackage{booktabs}
\usepackage{graphicx}
\usepackage{graphics}
\usepackage{epsfig,latexsym,verbatim}
\usepackage{color}
\usepackage{multicol}
\usepackage{threeparttable}
\usepackage{float}
\usepackage{cancel}
\usepackage{url}
\usepackage{ulem}
\usepackage{rotating}
\usepackage{hyperref}
\usepackage{cleveref}
\crefname{algocf}{alg.}{algs.}
\Crefname{algocf}{Algorithm}{Algorithms}
\usepackage{upgreek}
\usepackage{enumitem}
\usepackage{tikz}
\usetikzlibrary{shapes,backgrounds}
\usetikzlibrary{arrows.meta, positioning, quotes}
\usepackage{placeins}
\usepackage{xr}
\usepackage{secdot}
\usepackage{subcaption}
\usepackage[lined,algonl,boxed]{algorithm2e}
\usepackage{cancel}

\definecolor{red}{rgb}{0,0,0}
\definecolor{green}{rgb}{0,0,0}
\definecolor{blue}{rgb}{0,0,0}

\def\bs{\boldsymbol}
\def\bb{\mathbf}

\newcommand{\R}{{\mathbb R}}
\newcommand{\E}{{\mathrm E}}
\newcommand{\ba}{{\bs\alpha}}
\DeclareMathOperator*{\argmin}{arg\,min}

\newcommand{\var}{{\rm Var}}

\floatstyle{ruled}
\theoremstyle{definition}
\newtheorem{theorem}{Theorem}
\newtheorem{lemma}{Lemma}

\newtheorem{remark}{Remark}
\newtheorem{prp}{Proposition}
\usepackage{appendix}

\begin{document}

\renewcommand{\baselinestretch}{1.2}
\markboth{\hfill{\footnotesize\rm Zhiling Gu, Shan Yu, Guannan Wang, Ming-Jun Lai and Li Wang}\hfill}
{\hfill {\footnotesize\rm TSSS: Triangulated Spherical Spline Smoothing} \hfill}
\renewcommand{\thefootnote}{}
$\ $\par \fontsize{10.95}{14pt plus.8pt minus .6pt}\selectfont
\vspace{0.8pc} \centerline{\large\bf TSSS: A Novel Triangulated Spherical Spline Smoothing}  \centerline{\large\bf for Surface-based Data}
\vspace{.4cm} \centerline{ 
Zhiling Gu$^{a}$, 
Shan Yu$^{b}$, 
Guannan Wang$^{c}$, 
Ming-Jun Lai$^{d}$ and  
Li Wang$^{e}$ 
\footnote{\emph{Address for correspondence}: 
Li Wang, Department of Statistics, George Mason University, Fairfax, VA 22030, USA. Email: lwang41@gmu.edu}} \vspace{.4cm} 
\centerline{\it 
$^{a}$Iowa State University, 
$^{b}$University of Virginia, 
 $^{c}$College of William \& Mary,
 } 
 \centerline{\it 
$^{d}$University of Georgia and
 $^{e}$George Mason University} \vspace{.55cm}
\fontsize{9}{11.5pt plus.8pt minus .6pt}\selectfont



\begin{quotation}

\noindent {\it Abstract:}
Surface-based data is commonly observed in diverse practical applications spanning various fields. In this paper, we introduce a novel nonparametric method to discover the underlying signals from data distributed on complex surface-based domains. Our approach involves a penalized spline estimator defined on a triangulation of surface patches, which enables effective signal extraction and recovery. The proposed method offers several advantages over existing methods, including superior handling of ``leakage'' or ``boundary effects'' over complex domains, enhanced computational efficiency, and potential applications in analyzing sparse and irregularly distributed data on complex objects. We provide rigorous theoretical guarantees for the proposed method, including convergence rates of the estimator in both the $L_2$ and supremum norms, as well as the asymptotic normality of the estimator. We also demonstrate that the convergence rates achieved by our estimation method are optimal within the framework of nonparametric estimation. Furthermore, we introduce a bootstrap method to quantify the uncertainty associated with the proposed estimators accurately. The superior performance of the proposed method is demonstrated through simulation experiments and data applications on cortical surface functional magnetic resonance imaging data and oceanic near-surface atmospheric data.

\vspace{9pt}
\end{quotation}

\fontsize{10.95}{14pt plus.8pt minus .6pt}\selectfont
\thispagestyle{empty}
\setcounter{equation}{0}

\section{Introduction} \label{SEC:introduction}
	Surface-based data is widely observed in various fields, and extracting useful information from data distributed on surfaces is of great significance. For example, in planetary science, scientists are interested in the movement of tectonic plates \citep{chang2000regression, marzio2019} observed on the surfaces of celestial bodies. In cosmology, completing the missing data and correcting the noisy observations for the cosmic microwave background radiation are problems of interest \citep{abrial2008cmb}. In meteorology, hourly surface-based data is the most used and requested type of climatology data, and efforts have been made to integrate data from different stations across various repositories \citep{smith2011integrated}. Additional examples include the recovery of high-resolution time series of aerosol optical depth \citep{Zhang:etal:2022} and total electron content \citep{Sun:etal:2022} from surface-based measurements.
	
\begin{figure}[!ht]
		\centering
		\includegraphics[width = 0.9\textwidth]{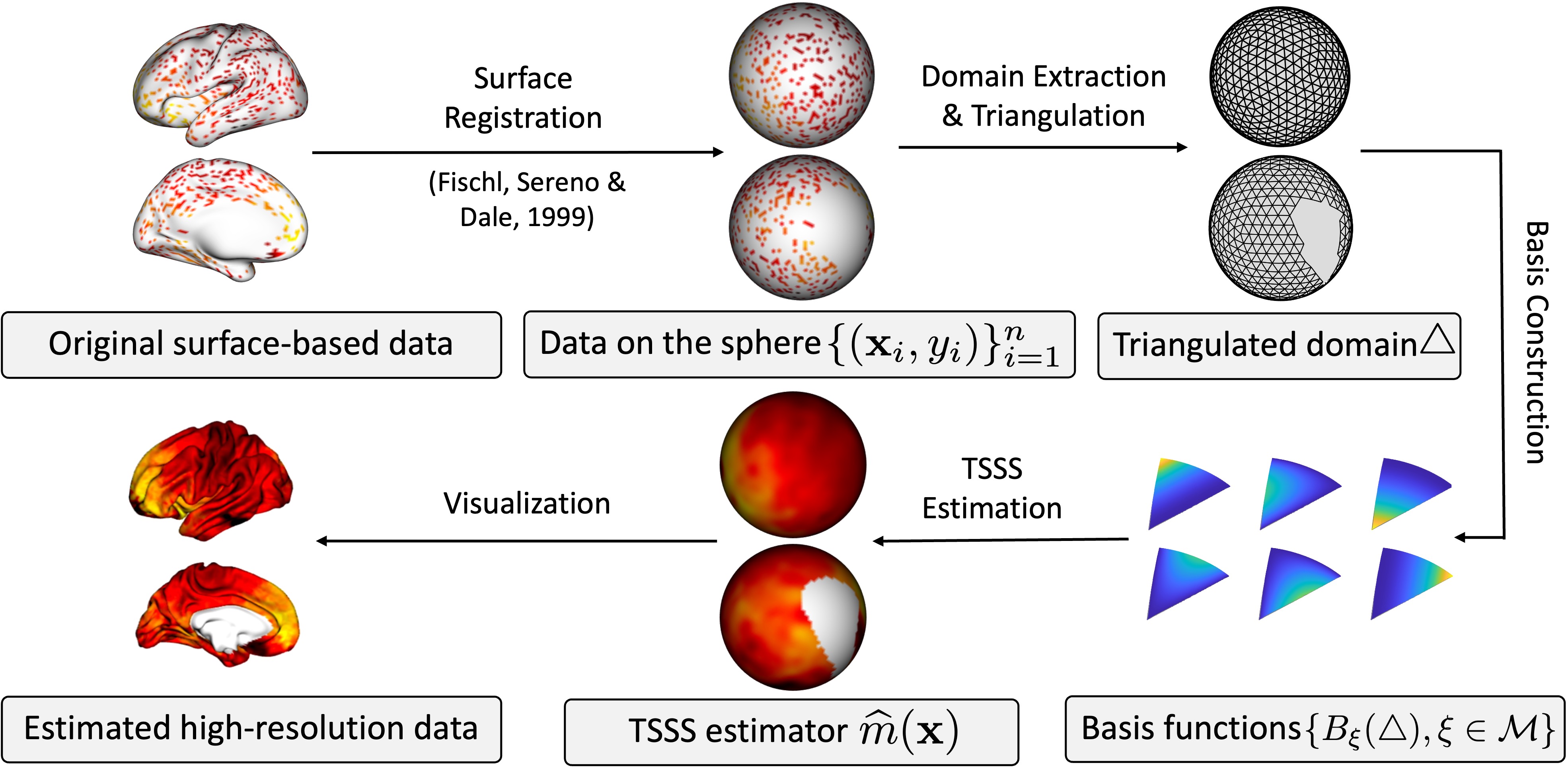}
		\caption{Workflow of the TSSS estimation method for brain surface data.} 
		\label{fig:hcpflow}
	\end{figure}

	Furthermore, in biomedical research, efforts have been made to explore the functional properties of proteins using molecular surface data \citep{kinoshita2003identification}. Another active area of investigation is cortical and cortical surface functional magnetic resonance imaging (fMRI) in neuroimaging. This technique offers the advantage of recovering anatomical structures with higher precision compared to volumetric fMRI methods \citep{brodoehl2020surface, Mejia:2020, lila:2020, cole2021surface, zhang2022lesa}. To analyze data distributed on general surfaces, it is common to employ mappings that project the data onto a unit sphere \citep{gu2004genus, fischl1999cortical}, as depicted in Figure \ref{fig:hcpflow}.  
	
	To model and denoise the spherical data, spherical harmonics (SH) \citep{seeley1966spherical, abrial2008cmb} are widely used as global basis methods to interpolate functions on the whole sphere. Later, thin plate splines on the sphere (TPSOS) \citep{Wahba:1981} were proposed as a special case of SH while accounting for the smoothness of functions. However, in some instances, they may lack the desired flexibility needed to capture the complexity of the underlying signal accurately. SH relies on a fixed set of basis functions that span the entire sphere. While it has the ability to represent a wide range of functions, it may encounter challenges when dealing with highly localized or rapidly changing functions. A high number of basis functions is often required to achieve a good approximation power in such situations. This increased complexity can pose computational challenges and limit the practicality of using spherical harmonics. Similarly, TPSOS incorporates the concept of function smoothness but may still face difficulties when fitting complex functions. The flexibility of TPSOS is limited by its basis functions, which are constructed based on polynomial functions of trigonometric functions of longitudes and co-latitudes. 
	
	Another popular method is the kernel-based approach \citep{Cao:2013}; however, it becomes computationally intensive for large datasets primarily due to the computation and storage requirements of kernel matrices and the need for solving optimization problems involving these matrices. The size of the kernel matrix is proportional to the number of data points, which makes the computation and storage requirements overgrow with the dataset size. The limitations of existing methods motivate the need for alternative methods that can offer enhanced flexibility in capturing the complexity of the underlying signal for spherical data.
	
	In addition, it is common for data to be collected over complex patches/regions of surfaces with irregular boundaries or holes, such as the domain of the oceans shown in Figure \ref{fig:triangulation}(e). It is well known that all aforementioned conventional tools for analyzing spherical data suffer from the problem of ``leakage'' across irregular-shaped domains due to inappropriately linking parts of the domain, as discussed in literature \citep{Ramsay:02, Wood:Etal:08}. Motivated by recent works, such as \cite{Lai:Wang:13} and \cite{WWLG20}, Bernstein-B\'{e}zier polynomials defined on triangles are useful tools over planar domains with irregular shapes. Given the similarity and common properties between classical polynomial splines over planar triangulation and the spherical setting, spherical Bernstein-B\'{e}zier splines \citep{Alfeld:Etal:96, Alfeld:Etal:96b} could be well suited for spherical data interpolation and approximation problems for complex sphere-like surfaces. The researchers in \cite{BL05}, \cite{Baramidze:Etal:06}, \cite{LaiEtal:2008}, and \cite{BL11} have employed spherical Bernstein-B\'{e}zier splines for data interpolation and fitting, particularly in the context of geo-potential reconstruction, demonstrating their efficacy in modeling the geo-potential function. In this paper, we introduce a novel application of spherical splines specifically tailored to analyze data distributed on complex surface-based domains such as surface patches with complex boundaries. This innovative approach, referred to as triangulated spherical spline smoothing (TSSS), enables the extraction of underlying signals from intricate data structures on surface patches in the presence of noise. Our TSSS approach provides a robust and computationally efficient method for accurately estimating and modeling complex surface patterns.
	
	The proposed TSSS method overcomes the limitations of existing methods and offers several advantages. Firstly, it effectively addresses the problem of ``leakage'' across complex domains by utilizing information from neighboring triangles to accurately denoise or deblur data while preserving geometric features and spatial structures. This makes it ideal for analyzing data on complex patches or regions of surfaces with irregular boundaries or holes. Secondly, TSSS employs compactly supported basis functions and a sparse or roughness penalty, similar to P-splines \citep{eilers2021practical} and bivariate penalized splines on triangulation \citep[BST;][]{Lai:Wang:13, Yu:Wang:Wang:Liu:Yang:20, Wang:Wang:Zhao:Cao:Li:23}. This enables the model to capture intricate patterns in the data without becoming overly complex, mitigate boundary effects, and facilitate robust smoothing of unevenly distributed data, such as sparse and irregular data on surface-based complex domains. In addition, TSSS offers computational efficiency as a global estimation method with an explicit model expression. Its efficiency stems from the need to solve only a single linear system, which reduces the computational complexity and makes it suitable for handling large datasets.
	
	Furthermore, we investigate the statistical properties of the proposed TSSS method. Specifically, we establish the convergence rate of the TSSS estimator, which is governed by the fineness of the triangle mesh, the spherical spline degree and smoothness, the penalty parameter, and the smoothness of the mean function. We further provide the conditions to achieve the optimal nonparametric convergence rate \citep{stone1982optimal}. We also derive the asymptotic normality for the TSSS estimator. However, due to the complexity of the spline basis functions, obtaining the exact form of the standard error can be challenging. Therefore, we propose a wild-bootstrap-based method to estimate the standard error.
	
	The rest of this paper is organized as follows. Section \ref{sec:prelim} introduces the triangulation of a sphere and the spline space defined on the triangulation. In Section \ref{sec:theory}, we present the asymptotic properties of the proposed TSSS estimator, including convergence rates and asymptotic normality. Additionally, we introduce a bootstrap method to effectively quantify the uncertainty associated with the TSSS estimators. Section \ref{sec:implementation} outlines the implementation details of the TSSS method, including the selection of triangulation, spline basis, and penalty parameters. In Section \ref{sec:simulations}, we present simulation studies to evaluate the finite sample performance of the TSSS estimator for functions observed on the grid of the whole sphere and functions observed on a complex spherical domain, and the uncertainty quantification of the TSSS estimators. In Section \ref{sec:data-app}, TSSS is applied to cortical surface fMRI data and near-surface ocean-atmospheric data. Section \ref{sec:conclusion} summarizes the main contributions of this paper and concludes with some remarks. Technical details are provided in the Supplementary Materials.
	
	\section{Penalized Spline Estimators on Triangulated Spheres}
	\label{sec:prelim}
	
	Consider a set of observations $\{(\theta_i, \phi_i, Y_i)\}_{i = 1}^n$, where $Y_i\in \R$ represents the response variable, $\theta_i \in [0, \pi]$ and $\phi_i \in [0, 2\pi)$ denote the corresponding colatitude and longitude as illustrated in Figure \ref{fig:triangulation}(a). It is worth pointing out that a sphere $\mathbb{S}^2$ is a 2-dimensional (2D) object, which, for the convenience of analysis, is embedded in a 3-dimensional (3D) Euclidean space $\R^3$. Consequently, it is equivalent to consider any point on $\mathbb{S}^2$ in the following form $\mathbf{X}_i = (X_{i1}, X_{i2}, X_{i3})^\top \in \R^3$, where $\|\mathbf{X}_i\|_2^2 = X_{i1}^2 + X_{i2}^2 + X_{i3}^2 = 1$. Suppose $\{(\mathbf{X}_i, Y_i), \mathbf{X}_i\in \Omega\}_{i = 1}^n$ is a i.i.d sample of size $n$ observed on a domain $\Omega$. Here, $\Omega$ can be either the entire unit spherical domain $\mathbb{S}^2$ or a specific spherical patch or region within $\mathbb{S}^2$. We assume the sample is drawn from the distribution of $(\mathbf{X}, Y)$, $\mathbf{X}\in \Omega$, through 
	\begin{equation}
		Y_i = m(\mathbf{X}_i) + \sigma(\mathbf{X}_i) \epsilon_i,  ~i = 1, 2, \ldots, n, 
		\label{model}
	\end{equation}
	where $m(\cdot)$ and $\sigma(\cdot)$ are the conditional mean and standard deviation functions, $\epsilon_i$'s are i.i.d random errors with mean $\E(\epsilon_i) = 0$, and variance $\mathrm{Var}(\epsilon_i) = 1$. In addition, we assume $\epsilon_i$ and $\mathbf{X}_i$ are independent for all $i$. The problem of interest is to estimate the unknown mean function $m(\bb{x})$ given observations $\{(\mathbf{X}_i, Y_i), \mathbf{X}_i\in \Omega\}_{i = 1}^n$.

	\subsection{Triangulation and Spherical Bernstein-B\'{e}zier polynomials}
	\label{ssec:triangulation}
	
\begin{figure}[!ht]
		\centering
		\scalebox{0.9}{
			\begin{tabular}{ccccc}
				\includegraphics[width = 0.2\textwidth]{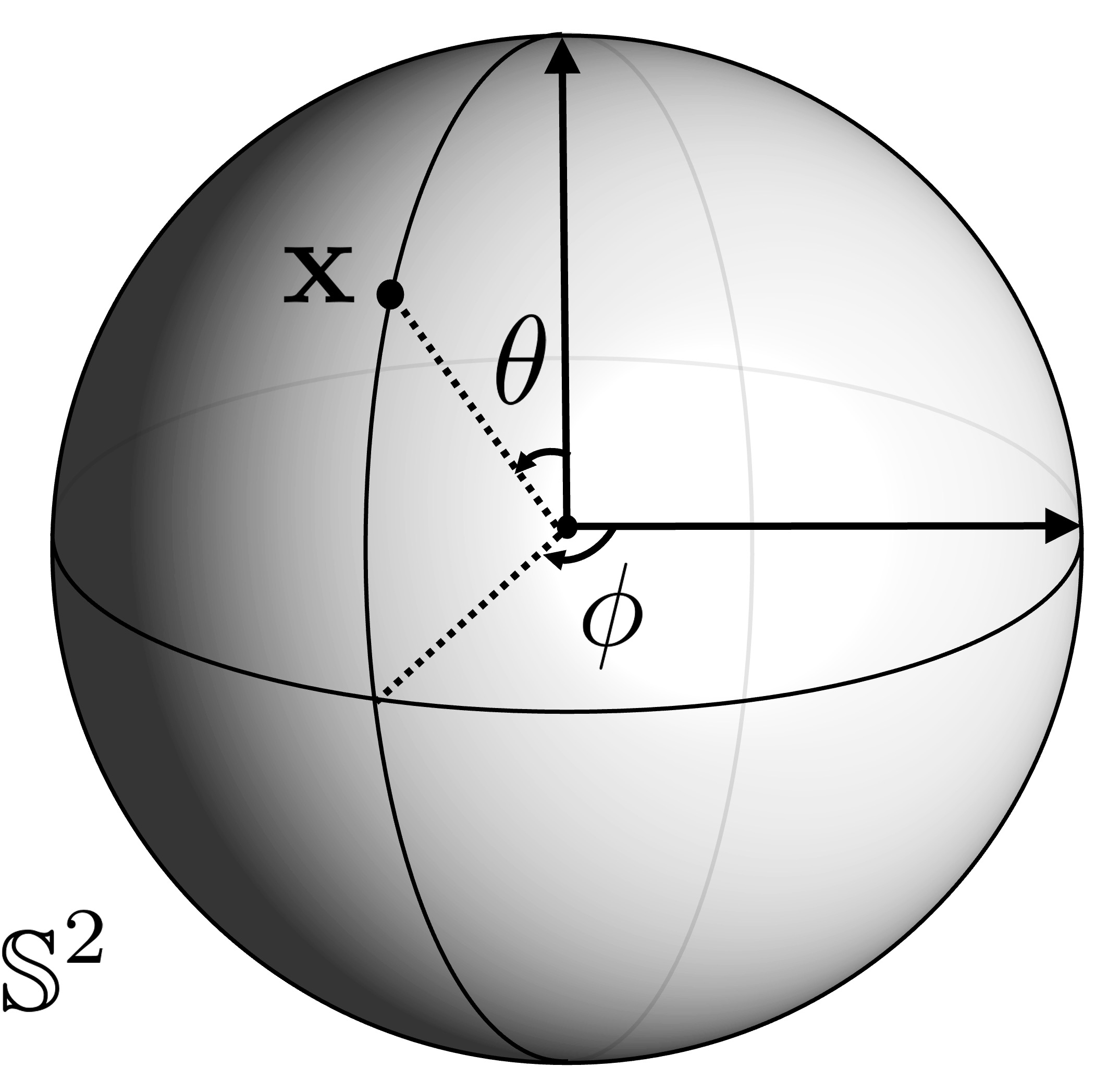} &
				\includegraphics[width = 0.18\textwidth]{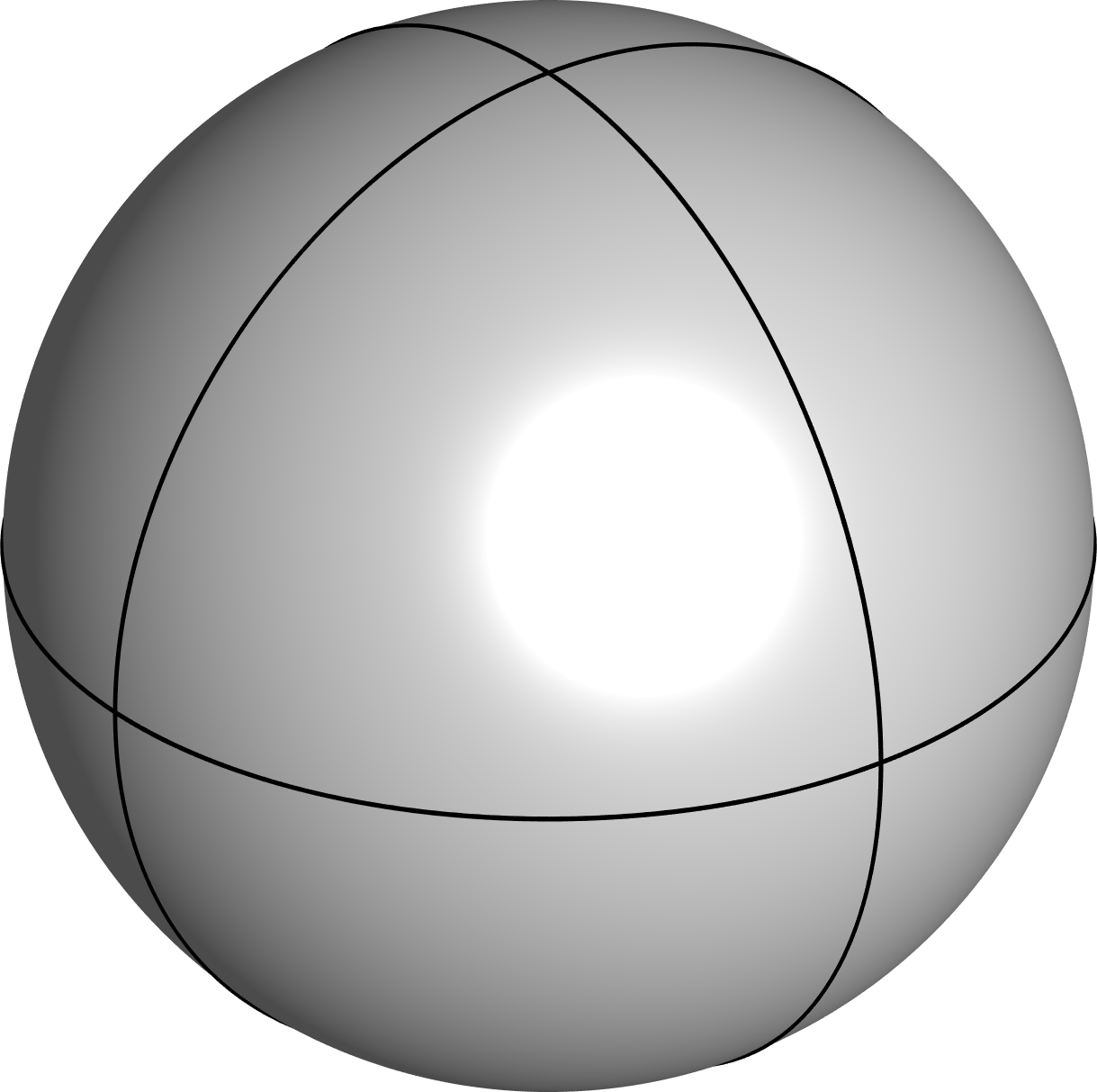} &
				\includegraphics[width = 0.18\textwidth]{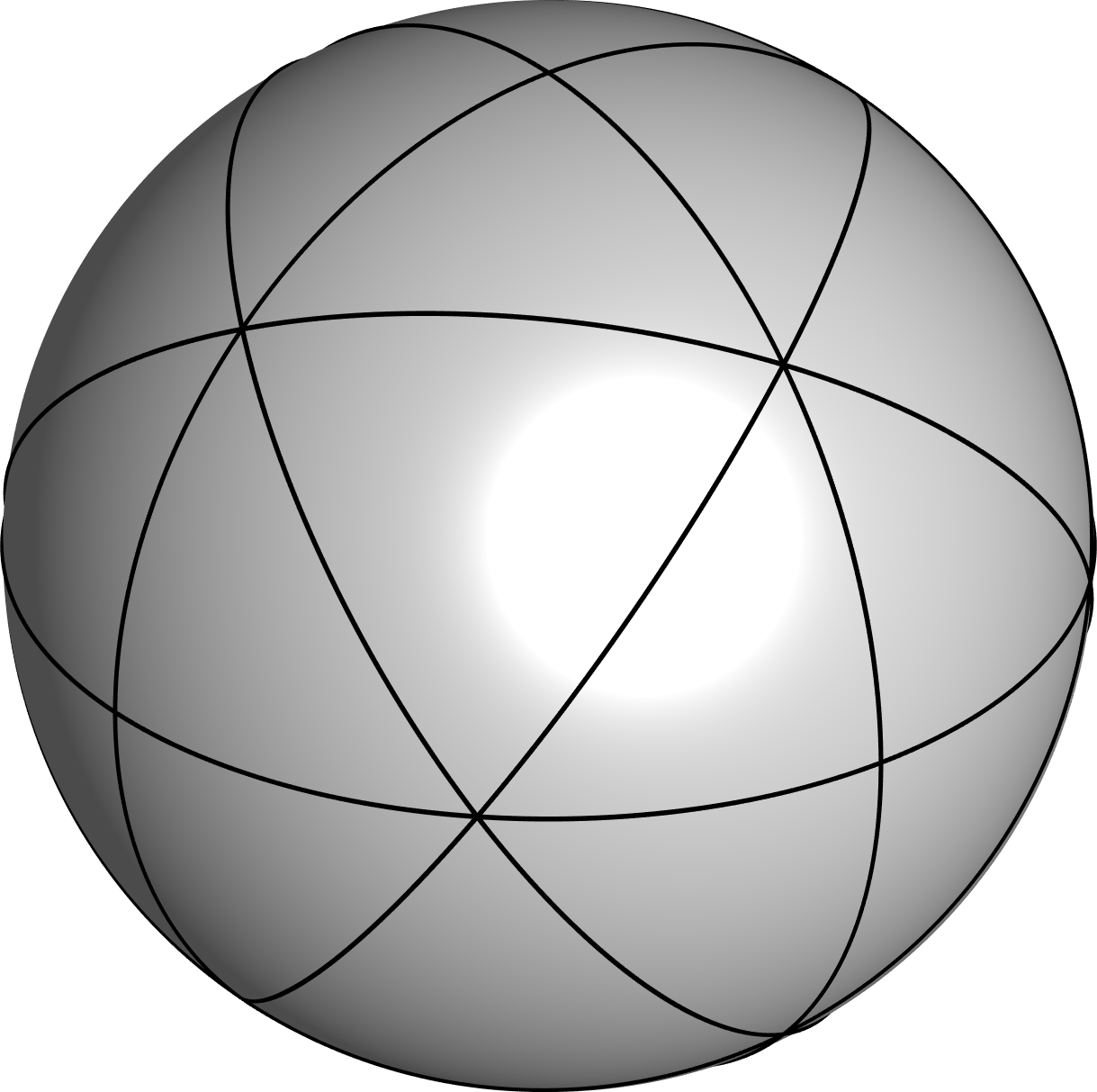} &
				\includegraphics[width = 0.18\textwidth]{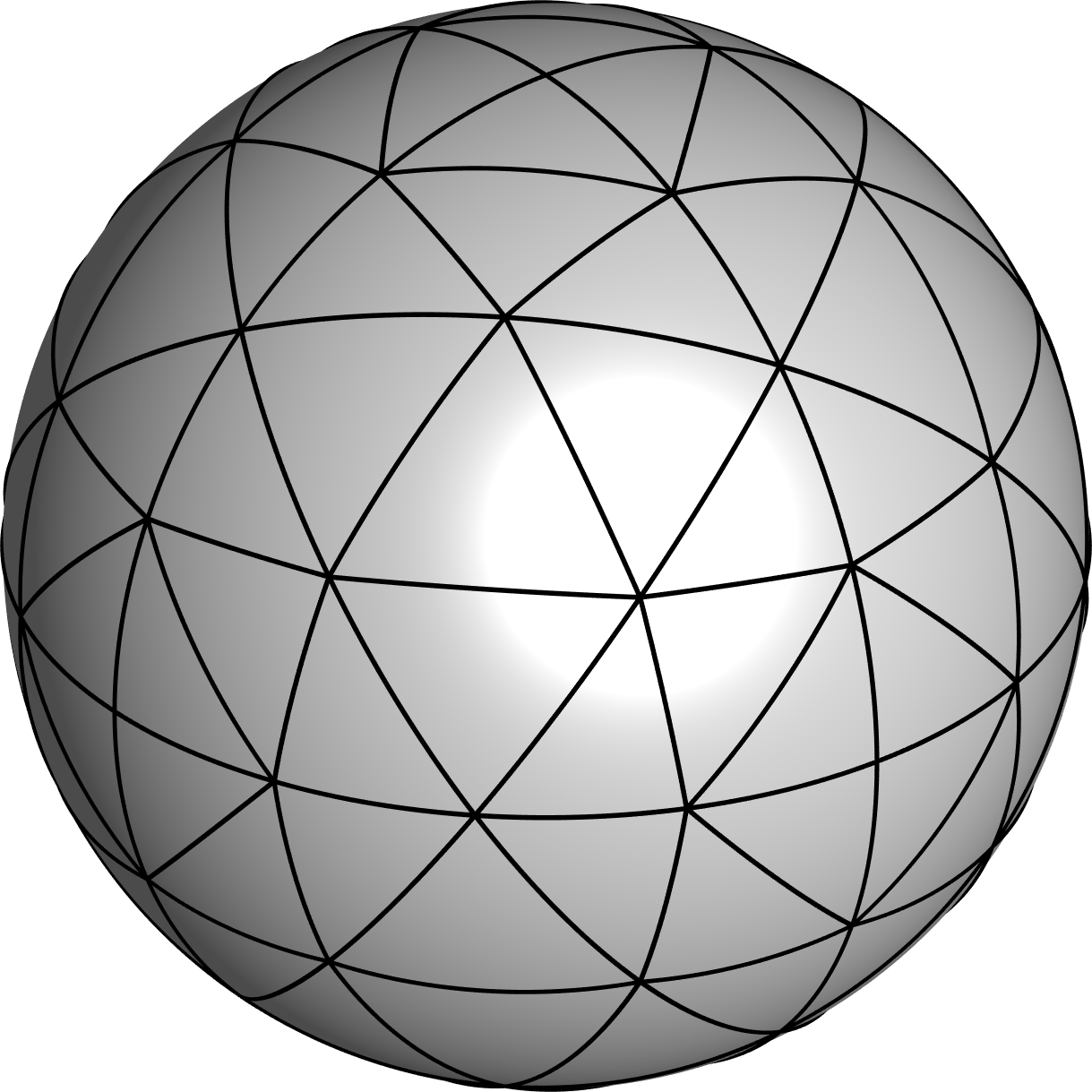} &
				\includegraphics[width = 0.18\textwidth]{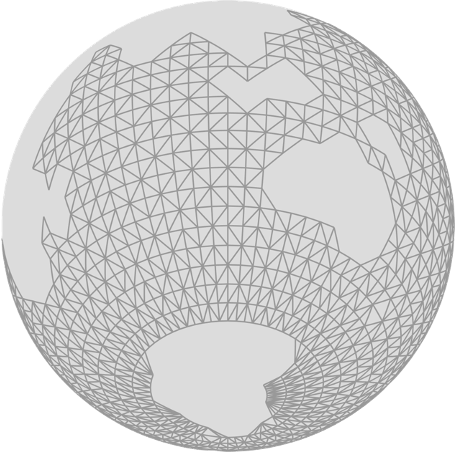}\\
				(a) & (b) & (c) & (d) & (e)
		\end{tabular}}
		\caption{Illustrations of (a) colatitude $\theta$ and longitude $\phi$, (b)-(d) spherical triangulations with varying levels of refinements, and (e) triangulation of the ocean surface. Refinement of the triangulation is achieved by connecting the center of each edge with the spherical edge with the shortest geodesic length.}
		\label{fig:triangulation}
	\end{figure}

	Triangulation has been widely used for irregular planar, spherical, and 3D domains due to its ability to effectively approximate complex geometries and simplify computational tasks \citep{mark2008computational, Lai:Schumaker:07, de2010triangulations}. 
	For domain $\Omega \subseteq \mathbb{S}^2$, a \textit{triangulation} $\triangle$ is defined as a collection of $N$ spherical triangles $\triangle = \{\tau_1, \ldots, \tau_N\}$, such that $\Omega = \cup_{i = 1}^N \tau_i$ \citep{Lai:Schumaker:07}. Each spherical triangle $\tau := \langle \mathbf{v}_1, \mathbf{v}_2, \mathbf{v}_3 \rangle$ comprises the set of points in $\Omega$ that lie within the region bounded by three circular arcs $\langle \mathbf{v}_i, \mathbf{v}_{i+1} \rangle, i = 1, 2, 3$. Here $\mathbf{v}_1, \mathbf{v}_2, \mathbf{v}_3$ are the three vertices of $\tau$, and $\mathbf{v}_4$ is identified as $\mathbf{v}_1$. These spherical triangles either share an edge (circular arc), share a vertex, or do not intersect each other. Figure \ref{fig:triangulation} provides an illustration of spherical triangulations.

\begin{figure}[!ht]
		\centering
		\scalebox{0.7}{
			\begin{tabular}{cccc}
				\includegraphics[width = 0.25\textwidth]{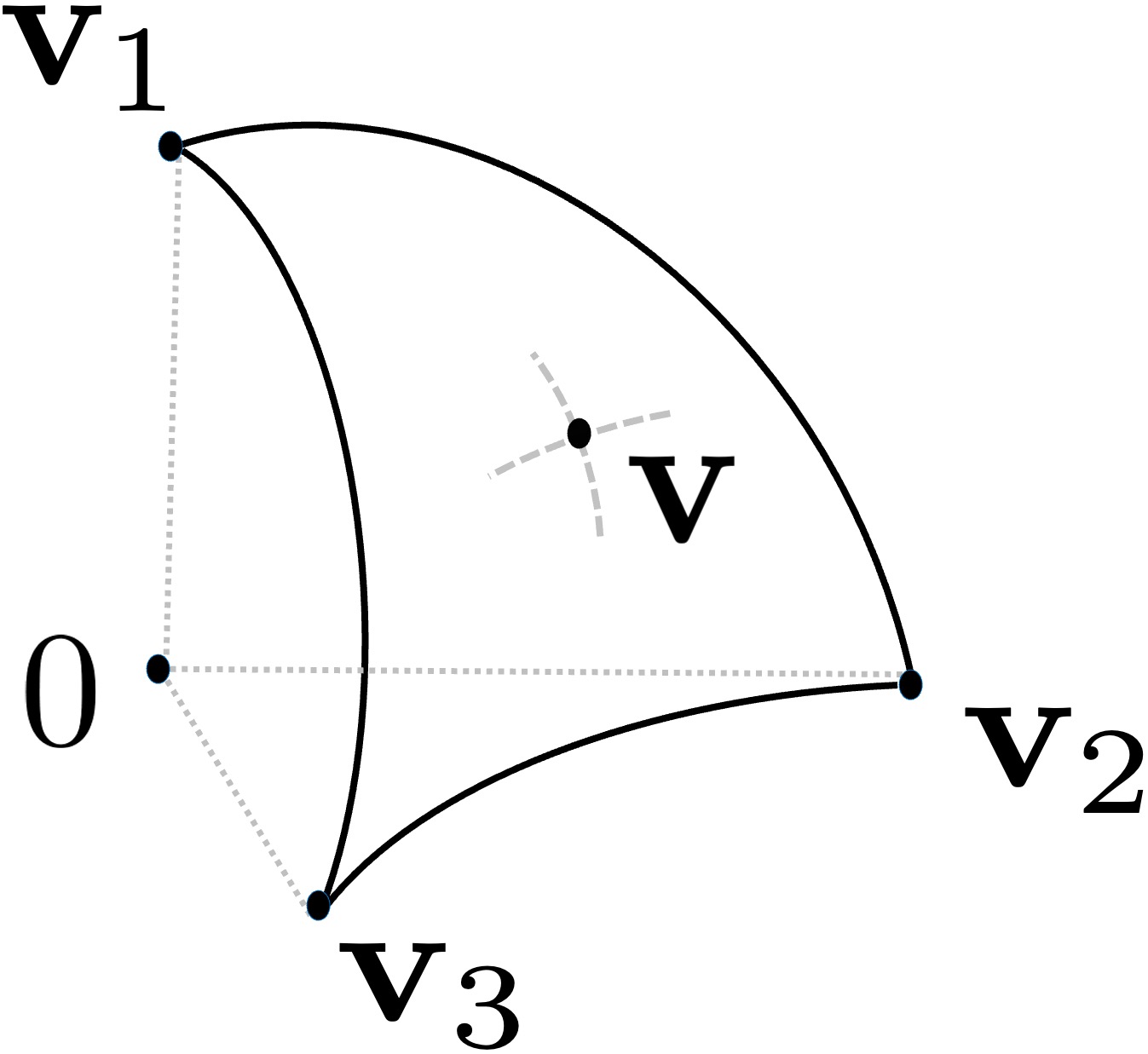} &
				\includegraphics[width = 0.25\textwidth]{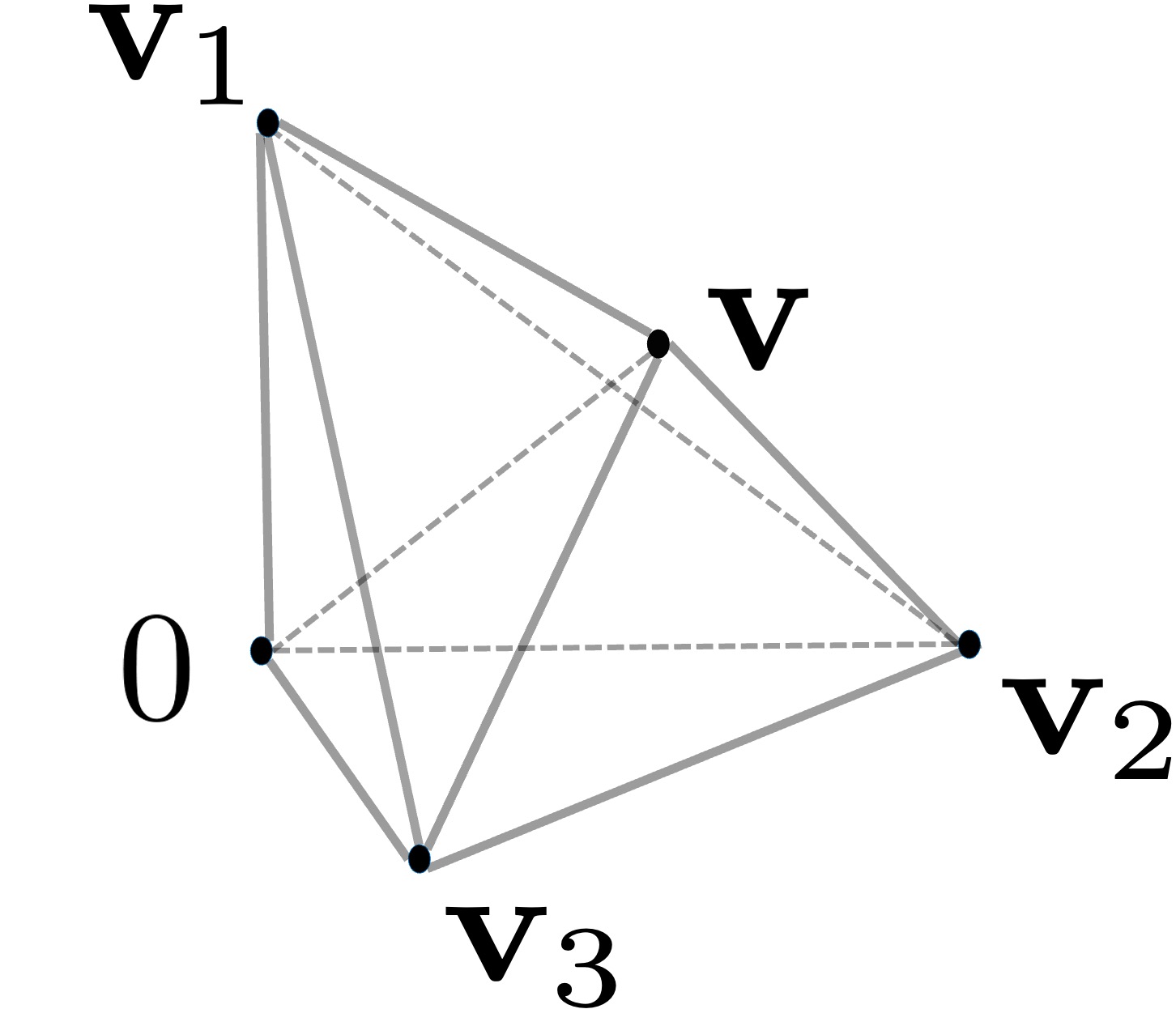}&
				\includegraphics[width = 0.25\textwidth]{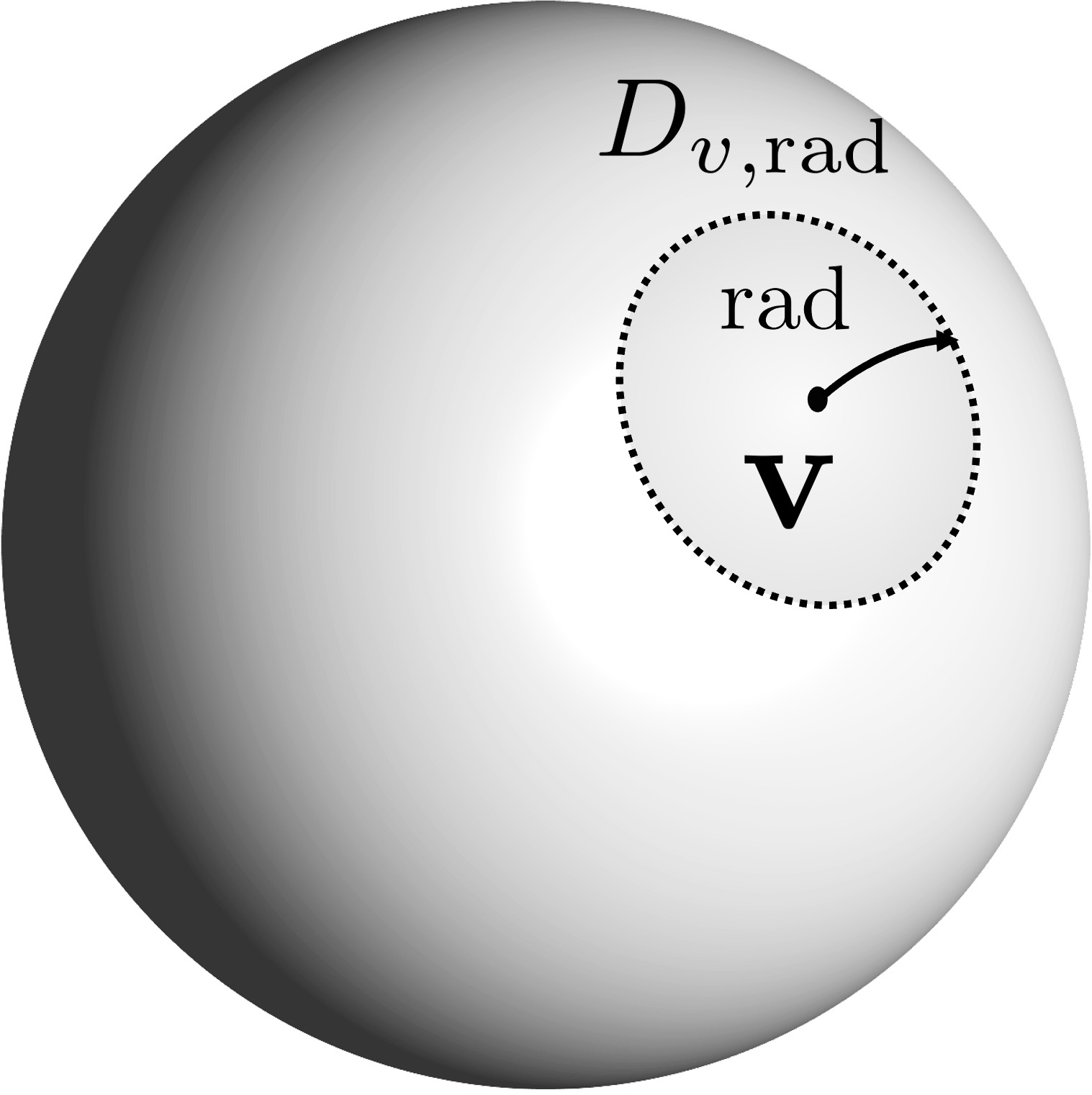}&
				\includegraphics[width = 0.25\textwidth]{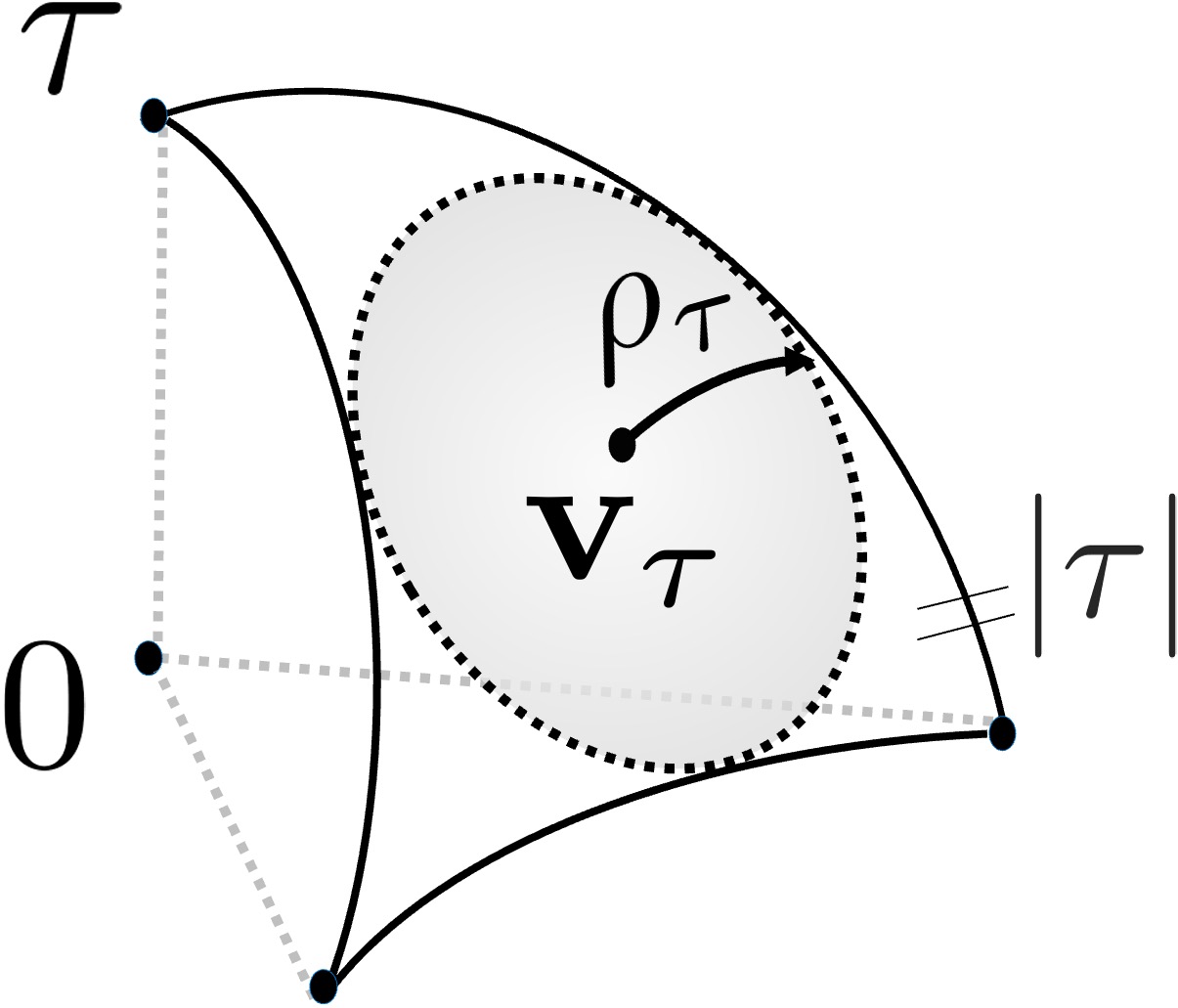}\\
				(a)& (b) & (c)& (d)
			\end{tabular}
		}
		\caption{Illustrations of (a) a spherical triangle $\tau$, (b) tetrahedrons related to $\mathbf{v}$, (c) a spherical cap $D_{v, \mathrm{rad}}$, and (d) an inscribed cap of spherical triangle $\tau$.}
		\label{fig:tetra}
	\end{figure}

	Further, we define the spherical barycentric coordinates $b_1(\mathbf{v}), b_2(\mathbf{v}), b_3(\mathbf{v})$ for any point $\mathbf{v} \in \mathbb{S}^2$ relative to a nondegenerate spherical triangle $\tau = \langle \mathbf{v}_1, \mathbf{v}_2, \mathbf{v}_3 \rangle$, such that $\mathbf{v} = b_1(\mathbf{v}) \mathbf{v}_1 + b_2(\mathbf{v}) \mathbf{v}_2 + b_3(\mathbf{v}) \mathbf{v}_3$. Note that $b_{\kappa}(\mathbf{v}) = \mathrm{vol}(\uptau_{\kappa})/\mathrm{vol}(\uptau)$ is unique, where $\uptau_{\kappa}$ is tetrahedron $\langle \mathbf{0}, \mathbf{v}, \mathbf{v}_{\kappa+1}, \mathbf{v}_{\kappa+2} \rangle$, and $\uptau$ is tetrahedron $\langle \mathbf{0}, \mathbf{v}_1, \mathbf{v}_2, \mathbf{v}_3 \rangle$. Here, $\mathbf{0}$ represents the center of a sphere; see Figure \ref{fig:tetra}. This holds for $\kappa = 1, 2, 3$, where $\mathbf{v}_2 = \mathbf{v}_5$. 
	
	For integers $d > r > 0$, let $C^r(\Omega)$ denote the space of $r$-th continuously differentiable functions, and $\mathcal{H}_d$ represent the space of homogeneous trivariate polynomials of degree $d$ \citep{alfeld1996}. Following \cite{Baramidze:Etal:06} and \cite{LaiEtal:2008}, we define the \emph{space of homogeneous spherical splines} as $\mathcal{S}_d^r(\triangle) = \{s \in C^r(\Omega), s|_\tau \in \mathcal{H}_d, \tau \in \triangle\}$, where $s|_\tau$ refers to the polynomial piece of spline $s$ restricted on triangle $\tau$. Using spherical barycentric coordinates, we can construct a basis for $\mathcal{H}_d$, called \emph{spherical Bernstein-B\'{e}zier (SBB) polynomials}: $B_{ijk}^{\tau}(\bb{x}) = \frac{d!}{i!j!k!} b_1(\bb{x})^i b_2(\bb{x})^j b_3(\bb{x})^k$, $i+j+k = d$, $\tau \in \triangle$. Consequently, any $s \in \mathcal{H}_d $ can be expressed as $s(\bb{x}) = \sum_{\tau\in\triangle} \sum_{i+j+k = d} \gamma_{ijk}^\tau B_{ijk}^\tau(\bb{x})$, where $\gamma_{ijk}^\tau$'s are referred to as \textit{B-coefficients}. Since $\gamma_{ijk}^\tau$'s are linearly independent, the dimension of $\mathcal{H}_d(\tau)$ is $(d+2)(d+1)/2$, as seen in Definition 13.17 of \cite{Lai:Schumaker:07}. Moreover, a stable local basis ${B_\xi, \xi \in \mathcal{M}}$ for $\mathcal{S}_d^r(\triangle)$ can be constructed, with $\mathcal{M}$ being the index set of SBB bases. For more details, see Theorem 14.25 in \cite{Lai:Schumaker:07}, and refer to the Supplementary Material \Cref{ssec:prelimSBB} for detailed illustrations.
	
	To describe the local properties of a domain on the sphere, we refer to the concept of  ``spherical cap'' as introduced in \cite{neamtu2004}. A spherical cap of radius $\mathrm{rad}$ associated with a point $\mathbf v \in \mathbb{S}^2$ is defined as the set of all points that has a geodesic distance to $\mathbf v$ at most $\mathrm{rad}$. For a spherical triangle $\tau\in\triangle$, we define the \emph{inscribed cap} as the largest spherical cap contained within $\tau$. The center $\mathbf{v}_\tau$ of the inscribed cap is then referred to as the \emph{incenter} of $\tau$, and the radius $\rho_\tau$ of the inscribed cap is denoted as the \emph{inradius} of $\tau$. The longest spherical edge of $\tau$ is represented by $|\tau|$; see Figure \ref{fig:tetra}. Denote by $|\triangle| = \max_{\tau\in\triangle} |\tau|$ the triangulation size corresponding to the longest spherical edge of all the triangles in $\triangle$. Let $\rho_\triangle$ denote the smallest inradius in $\triangle$, i.e., and $\rho_\triangle = \min_{\tau\in\triangle}\rho_\tau$. In addition, we define $A_\tau$ as the area of the spherical triangle $\tau$. To ensure all triangles in $\triangle$ have comparable sizes, we assume $\triangle$ satisfies Assumption \ref{C2} described in Section \ref{sec:theory} below.
	
	\subsection{Penalized spline estimators} \label{ssec:PSE}
	
	In this section, we present a penalized spline estimator of the regression function $m(\cdot)$ in Model (\ref{model}). To construct the penalized spline estimator, we first quantify the roughness of $f: \Omega\to\R$. Let $D^{\bs{\alpha}} f:= D_{x_1}^i D_{x_2}^j D_{x_3}^k f$ be a partial derivative of degree $|\bs{\alpha}| = i + j + k$ for $f$. The homogeneous extension of $f: \Omega\to\R$ of degree $p$, denoted as $f_p: \R^3 \to \R$, is defined as \begin{equation}
		f_p(\bb{x}) := \|\bb{x}\|^p f(\bb{x}/\|\bb{x}\|), ~\bb{x} \in \R^3 \backslash\{\mathbf{0}\}
		\label{EQN:fextension}
	\end{equation}
	for any integer $p$, where $\|\bb{x}\|$ is the Euclidean norm of $\bb{x}$. We define energy functional of $f$ as 
	\begin{equation}
		\mathcal E(f) = \sum_{|\bs\alpha| = 2}\int_{\Omega}|D^{\bs\alpha} f_{p}|^2 d\mu ,   
		\label{EQN:energy1}
	\end{equation}
	where $\mu$ is Lebesgue measure on $\mathbb{S}^2$ and the extended function $f_p$ is restricted to spherical domain $\Omega$ for integration. It is worth noting that any fixed value of $p$ can be chosen as the degree of extension. In our implementation, following \cite{LaiEtal:2008}, we set $p = 1$ for odd $d$ and $p = 0$ for even $d$.
		
	We propose the following \emph{Triangulated Spherical Spline Smoothing (TSSS)} estimator for $m$, defined as the minimizer of the following objective function: 
	\begin{equation}\label{EQN:tsss}
		\widehat m_{\lambda} = \argmin_{s\in\mathcal{S}_d^r(\triangle)} \sum_{i = 1}^n \{Y_i - s(\mathbf{X}_i)\}^2 + \lambda \mathcal{E}(s), 
	\end{equation}
	where $\mathcal{S}_d^r(\triangle)$ is the space of homogeneous spherical splines defined in Section \ref{ssec:triangulation}. The parameter $\lambda$ is a non-negative tuning parameter that balances the trade-off between the goodness of fit to the data and the smoothness of the estimate. Penalizing the energy functional with $\lambda \mathcal{E}(s)$ for some $\lambda > 0$ reduces the roughness of the estimator, since higher values of $\mathcal{E}(s)$ indicate less smoothness.

	Note that every function $s \in \mathcal{S}_d^r(\triangle)$ defined on a spherical triangle $\tau$ can be expressed as a linear combination of the SBB basis functions $\mathbf{B}^\tau(\bb{x}) = (B_{\xi}^\tau(\bb{x}), \xi\in\mathcal{M})^\top$ using the coefficients $\bs{\gamma}^\tau = (\gamma_{\xi}^\tau, \xi\in\mathcal{M})^\top$. Specifically, for any point $\bb{x} \in \tau$, we have $s(\bb{x}) = \bs{\gamma}^{\tau\top} \mathbf{B}^\tau(\bb{x})$. Then, the energy functional $\mathcal{E}(s)$ in (\ref{EQN:energy1}) can be rewritten as 
	\begin{equation}
		\mathcal{E}(s) = \sum_{\tau \in \triangle}
		\int_{\tau} \left\{\diamondsuit(\bs{\gamma}^{\tau\top}\mathbf{B}^\tau(\bb{x}))\right\}^{\otimes 2}d\mu
		= \sum_{\tau \in \triangle}
		\int_{\tau} \bs{\gamma}^{\tau\top}\left\{\diamondsuit(\mathbf{B}^\tau(\bb{x}))\right\}^{\otimes 2}\bs{\gamma}^{\tau}d\mu, 
		\label{EQN:energy2}
	\end{equation}
	where $\diamondsuit f := (D^{\bs\alpha} f_p, |\bs{\alpha}|=2)$ is the second order derivative vector of $f \in \mathcal{S}$ given in (\ref{EQN:fextension}), $\bb{x}^{\otimes 2} = \bb{x} \bb{x}^\top$.
	
	Since the spline coefficients of $s \in {\cal{C}}^r(\triangle)$ need to satisfy some smoothness conditions across each interior face of $\triangle$, a smoothness constraint matrix $\mathbf{M}$ can be used to impose the required smoothness. Specifically, $\mathbf{M}\bs{\gamma} = \bs{0}$ is imposed as a smoothness condition to ensure that the estimated function is $r$-th differentiable on all edges of $\triangle$. By doing so, the smoothness of the TSSS estimator can be controlled and achieved for different values of $r$. For spline function $\mathbf{B}(\bb{x})^{\top}  \bs{\gamma}$ we have $\mathcal{E}(\mathbf{B}^{\top} \bs{\gamma}) = \bs{\gamma}^{\top} \mathbf{P} \bs{\gamma}$, where $\mathbf{P}$ is a $\{N(d+1)(d+2)/2\}\times \{N(d+1)(d+2)/2\}$ block diagonal penalty matrix and can be obtained from (\ref{EQN:energy2}). Thus, one can estimate the spline coefficients via minimizing
	\begin{equation*}
		\widehat{\bs{\gamma}} = \argmin_{\bs{\gamma}} \sum_{i=1}^n\{Y_i - \mathbf{B}(\mathbf{X}_i)^{\top} \bs{\gamma}\}^2 + \lambda \bs{\gamma}^{\top} \mathbf{P} \bs{\gamma}, \mathrm{~subject~to~} \mathbf{M}\bs{\gamma} = \mathbf{0}.
	\end{equation*}
	Then, we can obtain the TSSS estimator by $\widehat{m}_\lambda (\mathbf{x})= \mathbf{B}^\top (\mathbf{x}) \widehat{\bs{\gamma}}$.

	\section{Theoretical Results}
	\label{sec:theory}
	
	To discuss the asymptotics of the TSSS estimator, we first introduce some technical notations. Let $X_n$, $n \geq 1$, be a sequence of random variables. We denote $X_n = O_p(a_n)$ if $\lim _{c\to\infty}\limsup_{n\to\infty} P(|X_n| \geq c a_n) = 0$. Similarly, we write $X_n = o_p(a_n)$ if $\lim_{n\to\infty} P(|X_n| \geq c a_n) = 0$ holds for any positive constant $c$. In addition, we denote $a_n\asymp b_n$ if there exists two positive constants $c$ and $C$ such that $c|a_n| \leq |b_n|\leq C|a_n|$. These notations are useful in discussing the rate of convergence and consistency of the TSSS estimator below.
	
	For any measurable functions $f, f_1, f_2$ defined over the closure of $\Omega \subseteq \mathbb{S}^2$, we define the theoretical and empirical inner products: $\langle f_1, f_2 \rangle_\Omega = \E\{f_1(\bb{X})f_2(\bb{X})\}$, $\langle f_1, f_2 \rangle_{n, \Omega} = n^{-1}\sum_{i=1}^n f_1(\bb{X}_i)f_2(\bb{X}_i)$, respectively, with induced norms $\|f\|_{\Omega} = \{\langle f, f\rangle_{\Omega}\}^{1/2}$ and $\|f\|_{n, \Omega} = \{\langle f, f \rangle_{n, \Omega}\}^{1/2}$. We define the supremum norm $\|f\|_{\infty, \Omega} = \sup_{\bb{x} \in \Omega}|f(\bb{x})|$ and the $L^q$-norm $\|f\|_{L^q(\Omega)} = \{\int_{\Omega} | f(\mathbf{x})|^q d \mathbf{x}\}^{1/q}$. Furthermore, denote $\langle f_1, f_2 \rangle_{\mathcal{E}} = 
	\sum_{\tau\in\triangle} \int_\tau (\diamondsuit f_{1, p} )^\top \diamondsuit f_{2,p} d\mu$, where $\diamondsuit f = (D^{\bs{\alpha}} f_p, |\bs{\alpha}| = 2)^\top$ refers to the second-order partial derivatives of $f$. 
	
	Lemmas \ref{LEM:norm1}-\ref{LEM:Vn} presented below provide the relationships between the inner products and norms defined above. We first state some technical assumptions.
	
	\begin{enumerate}[label = (C\arabic*), ref = (C\arabic*), itemsep=-1ex]
		\item The density function of $\mathbf{X}$, $f_{\mathbf{X}}(\cdot)$, is bounded away from zero and infinity over $\Omega$. \label{C1}
		\item For any spherical triangle $\tau$ within $\triangle$, it is contained in a spherical cap of radius $1/6$. The minimal determining set has constant $\curlywedge = 3$; see Supplementary Material Section \ref{SSEC:MDS}.
		Assume triangulation $\triangle$ is $\varrho$-quasi-uniform, that is, there exists a positive constant $\varrho$ such that $|\triangle|/\rho_{\tau} \leq \varrho$, for all $\tau  \in \triangle$. \label{C2}
		\item The number of triangles $N$ and sample size $n$ satisfy $N = Cn^{\omega}$ for some $C>0$ and $0<\omega <\eta/(\eta+2)$, where $\eta$ is defined in \ref{C5}.   \label{C3}
	\end{enumerate}
	
	Assumptions \ref{C1}--\ref{C3} are standard in nonparametric literature \citep{Huang:03b, Lai:Wang:13, Yu:etal:20}. Assumption \ref{C2} suggests the use of more uniform triangulations with smaller shape parameters \citep{BL11}. Assumption \ref{C3} can be handled via the adjustment of triangulation. Considering the area of domain $\Omega$ as fixed, we can interchange $N$ and $|\triangle|^{-2}$ in all results. 
	
	\begin{lemma}\label{LEM:norm1}
		Let $\{B_\xi\}_{\xi\in\mathcal M}$ be the basis for $\mathcal S_d^r(\triangle)$. Let $p_1 = \sum_{\xi\in\mathcal M} \gamma^{(1)}_{\xi} B_{\xi}$, $p_2 = \sum_{\xi\in\mathcal M} \gamma^{(2)}_{\xi} B_{\xi}$. Under Assumptions \ref{C1} and \ref{C3}, we have 
		\[
		R_n = \sup_{p_1, p_2 \in \mathcal S_d^r(\triangle)} \left|\frac{\langle p_1, p_2\rangle_{n,\Omega} - \langle p_1, p_2\rangle_{\Omega}}{\|p_1\|_\Omega \|p_2\|_\Omega} \right| = O_p \left\{ (N\log n)^{1/2} n^{-1/2} \right\}.
		\]
	\end{lemma}

	\begin{lemma} \label{LEM:Vn}
		Under Assumptions \ref{C1}--\ref{C3}, we have
		\[
		V_n \!=\! \!\sup_{p\in S_d^r(\triangle)} \!\!\left\{
		\frac{\|p\|_{\infty, \Omega}}{\|p\|_{n, \Omega}},
		\|p\|_{n, \Omega}\neq 0
		\right\}
		{= O_p(\sqrt{N})}, 
		\bar V_n \!=\! \! \sup_{p\in S_d^r(\triangle)} \!\!\left\{
		\frac{\|p\|_{\mathcal E}}{\|p\|_{n, \Omega}},
		\|p\|_{n, \Omega}\neq 0
		\right\}
		{=O_p(N).}
		\]
	\end{lemma}

	\subsection{Convergence Rate of TSSS Estimator} 
	\label{SEC:convergence}
	
	Before presenting the convergence results of the TSSS estimator, it is important to discuss the construction of the spherical Sobolev space and its related norms. The convergence rate of the estimator depends on the smoothness of the underlying regression function, and understanding the properties of the spherical Sobolev space is essential for understanding this relationship.
	
	For an open set $\bar \Omega \subseteq \mathbb R^2$, let $W^{\ell}_q(\bar \Omega)$ denote the classical Sobolev space as in Section 1.6 of \cite{Lai:Schumaker:07}, and $\|f\|_{W^{\ell}_q(\bar \Omega )} = \sup_{|\ba| =\ell}\|D^{\ba} f\|_{L^q(\bar \Omega)}$ denote the classical Sobolev norm. To work on an appropriate space for functions defined on $\Omega \subseteq \mathbb{S}^2$, we consider the following Sobolev space on the sphere 
	$W^{\ell}_q(\Omega) := \left\{
	f: (\alpha_j^* f)\circ \phi_j^{-1} \in W^{\ell}_q(\bar \Omega_j), \forall j\right\}, \, \ell >0$, 
	where $f: \Omega \to \mathbb R$ is a spherical function; $\phi_j^{-1}:\mathbb R^2\to {\mathbb{S}^2}$ is a smooth mapping; $\bar \Omega_j \subseteq \mathbb R^2$ is an open set and the support of $\phi_j^{-1}$; and $\alpha_j^*: {\mathbb{S}^2}\to \mathbb R$ is a mapping to indicate the partition of $\mathbb{S}^2$.
	Thus, $(\alpha_j^* f)$ and $\phi_j^{-1}$ bridge $\mathbb R^2$ and $\mathbb R$,
	and induce the \emph{spherical Sobolev norm} 
	$\|f\|_{\ell,q,\Omega} = \sum_j\|(\alpha_j^* f)\circ \phi_j^{-1}\|_{W^{\ell}_q(\bar \Omega_j)}$ for $\ell >0$. In addition, the \emph{spherical Sobolev seminorm} is defined as
	$|f|_{\ell,q, \Omega} = \sum_{|\ba| = \ell} \|D^{\ba} f_{\ell-1}\|_{L^q(\Omega})$, where $\|D^\ba f_{\ell-1}\|_{L^q(\Omega)}$ is the $L^q$-norm of the extended trivariate function $f_{\ell-1}$ restricted to $\Omega$.
	When $\ell = 0,$ the seminorm $|f|_{0, q,\Omega}$ reduces to $L^q$-norm $\|f\|_{L^q(\Omega)}$. 
	
	We introduce two additional assumptions. 
	\begin{enumerate}[label = (C\arabic*), ref = (C\arabic*), itemsep=-1ex]
		\setcounter{enumi}{3}
		\item The spherical mean function $m\in W^{\ell+1}_q (\Omega)$ for some integer $0 \leq \ell \leq d, \ell = d(\mathrm{mod}2)$. \label{C4}
		
		\item The noise $\epsilon$ satisfies $\lim_{\eta\to \infty} \E\{\epsilon^2 I(\epsilon>\eta)\}=0$, and $\E|\epsilon_i^{2+\eta}| \leq v_{\eta}$, $\eta>0$. The standard deviation function $\sigma(\bb x)$ is continuous, and $0<c_\sigma \leq \inf_{\mathbf x\in\Omega} \sigma(\mathbf x)\leq \sup_{\mathbf x\in\Omega} \sigma(\mathbf x)\leq C_\sigma <\infty$. \label{C5}
	\end{enumerate}
	Assumption \ref{C4} specifies the degree of smoothness required for $m$ \citep{BL05, BL11}. By characterizing the smoothness of the function using the Sobolev space and its norms, we can obtain the convergence rate of the estimator for different smoothness levels of the regression function. Assumption \ref{C5} places constraints on the behavior of the noise and the standard deviation function, which helps establish the asymptotic normality of the TSSS estimator \citep{Huang:03b, Lai:Wang:13}.
	
	Define $s_{\lambda, m}$ and $s_{\lambda,\epsilon}$ as the penalized spline estimators based on $\{m(\mathbf X_i)\}_{i=1}^n$ and $\{\sigma(\mathbf X_i)\epsilon_i\}_{i=1}^n$, respectively. Let $\mathbb{B}= \sum_{i=1}^n \mathbf{B}(\mathbf{X}_i) \mathbf{B}(\mathbf{X}_i)^{\top}$, then
	\begin{align}
		s_{\lambda, m} (\bb{x})&= \mathbf{B}(\bb{x})^{\top}\bs{\gamma}_{\lambda, m}, ~\bs{\gamma}_{\lambda, m} = \left(\mathbb{B} + \lambda \mathbf P \right)^{-1} \sum_{i=1}^n\mathbf{B}(\mathbf{X}_i) m(\mathbf{X}_i), \label{DEF:slamm}\\
		s_{\lambda, \epsilon} (\bb{x})&= \mathbf{B}(\bb{x})^{\top}\bs{\gamma}_{\lambda, \epsilon}, ~\bs{\gamma}_{\lambda, \epsilon} =  \left(\mathbb{B} + \lambda \mathbf P \right)^{-1}  \sum_{i=1}^n\mathbf{B}(\mathbf{X}_i) \sigma(\mathbf X_i) \epsilon_i. \label{DEF:slame}
	\end{align}
	
	To evaluate the bias and variance of the proposed estimator, we decompose the estimation error as follows: $\widehat m _{\lambda}(\bb{x}) - m(\bb{x}) = \{s_{\lambda, m}(\bb{x}) - m(\bb{x})\} + s_{\lambda, \epsilon}(\bb{x})$, where the first term represents the size of bias and the second term represents the size of variance in the estimation. The following propositions provide bounds for the order of the bias and variance terms.

	\begin{prp}
		Under Assumptions \ref{C1}--\ref{C4}, if $d\geq 3r +2$ and $\triangle$ is a spherical triangulation such that $|\triangle|<1/6$, then  we have the following uniform convergence rate: 
		$
		\|s_{\lambda, m} -m \|_{\infty, \Omega} = O_p\{
		\lambda n^{-1}N^{3/2}|m|_{2,\infty,\Omega} +
		( 1+ \lambda n^{-1} N^{5/2} )N^{-(\ell + 1)/2} |m|_{\ell+1, \infty, \Omega} \}
		$.
		\label{prop1}
	\end{prp} 
	
	\begin{prp}
		Under Assumptions \ref{C1}--\ref{C3} and \ref{C5}, we have ${\|s_{\lambda, \epsilon}\|_{L^2(\Omega)} = O_p(n^{-1/2}N^{1/2})}$ and ${\|s_{\lambda, \epsilon}\|_{\infty,\Omega} 
			= O_p\{n^{-1/2}(N\log n)^{1/2} +  \lambda n^{-3/2} N^3 \}}$.
		\label{prop2}
	\end{prp} 
	
	Combining Propositions \ref{prop1} and \ref{prop2}, we can obtain the following convergence rates of the TSSS estimator in terms of both $L^2$ and supremum norms in Theorem \ref{THM:convergence}. These convergence rates are important for assessing the performance of the TSSS estimator under different smoothness assumptions and for guiding the choice of the triangulation and spline basis functions.
	
	\begin{theorem}
		\label{THM:convergence}
		Under Assumptions \ref{C1}--\ref{C5}, we have 
		\begin{align*}
			\|\widehat m_\lambda -m\|_{L^2(\Omega)} =
			&O_p\left\{
			\frac{\lambda N^{3/2}}{n}|m|_{2,\infty,\Omega} + 
			\left( 1+ \frac{\lambda{N^{5/2}}}{n} \right)N^{-(\ell + 1)/2} |m|_{\ell+1, \infty, \Omega} +{\sqrt{\frac{N}{n}}}
			\right\},\\
			\|\widehat m_\lambda -m\|_{\infty,\Omega} 
			=&O_p\left\{
			\frac{\lambda{N^{3/2}}}{n}|m|_{2,\infty,\Omega} +
			\left( 1+ \frac{\lambda{N^{5/2}}}{n} \right)N^{-(\ell + 1)/2} |m|_{\ell+1, \infty, \Omega}  \right.\\
			&\hspace{7cm} \left.+ {\sqrt{\frac{{N}\log n }{n}} +   \frac{\lambda {N^{3}}}{n^{3/2}}}\right\}.
		\end{align*}
	\end{theorem}

	According to Theorem \ref{THM:convergence}, the convergence rate of the TSSS estimators depends on sample size, triangulation size, and the roughness penalty parameter, as well as the characteristics of the underlying signal being estimated. The first and the third terms in the order of  $\|\widehat m_\lambda -m\|_{L^2(\Omega)}$ and $\|\widehat m_\lambda -m\|_{\infty,\Omega}$ show the bias brought by the roughness penalty. When the tuning parameter is small enough, the second term in the order of $\|\widehat m_\lambda -m\|_{L^2(\Omega)}$ and $\|\widehat m_\lambda -m\|_{\infty,\Omega}$ represent the bias when approximating an arbitrary function $m$ by a spherical spline. The last two terms are the estimation variance generated from the random noise.
	
	Theorem \ref{THM:convergence} also provides a guideline on how to choose the triangulation. As the sample size increases, a finer triangulation can be considered for a more accurate estimation of the mean function. In addition, a finer triangulation is needed when there are rapid changes in the mean function or when the domain is highly curved or complex. A more detailed discussion of triangulation selection is given in Section \ref{sec:implementation}.
	
	It can be demonstrated that the TSSS estimator achieves the optimal convergence rate under certain regularity conditions. As indicated in \cite{stone1982optimal}, the optimal global approximation rate for a nonparametric estimator can be expressed as follows:  for the $L^q$-norm ($0<q<\infty$) of the approximation error, the rate is $n^{-r^*}$, where $r^* = p^*/(2p^*+d^*)$, for a $p^*$-times differentiable function of a $d^*$-dimensional measurement variable; for $q = \infty$, the optimal convergence rate becomes $(n^{-1}\log n)^{r^*}$.  In the case of our problem,  where $\bb x\in \Omega \subseteq \mathbb{S}^2$, we have $d^*=2$. Furthermore, the mean function $m\in W_q^{\ell+1}(\triangle)$ implies that it is $p^*=(\ell+1)$-times differentiable. Therefore, the optimal global approximation rate is $n^{-(\ell+1)/(2\ell +4)}$ when $0<q<\infty$, and  $(n/\log n)^{-(\ell+1)/(2\ell +4)}$ when $q=\infty$.  When $\lambda = 0$, the TSSS achieves the optimal convergence rate in $L^2$-norm when the number of triangles satisfies the condition $N\asymp n^{1/(\ell + 2)}$, and it achieves the optimal convergence rate in the supremum norm when ${N\asymp (n/\log n )^{1/(\ell + 2)}}$. When $\lambda>0$, the optimal convergence rate in $L^2$-norm is achieved under the conditions ${\lambda = O\{n^{\ell/(2\ell + 4)}\}}$  and {$N \asymp n^{1/(\ell + 2)}$} . Furthermore, when ${\lambda = o\{n^{\ell/(2\ell + 4)}\}}$ and $N \asymp (n/\log n)^{1/(\ell + 2)}$, the TSSS estimator achieves the optimal convergence rate in the supremum norm.

	\subsection{Asymptotic Normality}
	\label{SEC:normality}
	
	To derive the asymptotic distribution of the TSSS estimator, we further assume the following: 
	\begin{enumerate}[start = 5, label = (C\arabic*'), ref = (C\arabic*')]
		\item \label{C5'} The number of spherical triangles $N = Cn^\omega$, for some $C>0$ and $1/(\ell + 2) < \omega< \eta/(\eta+2)$.
	\end{enumerate} 
	\begin{enumerate}[start = 6, label = (C\arabic*), ref = (C\arabic*)]
		\item The roughness parameter satisfies that 
		$\lambda = o(n^{1/2}N^{-1} \wedge n N^{-2})$. 
		\label{C6} 
	\end{enumerate} 
	
	Compared with Assumption \ref{C3} in Section \ref{SEC:convergence}, \ref{C5'} further assumes that the number of spherical triangles needs to be greater than a lower bound, which depends on the degree of the function. A similar assumption for the univariate case and bivariate case has been discussed \citep{Li:Ruppert:08, Lai:Wang:13}. Meanwhile, Assumption \ref{C6} requires smaller $\lambda$, which reduces the bias through undersmoothing. 
	
	Theorem \ref{THE:normal} below states the asymptotic normality of the proposed TSSS estimator. 
	\begin{theorem}
		\label{THE:normal}
		Under Assumptions \ref{C1},\ref{C2}, \ref{C4}, \ref{C5}, \ref{C5'} and \ref{C6}, as $n\to \infty$, for each $\bb{x}\in \Omega$, 
		$\left[\var\left\{\widehat m_{\lambda}(\bb{x})|\mathbb X\right\}\right]^{-1/2} \left\{\widehat m_\lambda (\bb{x})- m(\bb{x})\right\} \stackrel{d}{\to}N(0,1)$, 
		where $\mathbb X$ is the collection of observed $\bb{X}_1, \ldots, \bb{X}_n$.
	\end{theorem}
	
	Theoretically, the above asymptotic distribution result can be used to construct asymptotic confidence intervals. However, it is challenging to obtain the exact form of the standard error for general TSSS estimators due to the characteristics of the trivariate spline basis functions. To overcome this, we propose using a wild bootstrap method \citep{mammen:1993, hall:2013} to estimate the standard errors and quantify the uncertainty of the estimators, as outlined in Algorithm \ref{alg:uncertain}. Note that in the algorithm, $\delta_i$ has mean zero and enforces $\epsilon_i^*$ to have mean zero. In addition, $\delta_i$ introduces a controlled level of randomness to the process of bootstrap residuals and improves the convergence rate, as demonstrated in  Table 1 of \cite{mammen:1993}.
	
	\begin{algorithm}
		\caption{Bootstrap estimation of TSSS estimator's standard error}
		\label{alg:uncertain}
		\textbf{Input: } $\{(\mathbf{X}_i, Y_i)\}_{i = 1}^{n}$.
		\textbf{Output: } The standard deviation of  TSSS estimator $\widehat m(\bb x), \bb x\in \Omega$.\\
		\textbf{Step 1:} Generate TSSS estimator $\widehat m$, and calculate residuals $\{\widehat \epsilon_i = m(\bb X_i) - \widehat m(\bb X_i)\}_{i = 1}^n$. \\
		\textbf{Step 2:} Generate bootstrap samples of residuals $\{\epsilon_i^*  = \delta_i \widehat \epsilon_i\}_{i = 1}^n$, where
		\[
		\delta_i = \begin{cases}
			\frac{1-\sqrt{5}}{2} & w.p. ~ \frac{5+ \sqrt{5}}{10}\\
			\frac{1+\sqrt{5}}{2} & w.p. ~\frac{5- \sqrt{5}}{10},
		\end{cases}
		\] and define $Y^*_i = \widehat Y_i + \epsilon^*_i.$\\
		\textbf{Step 3:} Estimate TSSS estimator $\widehat m^*$ from $\{(\mathbf X_i, Y_i^*\}_{i =1}^n$. \\
		\textbf{Step 4:} Repeat Step 2 and Step 3 for $B$ times, and denote the TSSS estimators using bootstrap samples as $\{\widehat m_b^*\}_{b =1}^B$. Calculate the standard deviation of $\widehat m$ at $\bb x\in\Omega$ as 
		\[
		\left[\sum_{b=1}^B\frac{1}{B}\{\widehat m^*_b(\bb x) - \overline m^*(\bb x)\}^2\right]^{1/2},
		\]
		where $\overline m^*(\bb x) = B^{-1} \sum_{b =1}^B \widehat m^*(\bb x)$.
	\end{algorithm}
	
	\section{Implementation Details} 
	\label{sec:implementation}
	
	The selection of triangulation, spline basis, and penalty parameters is a critical step in the proposed TSSS method. In this section, we will discuss the strategies for selecting these parameters systematically and effectively.
	
	\subsection{Selection of Triangulation} 
	\label{ssec:implementation-1}
	
	Triangulation, which constructs a mesh that accurately captures the shape of the domain, is a critical step of the proposed TSSS method. Several factors influence the goodness level of triangulation required to estimate the regression function accurately. Firstly, the complexity of the domain plays a significant role in determining the fineness of triangulation. Highly curved or complex domains typically require finer triangulation to capture the underlying pattern of the function accurately. However, constructing a suitable triangulation for a complex domain can be challenging, and generating a fine triangulation can be computationally complicated and time-consuming. Usually, triangulation algorithms automatically generate interior vertices of the triangulation. Our triangulation algorithm is able to take any given vertices (both interior and boundary vertices) to form a good triangulation.
	
	Secondly, the characteristic of the underlying function is an essential consideration when selecting the goodness level of triangulation. A function with rapid changes or highly localized features may require a finer triangulation to capture these features accurately. However, to establish a function-dependent triangulation, it is often necessary to have knowledge of the true signal at a substantial number of locations across the domain of interest, which may not be readily available in practical applications. 
	
	Finally, the sample size $n$ can also influence the appropriate goodness level of triangulation. As $n$ increases, more data points become available, and practitioners can consider finer triangulation and/or increase the degree of spline space to obtain a more accurate estimation of the mean function. However, based on our computational experience, it is recommended that each triangle should contain at least a few data observations to ensure a reliable estimation.
	
	Ultimately, the selection of triangulation should balance the need for accuracy with computational efficiency. While it is essential to construct a suitable triangulation for surface-based data analysis, practitioners can have some flexibility in their choice of triangulation without sacrificing estimation performance. Assumption (C3) from Section \ref{sec:theory} indicates that the choice of triangulation has a minimal effect on the performance of the TSSS estimator as long as the triangulation is fine enough to capture the underlying pattern. The TSSS estimator can accurately estimate the regression function even with relatively coarse triangulations, provided that the regression function is not very oscillated, e.g., a  linear polynomial. More complicated regression functions require more complicated triangulation into to approximate them well. Therefore, practitioners need to choose an appropriate level of triangulation that accurately captures the underlying pattern of the function without being too computationally expensive if the regression to be estimated is not very complicated.
	
	In practice, practitioners can make use R packages ``Triangulation'' \citep{Triangulation} and ``INLA'' \citep{lindgren:2015}, and MATLAB algorithm ``DistMesh'' \citep{persson2004} for 2D planar patches treating the boundary of a spherical domain as the boundary of a planar domain $\subseteq [0,\pi] \times [0,2\pi)$, while ensuring boundary points that wrap around the sphere are identical, see triangulation examples constructed using ``Triangulation'' in Simulation Study 2 in Section \ref{sec:simstudy2}. Practitioners can also utilize the Python package ``Differentiable Surface Triangulation'' developed in \cite{rakotosaona2021} for surface-based domains. Another simple approach is to create a uniform triangulation over the whole sphere and select only the spherical triangles that contain observations while ensuring the collection of triangles forms a triangulation satisfying the definition of triangulation in \cite{Lai:Schumaker:07}. Examples of such triangulations are illustrated in Section \ref{sec:data-app}.
	
	\subsection{Selection of Spline Basis Functions and the Roughness Penalty Parameter}
	\label{ssec:implementation-2}
	
	Unlike the selection of triangulation, choosing the spline basis is generally easier to control. The parameters for the spline space $\mathcal{S}_d^r(\triangle)$, namely $d$ and $r$, can be predetermined or selected by the user. As demonstrated in Simulation Study 1 in Section \ref{sec:simstudy1}, a higher degree of spline basis $d$ leads to a more flexible estimator but may also result in overfitting. On the other hand, a lower degree of spline basis may lead to underfitting. In practice, the choice of $d$ and $r$ is closely tied to the intended interpretation of the estimated function. If the goal is to enhance the signal-to-noise ratio for visualization or to suggest a simple parametric model, then a slightly over-smoothed function with a subjectively chosen parameter may be appropriate. However, if the focus is on accurately estimating the regression function and preserving local structures, then a slightly under-smoothed function may be more suitable. 
	
	In practice, we select the degree of spline basis $d$ based on cross-validation (CV), such as $K$-fold CV \citep{Lai:Wang:13, Mu:Wang:Wang:18}, which is a widely used technique for model selection. The smoothing penalty parameter $\lambda$ can also be selected through $K$-fold CV. In $K$-fold CV, the original data is randomly partitioned into $K$ folds, with one fold reserved for validation and the remaining $(K-1)$ folds used to train the model. This process is repeated $K$ times so that each fold is used once for validation. The CV score is then computed as $\mathrm{CV}(d, \lambda) = \sum_{i=1}^n \{Y_i - \widehat{m}_\lambda ^{-k[i]}(\bb{X}_i)\}^2$, where $k[i]$ is the index of the fold that contains the $i$-th observation and $\widehat{m}_{\lambda}^{-k[i]}$ is the estimate of $m$ given $\lambda$ and the data without the $k[i]$-th fold. The value of $d$ and $\lambda$ that minimize $\mathrm{CV}(d,\lambda)$ is then selected. We adopt the 5-fold CV approach throughout our numerical studies.
	
	\section{Simulation Studies}
	\label{sec:simulations}
	
	This section presents an empirical evaluation of the proposed TSSS method and compares its performance to three alternative methods: the ridge kernel regression (Kernel) \citep{Cao:2013}, thin plate splines on the ordinary sphere (TPSOS) \citep{Wahba:1981, Wahba:1982}, and tensor product splines \citep{wood2006low, LycheSchumaker:2000} using spherical coordinates as input (Tensor-Sphere) through three simulation studies. Simulation Study 1 investigates the performance of the proposed TSSS method for functions defined on the whole sphere and explores the effects of different triangulations and smoothing parameters. Simulation Study 2 examines the performance of the TSSS method in addressing the ``leakage'' problem and handling complex domains. Simulation Study 3 focuses on the performance of the proposed bootstrap variance estimator for the TSSS estimator.
	
	We assess the performance of our simulation under various signal-to-noise ratios (SNRs). When considering a fixed standard deviation, we define SNR as the ratio of the sample variance of $\{m(\bb x_j)\}_{j = 1}^{N_g}$ to $\sigma^2$ \citep{Ruppert2002}, where $\{\bb x_j\}_{j=1}^{N_g}$ are grid points defined on domain $\Omega$. For varying standard variation function $\sigma(\bb x)$, we define the SNR as follows:
	$$
	\frac{\int_\Omega \widehat{\text{Var}}(m)d\bb x}{\int_\Omega \sigma(\bb x)^2 d\bb x}
	= \frac{A_\Omega \widehat{\text{Var}}(m)}{\int_\Omega \sigma(\bb x)^2 d\bb x} 
	\approx \frac{\widehat{\text{Var}}(m)}{N_g^{-1}\sum_{j=1}^{N_g} \sigma(\bb x_j)^2},
	$$
	where $\widehat{\text{Var}}(m)$ represents the sample variance of $\{m(\bb x_j)\}_{j = 1}^{N_g}$, and $A_{\Omega}$ denotes the area of $\Omega$. In general, the SNRs in the simulation studies vary from 2 to 8; see Table \ref{tab:snr1} in Supplementary Material for the specific SNRs in each simulation setting.
	
	Throughout all numerical experiments, the TSSS method is implemented in MATLAB, and the smoothness parameter $r$ is fixed to be one while the penalty parameter $\lambda$ is selected by $K$-fold CV introduced in Section \ref{ssec:implementation-2}. The Kernel, TPSOS, and Tensor-Sphere are conducted using the R package ``mgcv''  \citep{Wood:03, mgcv} based on the functions \texttt{magic()}, \texttt{s()} and \texttt{te()}, respectively, along with \texttt{gam()}, and $\lambda$ is selected by generalized cross-validation (GCV) for the competing methods. In addition, for the TPSOS, we choose $m = 2$ to penalize the second derivative smoothness. For both TPSOS and Tensor-Sphere, we fix the dimension for the smooth term as $k = 100$ and $k = 64$, respectively. These dimensions are selected to align closely with the dimension utilized in the TSSS method to ensure a fair comparison.

	\subsection{Simulation Study 1: Functions on Whole Sphere}
	\label{sec:simstudy1}
	
\begin{figure}[!ht]
		\centering
		\begin{subfigure}[b]{0.31\linewidth}
			\centering
			\includegraphics[width = 1 \textwidth]{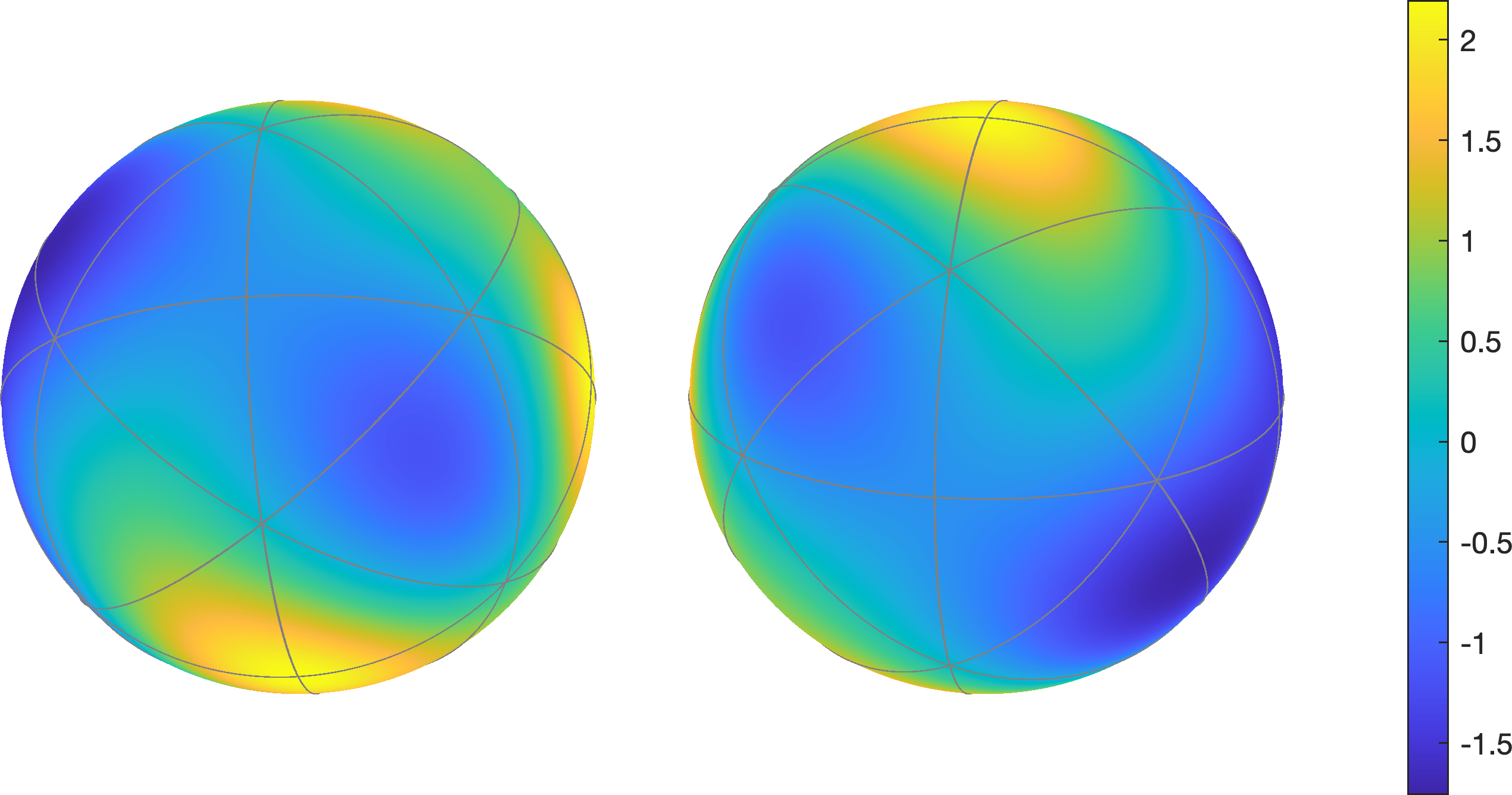}
			\caption{$m_1|{\mathbb{S}^2}$}
		\end{subfigure}
		\begin{subfigure}[b]{0.31\linewidth}
			\centering
			\includegraphics[width = 1 \textwidth]{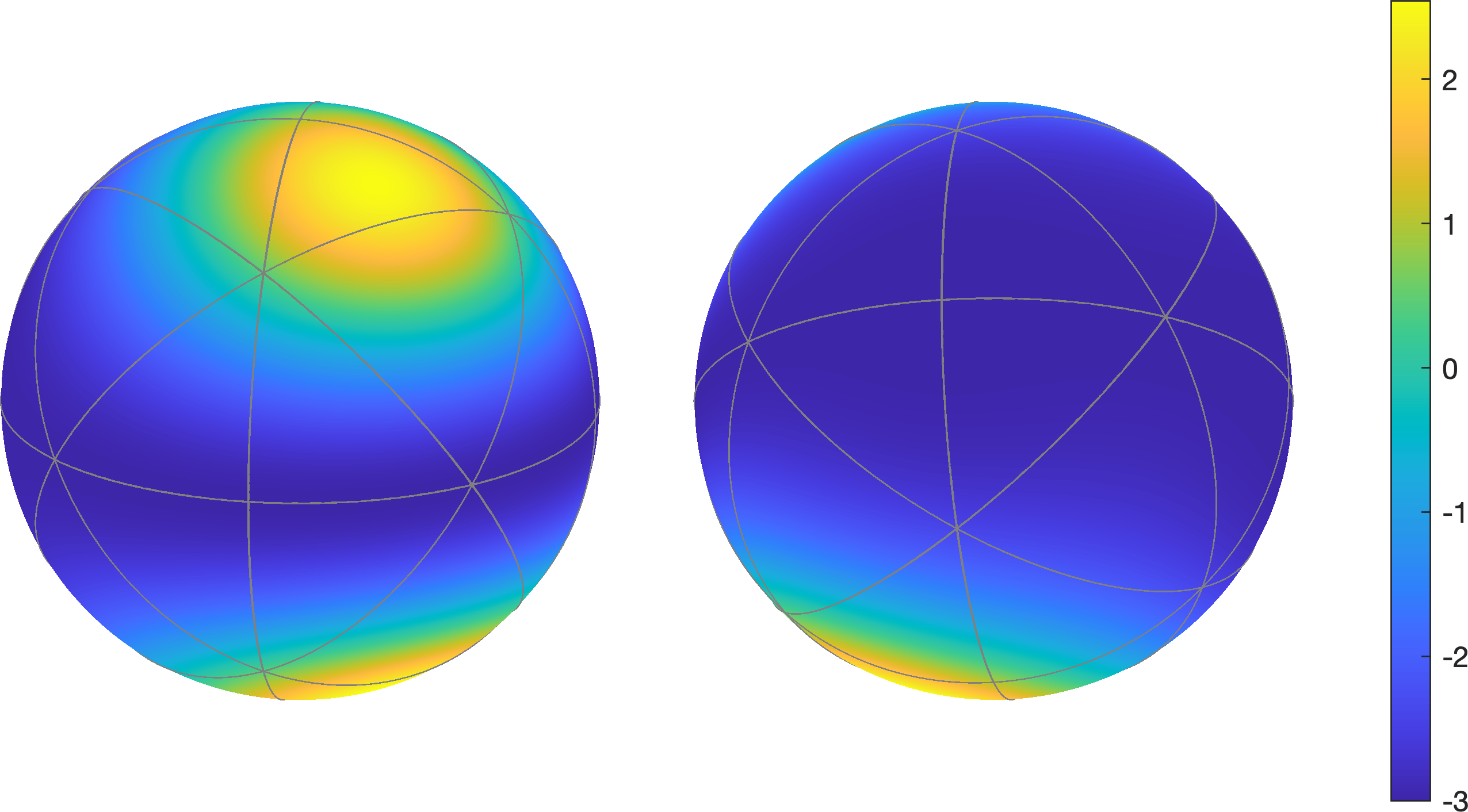}
			\caption{$m_2|{\mathbb{S}^2}$}
		\end{subfigure}
		\begin{subfigure}[b]{0.34\linewidth}
			\centering
			\includegraphics[width = 1 \textwidth]{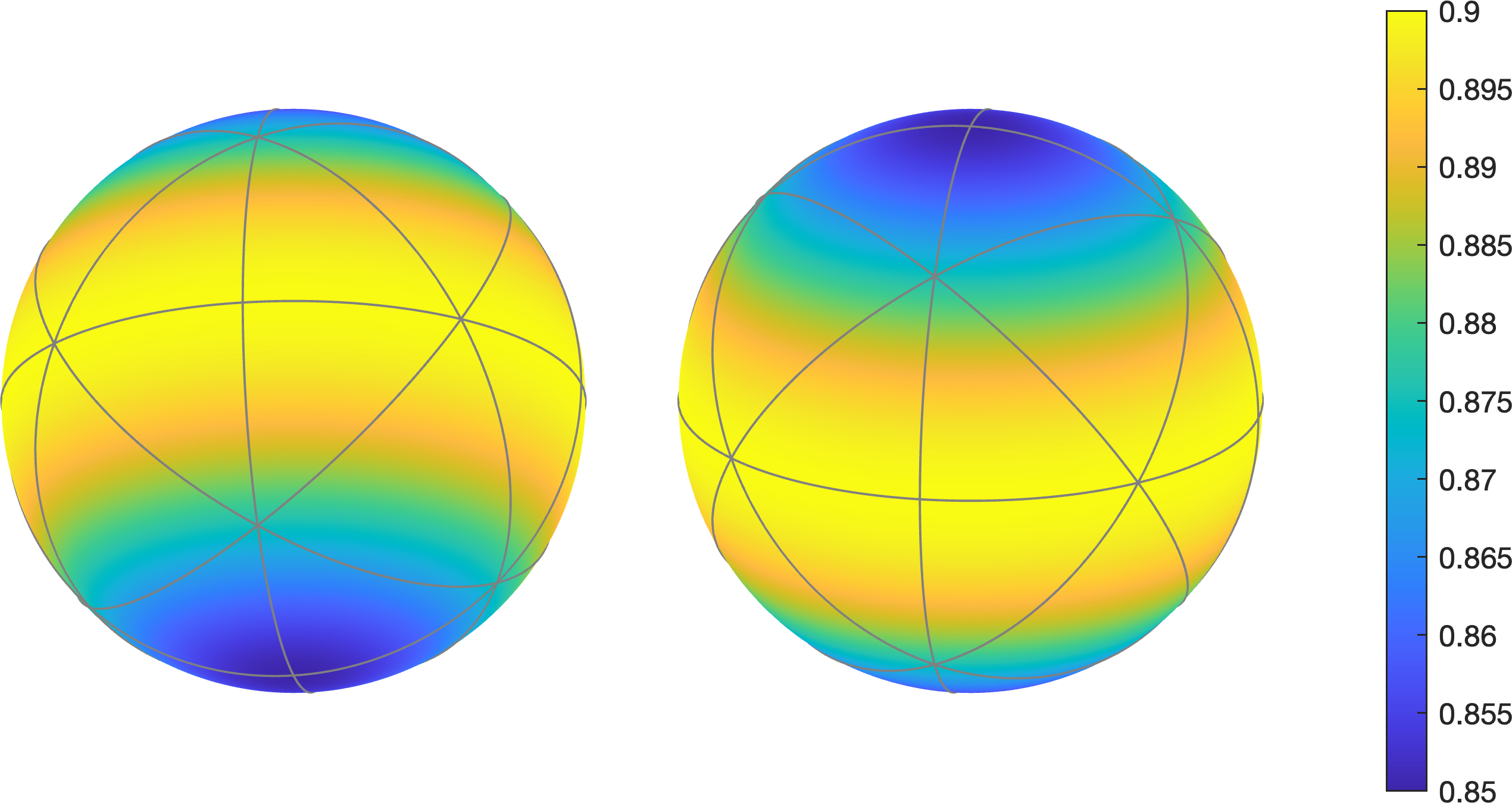}
			\caption{$\sigma_1|\mathbb S^2$}
		\end{subfigure}
		\begin{subfigure}[b]{0.32\linewidth}
			\centering
			\includegraphics[width = 1 \textwidth]{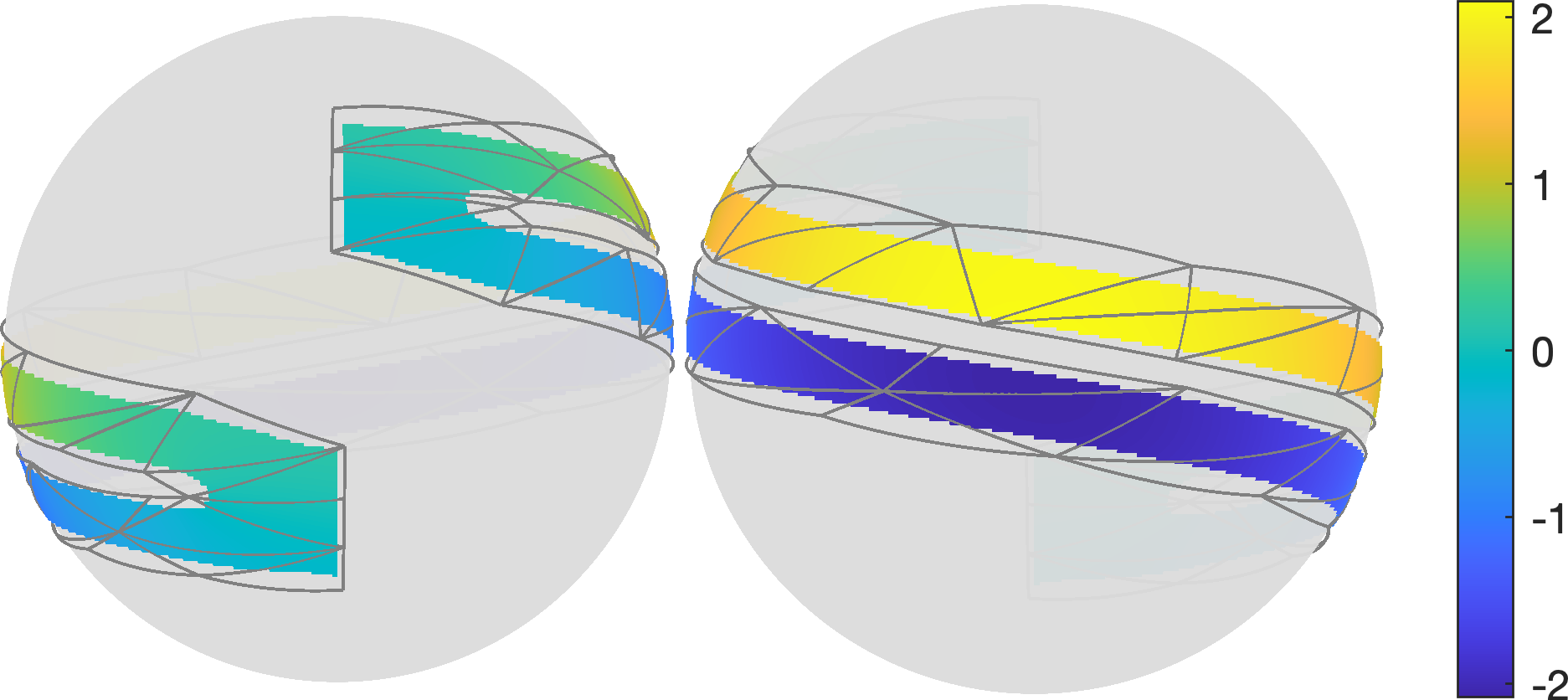}
			\caption{$m_3|{\Omega}$}
		\end{subfigure}
		\begin{subfigure}[b]{0.34\linewidth}
			\centering
			\includegraphics[width = 1 \textwidth]{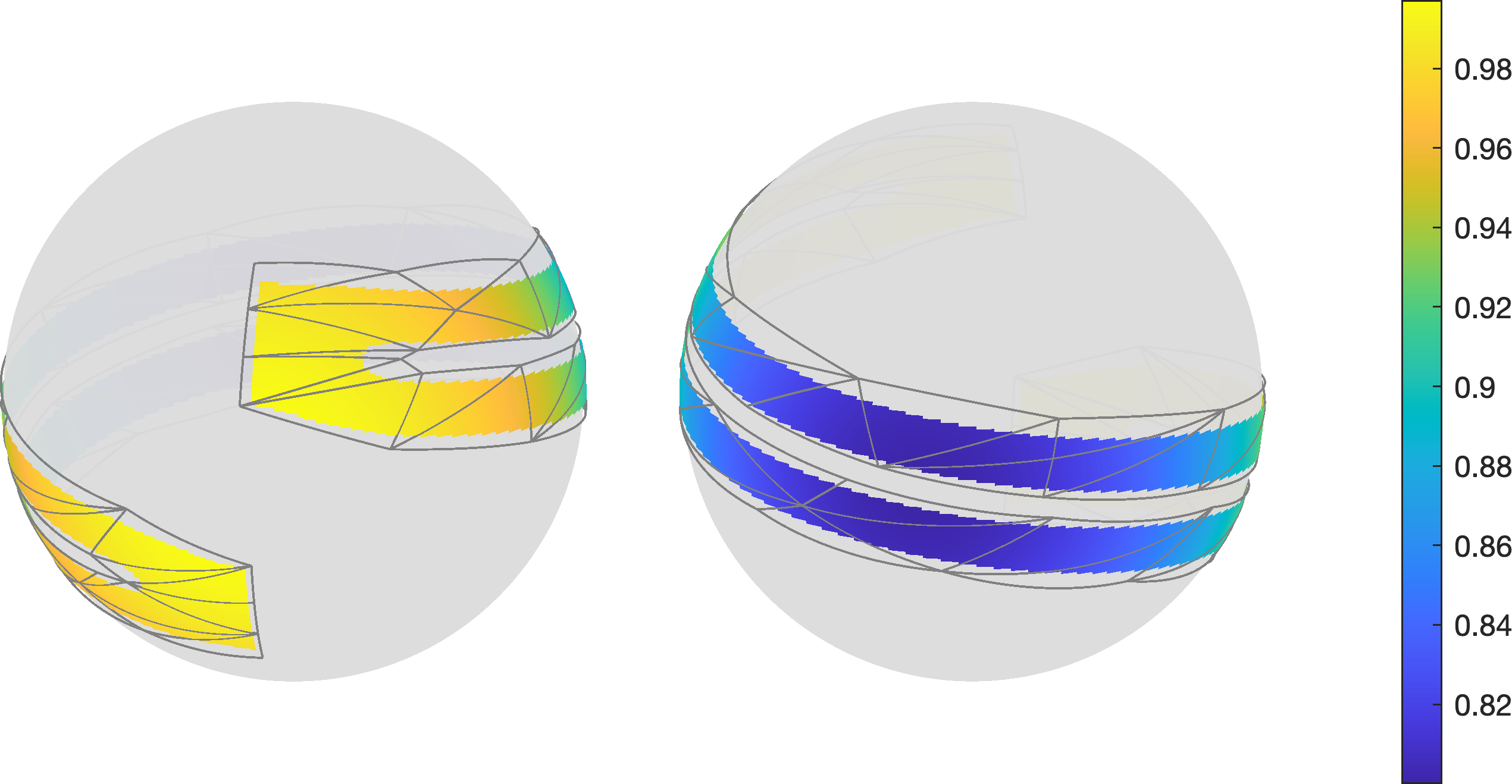}
			\caption{$\sigma_2|\Omega$}
		\end{subfigure}
		\caption{Plots of the true mean and standard deviation functions in simulation studies: (a) $m_1|{\mathbb{S}^2}$ with triangulation ($N=32$) from two viewpoints; (b) $m_2|{\mathbb{S}^2}$ with triangulation ($N=32$) from two viewpoints; (c) $\sigma_1|\mathbb S^2$ with $c_{\sigma}=1$ from two viewpoints; and (d) $m_3 | \Omega$ with triangulation ($N = 58$); and (e) $\sigma_2|\Omega$, with $c_\sigma=1$ from two viewpoints.}
		\label{fig:eg12}
	\end{figure}

	In this simulation, we generate $n$ design points $\bb{X}_i  = (X_{i1}, X_{i2}, X_{i3})^\top \in \mathbb{S}^2$ with $\|\bb{X}_i\| = 1$, for $i=1, \ldots, n$. We generate the response variable $Y_i$ according to the following model:
	\begin{equation}
		Y_i = m(\bb{X}_i) + \sigma_1(\bb{X}_i)\epsilon_i, ~\epsilon_i \overset{i.i.d}{\sim} N(0, 1), ~ i = 1, \ldots, n, 
		\label{SIM:model}
	\end{equation}
	where $m(\bb{x}): \mathbb S^2\to \mathbb R$ represents the mean function, and we consider two mean functions:
	(i) $m_1(\bb{x}) = -2 + 1/2 \{x_1^2 + \exp(2x_2^3) + \exp(2x_3^2) + 10x_1x_2x_3\}$;
	(ii) 
	$m_2(\bb{x}) =2.5\{ (x_1 - 1) (x_2 - 1) x_3^2\} - 3$. 
	The illustration of $m_1$ and $m_2$ is presented in Figure \ref{fig:eg12} (a) and (b). We generate $\bb{X}_i$'s on $n =$ 400, 900, and 2,500 grid locations on a unit sphere for training and evaluate the performance of the estimators at $N_g =$  10,201 grid points $\{\bb{x}_j\}_{j=1}^{N_g}$ on the sphere for each simulation setting, which is fixed across all 100 replications. We consider both constant standard deviation $\sigma_1(\bb{x}) \equiv \sigma$ ($\sigma = 0.50,~0.75$), and heterogeneous standard deviation function $\sigma_1(\bb{x})$, which varies across different spatial locations. Specifically, we define $\sigma_1(\bb{x}) = c_{\sigma} \{1 - (x_1^2 + x_2^2 + 1.5x_3^2)/10\}$, where $c_{\sigma}$ takes values of $0.50$ and $0.75$. A visual representation of the standard deviation function $\sigma_1(\mathbf{x})$ can be found in Figure \ref{fig:eg12} (c). 
	
	\subsubsection{Comparative analysis of TSSS with competing methods}
	\label{ssec:comparison}
	
	In this section, we conduct a comparative analysis between the proposed TSSS method and the Kernel and TPSOS methods. To evaluate the performance of different estimators, we use two metrics: the predicted mean squared error (PMSE) and the training mean squared error (TMSE). The PMSE is calculated on a set of grid points $\{\bb{x}_j\}_{j = 1}^{N_g}$ on the sphere, and is defined as $N_g^{-1} \sum_{j = 1}^{N_g} \{m(\bb{x}_j) - \widehat{m}(\bb{x}_j)\}^2$, where $m$ and $\widehat{m}$ are the true and generic estimated functions. The TMSE is calculated on the training data locations $\{\bb{X}_i\}_{i=1}^n$ and is defined as  $n^{-1}\sum_{i = 1}^n \{m(\bb{X}_i) - \widehat{m}(\bb{X}_i)\}^2$. In addition, we report the dimension of the design parameters (``Dim'') and the average computation time per replication (``Time''), measured in seconds.
	
	Due to the similar results, we only report the results in the heterogeneous standard deviation setting in Table \ref{tab:sim1_vary} and defer the results in the constant standard deviation setting to the Supplementary Material. \Cref{tab:sim1_vary} reports the estimation results for $m_{1}|\mathbb S^2$ and $m_{2}|\mathbb S^2$ when the errors are heterogeneous with varying standard deviation function: $\sigma_1(\bb x)= c_\sigma \{1-(x_1^2 + x_2^2 + 1.5x_3^2)/10\}$. For TSSS, we utilize the triangulation mesh with $N = 32$ triangles, illustrated in Figure \ref{fig:eg12}, along spherical spline basis, which are based on either a fixed degree ($d = 3$) or a degree selected via the CV ($d_{\text{CV}}$), as described in Section \ref{ssec:implementation-2}.
	
	In Table \ref{tab:sim1_vary}, it is evident that at a lower noise level ($\sigma= 0.5$), TSSS with degree $d=3$ outperforms others with the smallest PMSE and TMSE across all sample sizes, suggesting its robustness in precision. For a higher noise level ($\sigma= 0.75$), the TSSS with CV-selected degree $d_{\text{CV}}$ consistently exhibits strong performance, often achieving or competing for the lowest errors, which reflects its effective handling of increased variability. Conversely, Tensor-Sphere shows relatively poor performance, with the highest PMSE and TMSE values under both noise conditions, marking it as the least accurate. Kernel and TPSOS generally achieve moderate error rates, neither excelling nor performing the poorest across the different scenarios. These patterns remain consistent as the sample size increases from 400 to 2500, indicating that the relative performance of these methods is stable across different amounts of data.
	
	Regarding computational efficiency, as indicated in Table \ref{tab:sim1_vary}, the Kernel method is the most time-consuming among the methods considered. Comparatively, Tensor-Sphere stands out due to its minimal computational time. The TSSS with a fixed degree $d$ and the TPSOS estimators exhibit similar computational performance. The TSSS with a CV-selected degree $d$ exhibits a slightly slower computational pace, as anticipated, but remains significantly faster, by hundreds or thousands of times, than the Kernel method. Therefore, TSSS achieves a favorable balance between estimation efficacy and computational efficiency for functions defined on the entire sphere.
	
	\begin{table}[htbp]
		\centering
		\caption{Simulation Study 1 results for different estimation methods on two mean functions $m_{1}|\mathbb S^2$ and $m_{2}|\mathbb S^2$ with varying standard deviation function {$\sigma_1(\bb x)$}: the average (and standard deviations) of predicted mean squared error (PMSE), training mean squared error (TMSE), dimension of the design parameters (Dim), and computation time per iteration in seconds (Time). The results for TSSS are based on a triangulation with $N = 32$ triangles and spline basis functions with either a fixed degree $d=3$ or a CV-selected degree $d_{\text{CV}}$.  TPSOS and Tensor-Sphere have dimensions $k = 100$ and $k =64$, respectively. Kernel method results have only $30$ iterations instead of $100$ due to computation inefficiency. A factor of $10^3$ scales the reported average (and standard deviations) of PMSEs and TMSEs. Results for $m_1|\mathbb S^2$ and $m_2|\mathbb S^2$ with constant standard deviation function can be found in Table \ref{tab:sim1_const} in Supplementary Material.}
		\label{tab:sim1_vary}
		\scalebox{0.57}{
			\begin{tabular}{c|l|cccccc|cccccc|cccccc}
				\toprule
				\multicolumn{20}{c}{$m_1|\mathbb S^2$} \\
				\midrule
				&       & \multicolumn{6}{c|}{$n=400$}                  & \multicolumn{6}{c|}{$n=900$}                  & \multicolumn{6}{c}{$n=2500$} \\
				\midrule
				$c_\sigma$ & \multicolumn{1}{c|}{Method} & \multicolumn{2}{c}{PMSE} & \multicolumn{2}{c}{TMSE} & Dim   & Time(s) & \multicolumn{2}{c}{PMSE} & \multicolumn{2}{c}{TMSE} & Dim   & Time(s) & \multicolumn{2}{c}{PMSE} & \multicolumn{2}{c}{TMSE} & Dim   & Time(s) \\
				\midrule
				\multirow{5}[2]{*}{0.5} & Kernel & 22.61 & (4.6) & 19.71 & (3.9) & 400   & 4901  & 10.29 & (2.1) & 9.95  & (2.1) & 900   & 8325  & 5.28  & (0.8) & 5.17  & (0.7) & 2500  & 23207 \\
				& TPSOS & 25.59 & (4.7) & 23.11 & (4.2) & 100   & 2     & 13.33 & (2.2) & 12.72 & (2.1) & 100   & 4     & 6.26  & (1.0) & 6.12  & (0.9) & 100   & 8 \\
				& Tensor-Sphere & 50.41 & (11.4) & 38.64 & (6.7) & 64    & 0     & 39.18 & (4.5) & 33.70 & (4.0) & 64    & 0     & 35.97 & (1.9) & 32.20 & (2.3) & 64    & 0 \\
				& TSSS, $d=3$ & 21.78 & (4.5) & 19.43 & (3.9) & 57    & 3     & 12.97 & (2.0) & 12.15 & (1.9) & 57    & 3     & 7.85  & (0.9) & 7.58  & (0.8) & 57    & 3 \\
				& TSSS, $d_{\text{CV}}$ & 23.07 & (6.0) & 20.44 & (4.8) & 102   & 19    & 11.19 & (2.4) & 10.58 & (2.3) & 129   & 24    & 4.56  & (0.9) & 4.48  & (0.8) & 151   & 41 \\
				\midrule
				\multirow{5}[2]{*}{0.75} & Kernel & 47.27 & (9.9) & 41.51 & (8.4) & 400   & 4893  & 20.04 & (4.7) & 19.52 & (4.6) & 900   & 8252  & 8.89  & (1.8) & 8.74  & (1.7) & 2500  & 23583 \\
				& TPSOS & 45.72 & (8.9) & 42.57 & (8.5) & 100   & 2     & 24.55 & (4.4) & 23.75 & (4.2) & 100   & 3     & 11.50 & (1.9) & 11.32 & (1.9) & 100   & 8 \\
				& Tensor-Sphere & 74.20 & (13.1) & 60.24 & (10.0) & 64    & 0     & 52.10 & (6.7) & 45.51 & (5.7) & 64    & 0     & 41.24 & (2.7) & 37.08 & (2.9) & 64    & 0 \\
				& TSSS, $d=3$ & 39.82 & (9.3) & 36.23 & (8.1) & 57    & 3     & 22.72 & (4.3) & 21.50 & (4.1) & 57    & 3     & 11.94 & (2.0) & 11.63 & (1.9) & 57    & 3 \\
				& TSSS, $d_{\text{CV}}$ & 39.66 & (10.4) & 36.08 & (8.8) & 110   & 19    & 21.19 & (4.8) & 20.32 & (4.7) & 115   & 23    & 9.92  & (1.9) & 9.74  & (1.8) & 132   & 41 \\
				\midrule
				\multicolumn{20}{c}{$m_2|\mathbb S^2$} \\
				\midrule
				&       & \multicolumn{6}{c|}{$n=400$}                  & \multicolumn{6}{c|}{$n=900$}                  & \multicolumn{6}{c}{$n=2500$} \\
				\midrule
				$c_\sigma$ & \multicolumn{1}{c|}{Method} & \multicolumn{2}{c}{PMSE} & \multicolumn{2}{c}{TMSE} & Dim   & Time(s) & \multicolumn{2}{c}{PMSE} & \multicolumn{2}{c}{TMSE} & Dim   & Time(s) & \multicolumn{2}{c}{PMSE} & \multicolumn{2}{c}{TMSE} & Dim   & Time(s) \\
				\midrule
				\multirow{5}[2]{*}{0.5} & Kernel & 21.26 & (5.1) & 18.72 & (4.3) & 400   & 4356  & 8.60  & (2.3) & 8.33  & (2.3) & 900   & 9771  & 3.11  & (0.8) & 3.07  & (0.8) & 2500  & 26619 \\
				& TPSOS & 22.18 & (4.7) & 20.82 & (4.3) & 100   & 9     & 11.57 & (2.4) & 11.12 & (2.2) & 100   & 19    & 5.42  & (0.9) & 5.29  & (0.9) & 100   & 39 \\
				& Tensor-Sphere & 25.68 & (6.9) & 19.94 & (4.5) & 64    & 0     & 12.81 & (3.4) & 10.40 & (2.3) & 64    & 0     & 5.48  & (1.0) & 4.72  & (0.8) & 64    & 0 \\
				& TSSS, $d=3$ & 16.73 & (4.2) & 15.37 & (3.7) & 57    & 3     & 8.61  & (2.0) & 8.27  & (1.9) & 57    & 3     & 3.59  & (0.9) & 3.55  & (0.9) & 57    & 4 \\
				& TSSS, $d_{\text{CV}}$ & 18.19 & (5.6) & 16.69 & (4.9) & 86    & 21    & 8.91  & (2.1) & 8.57  & (2.0) & 71    & 29    & 3.42  & (0.9) & 3.39  & (0.8) & 60    & 49 \\
				\midrule
				\multirow{5}[2]{*}{0.75} & Kernel & 46.94 & (10.9) & 41.54 & (9.3) & 400   & 4355  & 19.30 & (5.3) & 18.70 & (5.3) & 900   & 9763  & 6.99  & (1.9) & 6.90  & (1.8) & 2500  & 26642 \\
				& TPSOS & 40.41 & (9.4) & 38.67 & (8.8) & 100   & 8     & 21.39 & (4.7) & 20.74 & (4.5) & 100   & 18    & 9.94  & (1.8) & 9.76  & (1.8) & 100   & 40 \\
				& Tensor-Sphere & 47.97 & (12.6) & 38.93 & (9.1) & 64    & 0     & 24.51 & (6.4) & 20.41 & (4.7) & 64    & 0     & 10.42 & (2.1) & 9.04  & (1.8) & 64    & 0 \\
				& TSSS, $d=3$ & 36.40 & (9.3) & 33.58 & (8.2) & 57    & 3     & 19.04 & (4.5) & 18.30 & (4.3) & 57    & 3     & 7.96  & (2.0) & 7.88  & (2.0) & 57    & 4 \\
				& TSSS, $d_{\text{CV}}$ & 34.77 & (10.1) & 32.51 & (8.9) & 80    & 21    & 16.47 & (4.7) & 15.92 & (4.5) & 64    & 28    & 6.76  & (1.8) & 6.71  & (1.8) & 61    & 48 \\
				\bottomrule
			\end{tabular}%
		}
	\end{table}

	\subsubsection{Impact of $d$ and $\triangle$, and CV for $d$ and $\lambda$ on TSSS Estimation}
	\label{ssec:d_CV}
	
	This section investigates the effect of degree $d$ and triangulation $\triangle$ on the estimation performance for the TSSS method. We also investigate the behavior of CV-selected $d$ and $\lambda$, 
	and their impact on the TSSS estimation. Specifically, we first investigate the impact of $d$ and $\triangle$ on TSSS methods under different $\sigma$ and $n$ settings. In addition, we fix $\triangle$ and select the optimal combination of $d$ and penalty parameter $\lambda$ through the CV method and study the behavior of CV-selected $d_{\text{CV}}$ compared to fixed $d$'s.
	
\begin{figure}
		\centering
		\begin{subfigure}[b]{0.48\linewidth}
			\centering    \includegraphics[width=1\linewidth]{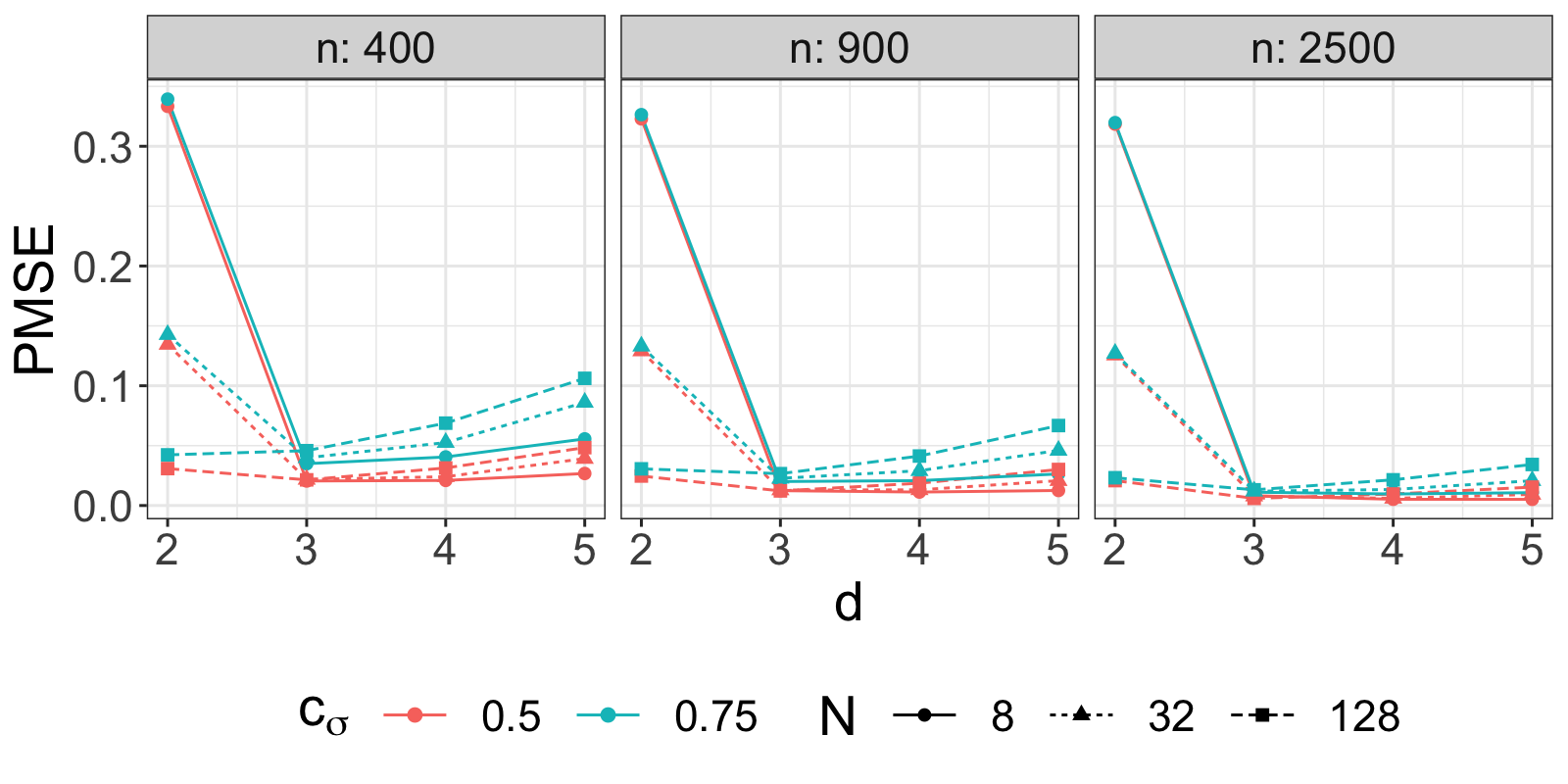}
			\caption{$m_{1}|{\mathbb{S}^2}$}
		\end{subfigure}
		\begin{subfigure}[b]{0.48\linewidth}
			\centering
			\includegraphics[width=1\linewidth]{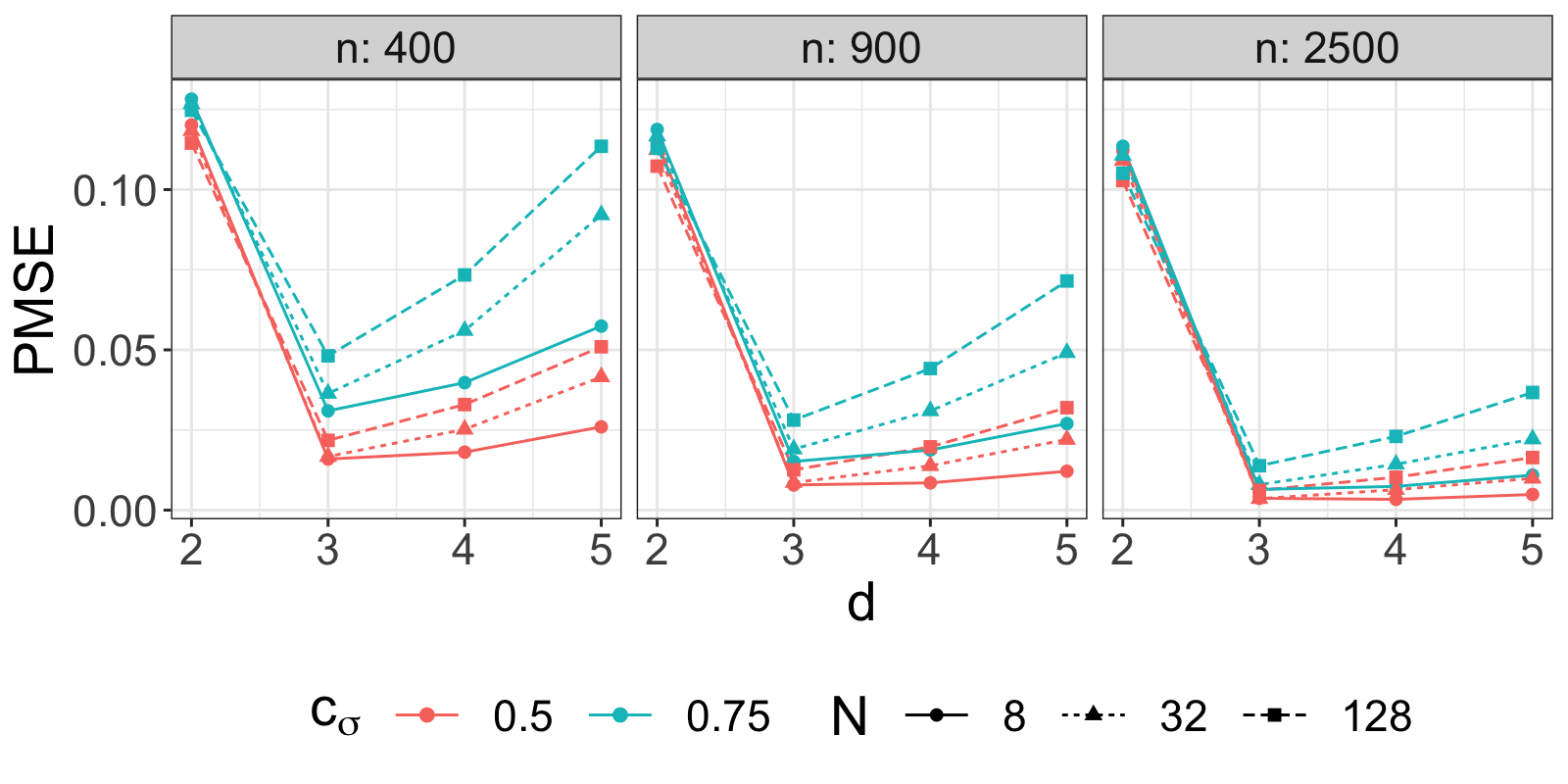}
			\caption{$m_{2}|{\mathbb{S}^2}$}
		\end{subfigure}
		
		\begin{subfigure}[b]{1\textwidth}
			\centering
			\includegraphics[width = 0.9\textwidth]{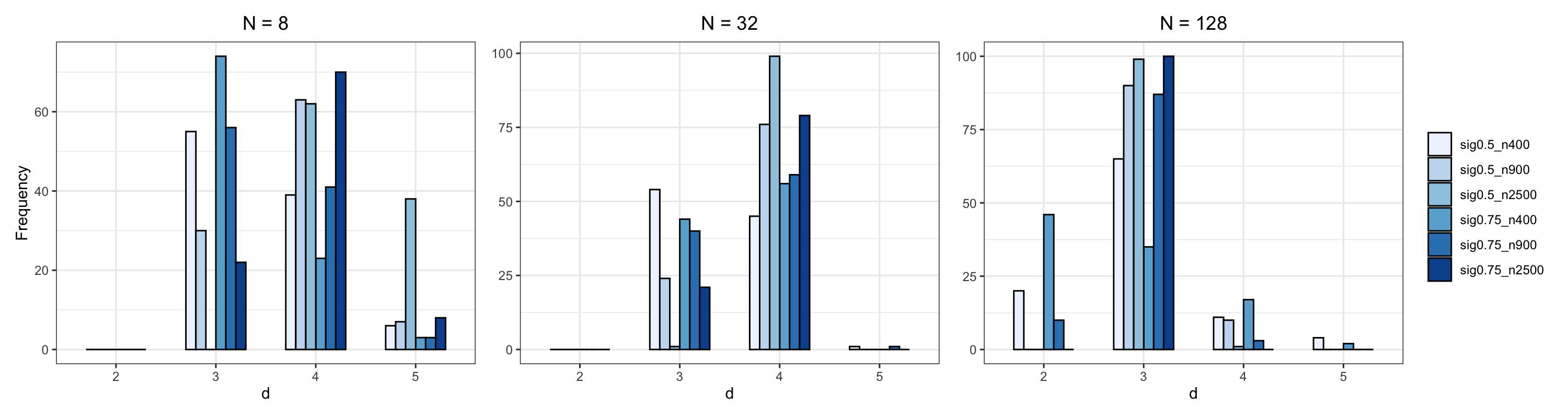}
			\caption{$m_{1}|\mathbb{S}^2$}
		\end{subfigure} 
		
		\noindent
		\begin{subfigure}[b]{1\textwidth}
			\centering
			\includegraphics[width = 0.9\textwidth]{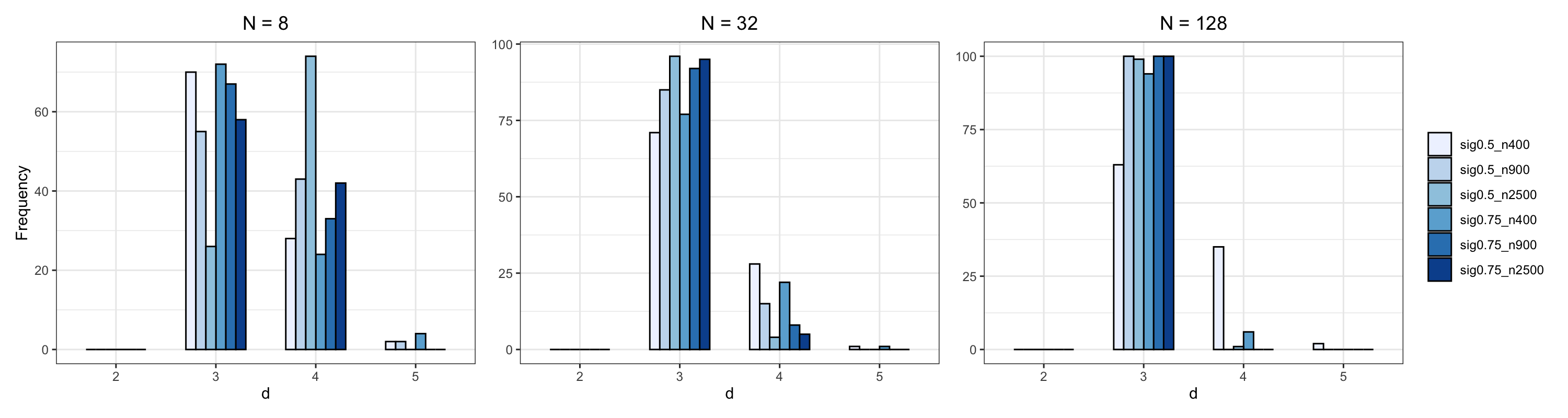}
			\caption{$m_{2}|\mathbb{S}^2$}
		\end{subfigure}  
		\caption{(a) \& (b): Performance of TSSS for $m_{1}|\mathbb S^2$ and $m_{2}|\mathbb S^2$ in Simulation Study 1 with different $n$, $d$, $c_\sigma$, and $N$. Additional results based on the constant standard deviation setting can be found in Figure \ref{fig:Y12TSSS-dtri_const} in Supplementary Material;
			(c) \& (d) Bar graphs of $d_{\text{CV}}$ under different combinations of $c_\sigma = 0.50,~0.75$ (sig), $n=400, ~900, ~ 2,500$, and $N = 8, ~32, ~ 128$ in Simulation Study 1. Results of the constant standard deviation setting can be found in Figure \ref{fig:d_hist_const} of Supplementary Material.}
		\label{fig:Y12TSSS-dtri}
	\end{figure}

	The results in Figure \ref{fig:Y12TSSS-dtri} reveal that different combinations of $d$ and triangulation can lead to varied TSSS performance. Firstly, when $d$ is small, a finer triangulation improves PMSE; see $d = 2$. In contrast, for larger $d$, a finer triangulation may not necessarily improve the PMSE, as observed in the cases of $d = 3, 4, 5$. This is because a larger degree $d$ and number of triangles $N$ can lead to over-parameterization of the mean function. Secondly, when $d$ is too small for a fixed triangulation, the mean functions may not be fully represented, leading to underfitting. Thirdly, as the sample size $n$ increases, both $d$ and the fineness of triangulation may need to be adjusted to achieve optimal estimation performance. When the noise is small, the estimated mean function can better approximate the true mean function, which allows for a higher degree of spline basis $d$ and a finer triangulation to be used without overfitting. However, the choice of $d$ and $N$ should still be balanced with computational efficiency, as both a very high $d$ and a very large $N$ can lead to high computational costs. In general, a careful trade-off between accuracy and computational efficiency must be considered when choosing the appropriate $d$ and $N$ for a given sample size and noise level.
	Readers are referred to Section \ref{sec:implementation} for a detailed discussion.

	Next, we assess the performance of the TSSS estimator employing spline basis functions with the selected degree $d_{\text{CV}}$. As before, we consider simulation settings with various combinations of parameters: $n = 400,~900, ~2500$; $d = 2,~3,~4,~5$; $c_\sigma = 0.50,~0.75$; and $N = 8,~32,~128$, applied to $m_{1}|\mathbb S^2$ and $m_{2}|\mathbb S^2$. For each setting, we simulate $\{(\mathbf{X}_i, Y_i)\}_{i=1}^n$ and compute the TSSS estimator using the selected $d_{\text{CV}}$. This procedure is repeated $100$ times for each setting. The bar graphs in \Cref{fig:Y12TSSS-dtri} and columns ``$f_{d_\text{CV}}$'' in \Cref{tab:CV_d} illustrate the empirical frequency of $d_{\text{CV}}$ across various settings. \Cref{tab:CV_d} also presents the PMSE of TSSS estimators for $d = 2,~3,~4,~5$ and $d_{\text{CV}}$. From the table and bar graphs, for fixed $n,~\sigma$, as $N$ increases, smaller $d_{\text{CV}}$'s are selected. This is consistent with our previous discussion about the balance between $d$ and $N$. In addition, when $n$ and $N$ are fixed, as $\sigma$ increases, a smaller $d_{\text{CV}}$ tends to be selected. This trend is reasonable because higher $d$ tends to pick up more noise, which does not benefit the overall fitting. Furthermore, compared to TSSS estimators with fixed $d = 2,~3,~4,~5$, the TSSS estimator with $d_{\text{CV}}$ is among the best in terms of PMSEs. Therefore, the proposed procedure of selecting $d$ and $\lambda$ by CV is validated.
    
	\begin{table}[htbp]
		\centering
		\caption{Performance comparison of TSSS Estimators for $m_{1}|\mathbb S^2$ and $m_{2}|\mathbb S^2$ in Simulation Study 1: $d_{\text{CV}}$ vs. fixed degree $d= 2,~3,~4,~5$. The frequency of selecting $d_{\text{CV}}$ among $100$ replications is denoted as $f_{d_{\text{CV}}}$. The reported average (and standard deviations) of PMSEs are scaled by a factor of $10^3$. Additional results for $m_1|\mathbb S^2$ and $m_2|\mathbb S^2$ based on the constant standard deviation function setting can be found in Table \ref{tab:CV_d_const} in Supplementary Material.}
		\scalebox{0.6}{
			\begin{tabular}{ll|rr|rr|rr|rr|rr|rr|rr|rr|rr}
				\toprule
				\multicolumn{20}{c}{$m_1|\mathbb S^2$} \\
				\midrule
				&       & \multicolumn{6}{c|}{$N=8$}                    & \multicolumn{6}{c|}{$N=32$}                   & \multicolumn{6}{c}{$N=128$} \\
				\cmidrule{3-20}          &       & \multicolumn{2}{c|}{$n=400$} & \multicolumn{2}{c|}{$n=900$} & \multicolumn{2}{c|}{$n=2500$} & \multicolumn{2}{c|}{$n=400$} & \multicolumn{2}{c|}{$n=900$} & \multicolumn{2}{c|}{$n=2500$} & \multicolumn{2}{c|}{$n=400$} & \multicolumn{2}{c|}{$n=900$} & \multicolumn{2}{c}{$n=2500$} \\
				\cmidrule{3-20}    $c_\sigma$ & $d$     & PMSE  & $f_{d_{\text{CV}}}$ & PMSE  & $f_{d_{\text{CV}}}$ & PMSE  & $f_{d_{\text{CV}}}$ & PMSE  & $f_{d_{\text{CV}}}$ & PMSE  & $f_{d_{\text{CV}}}$ & PMSE  & $f_{d_{\text{CV}}}$ & PMSE  & $f_{d_{\text{CV}}}$ & PMSE  & $f_{d_{\text{CV}}}$ & PMSE  & \multicolumn{1}{r}{$f_{d_{\text{CV}}}$} \\
				\midrule
				\multirow{10}[2]{*}{0.5} & 2     & 333.35 & 0     & 322.90 & 0     & 318.50 & 0     & 134.50 & 0     & 129.03 & 0     & 125.59 & 0     & 30.84 & 20    & 24.63 & 0     & 20.80 & 0 \\
				&       & (7.82) &       & (2.73) &       & (1.29) &       & (4.93) &       & (2.21) &       & (0.88) &       & (4.45) &       & (1.55) &       & (0.66) &  \\
				& 3     & 20.48 & 55    & 12.43 & 30    & 8.20  & 0     & 21.78 & 54    & 12.97 & 24    & 7.85  & 1     & 21.44 & 65    & 12.21 & 90    & 5.96  & 99 \\
				&       & (3.78) &       & (1.81) &       & (0.60) &       & (4.48) &       & (1.99) &       & (0.89) &       & (4.33) &       & (1.99) &       & (0.93) &  \\
				& 4     & 20.95 & 39    & 11.24 & 63    & 5.21  & 62    & 24.05 & 45    & 13.11 & 76    & 6.01  & 99    & 31.36 & 11    & 18.62 & 10    & 9.62  & 1 \\
				&       & (5.50) &       & (2.18) &       & (0.93) &       & (5.18) &       & (2.15) &       & (0.94) &       & (5.37) &       & (2.31) &       & (1.13) &  \\
				& 5     & 26.74 & 6     & 12.57 & 7     & 5.15  & 38    & 39.22 & 1     & 20.74 & 0     & 9.28  & 0     & 48.33 & 4     & 30.09 & 0     & 15.37 & 0 \\
				&       & (5.59) &       & (2.31) &       & (0.95) &       & (7.09) &       & (2.91) &       & (1.28) &       & (7.18) &       & (3.23) &       & (1.52) &  \\
				& $d_{\text{CV}}$ & 21.79 & --    & 12.04 & --    & 5.29  & --    & 23.07 & --    & 11.19 & --    & 4.56  & --    & 23.95 & --    & 11.34 & --    & 4.79  & -- \\
				&       & (4.95) &       & (2.36) &       & (0.96) &       & (6.02) &       & (2.44) &       & (0.85) &       & (5.83) &       & (2.16) &       & (0.89) &  \\
				\midrule
				\multirow{10}[2]{*}{0.75} & 2     & 339.35 & 0     & 326.40 & 0     & 319.74 & 0     & 142.69 & 0     & 133.00 & 0     & 127.03 & 0     & 42.30 & 46    & 30.63 & 10    & 23.20 & 0 \\
				&       & (9.91) &       & (4.08) &       & (1.72) &       & (8.59) &       & (3.82) &       & (1.44) &       & (8.87) &       & (3.19) &       & (1.29) &  \\
				& 3     & 34.93 & 74    & 20.03 & 56    & 11.10 & 22    & 39.82 & 44    & 22.72 & 40    & 11.94 & 21    & 45.79 & 35    & 26.61 & 87    & 13.11 & 100 \\
				&       & (7.84) &       & (3.72) &       & (1.29) &       & (9.25) &       & (4.32) &       & (2.04) &       & (9.54) &       & (4.45) &       & (2.06) &  \\
				& 4     & 40.56 & 23    & 20.79 & 41    & 9.63  & 70    & 52.58 & 56    & 29.05 & 59    & 13.39 & 79    & 68.78 & 17    & 41.31 & 3     & 21.46 & 0 \\
				&       & (11.16) &       & (4.75) &       & (1.95) &       & (11.34) &       & (4.80) &       & (2.10) &       & (11.94) &       & (5.18) &       & (2.55) &  \\
				& 5     & 55.64 & 3     & 26.37 & 3     & 10.76 & 8     & 86.11 & 0     & 46.00 & 1     & 20.72 & 0     & 106.21 & 2     & 66.79 & 0     & 34.26 & 0 \\
				&       & (12.20) &       & (4.90) &       & (2.08) &       & (15.65) &       & (6.46) &       & (2.87) &       & (16.06) &       & (7.26) &       & (3.41) &  \\
				& $d_{\text{CV}}$ & 36.85 & --    & 20.78 & --    & 10.12 & --    & 39.66 & --    & 21.19 & --    & 9.92  & --    & 41.01 & --    & 20.26 & --    & 8.99  & -- \\
				&       & (9.13) &       & (4.19) &       & (1.94) &       & (10.42) &       & (4.76) &       & (1.90) &       & (10.56) &       & (5.06) &       & (1.66) &  \\
				\midrule
				\multicolumn{20}{c}{$m_2|\mathbb S^2$} \\
				\midrule
				&       & \multicolumn{6}{c|}{$N=8$}                    & \multicolumn{6}{c|}{$N=32$}                   & \multicolumn{6}{c}{$N=128$} \\
				\cmidrule{3-20}          &       & \multicolumn{2}{c|}{$n=400$} & \multicolumn{2}{c|}{$n=900$} & \multicolumn{2}{c|}{$n=2500$} & \multicolumn{2}{c|}{$n=400$} & \multicolumn{2}{c|}{$n=900$} & \multicolumn{2}{c|}{$n=2500$} & \multicolumn{2}{c|}{$n=400$} & \multicolumn{2}{c|}{$n=900$} & \multicolumn{2}{c}{$n=2500$} \\
				\cmidrule{3-20}    $c_\sigma$ & $d$     & PMSE  & $f_{d_{\text{CV}}}$ & PMSE  & $f_{d_{\text{CV}}}$ & PMSE  & $f_{d_{\text{CV}}}$ & PMSE  & $f_{d_{\text{CV}}}$ & PMSE  & $f_{d_{\text{CV}}}$ & PMSE  & $f_{d_{\text{CV}}}$ & PMSE  & $f_{d_{\text{CV}}}$ & PMSE  & $f_{d_{\text{CV}}}$ & PMSE  & \multicolumn{1}{r}{$f_{d_{\text{CV}}}$} \\
				\midrule
				\multirow{10}[2]{*}{0.5} & 2     & 120.16 & 0     & 115.00 & 0     & 112.18 & 0     & 118.35 & 0     & 112.43 & 0     & 109.05 & 0     & 114.51 & 0     & 107.31 & 0     & 102.80 & 0 \\
				&       & (3.87) &       & (1.91) &       & (0.86) &       & (4.72) &       & (2.10) &       & (0.92) &       & (5.25) &       & (2.26) &       & (1.04) &  \\
				& 3     & 15.94 & 70    & 7.86  & 55    & 3.70  & 26    & 16.73 & 71    & 8.61  & 85    & 3.59  & 96    & 21.77 & 63    & 12.59 & 100   & 6.19  & 99 \\
				&       & (4.27) &       & (1.89) &       & (0.62) &       & (4.19) &       & (2.02) &       & (0.91) &       & (4.51) &       & (2.25) &       & (0.99) &  \\
				& 4     & 18.09 & 28    & 8.52  & 43    & 3.37  & 74    & 25.22 & 28    & 13.85 & 15    & 6.39  & 4     & 32.94 & 35    & 19.75 & 0     & 10.27 & 1 \\
				&       & (5.51) &       & (2.12) &       & (0.83) &       & (5.37) &       & (2.35) &       & (1.01) &       & (5.70) &       & (2.64) &       & (1.23) &  \\
				& 5     & 25.98 & 2     & 12.14 & 2     & 4.89  & 0     & 41.55 & 1     & 22.04 & 0     & 9.92  & 0     & 50.97 & 2     & 31.96 & 0     & 16.39 & 0 \\
				&       & (5.52) &       & (2.35) &       & (0.99) &       & (7.75) &       & (3.10) &       & (1.38) &       & (7.79) &       & (3.65) &       & (1.65) &  \\
				& $d_{\text{CV}}$ & 17.01 & --    & 8.43  & --    & 3.47  & --    & 18.19 & --    & 8.91  & --    & 3.42  & --    & 18.83 & --    & 8.64  & --    & 4.03  & -- \\
				&       & (4.80) &       & (2.31) &       & (0.84) &       & (5.62) &       & (2.07) &       & (0.86) &       & (5.14) &       & (2.03) &       & (0.79) &  \\
				\midrule
				\multirow{10}[2]{*}{0.75} & 2     & 128.22 & 0     & 118.76 & 0     & 113.52 & 0     & 126.73 & 0     & 116.68 & 0     & 110.67 & 0     & 124.75 & 0     & 112.93 & 0     & 105.09 & 0 \\
				&       & (8.28) &       & (3.51) &       & (1.49) &       & (8.64) &       & (3.67) &       & (1.58) &       & (9.64) &       & (4.02) &       & (1.77) &  \\
				& 3     & 30.97 & 72    & 15.18 & 67    & 6.48  & 58    & 36.40 & 77    & 19.04 & 92    & 7.96  & 95    & 48.14 & 94    & 28.09 & 100   & 13.86 & 100 \\
				&       & (9.17) &       & (3.99) &       & (1.34) &       & (9.27) &       & (4.46) &       & (2.04) &       & (10.13) &       & (5.00) &       & (2.22) &  \\
				& 4     & 39.77 & 24    & 18.76 & 33    & 7.41  & 42    & 56.02 & 22    & 30.96 & 8     & 14.31 & 5     & 73.37 & 6     & 44.18 & 0     & 23.00 & 0 \\
				&       & (12.39) &       & (4.70) &       & (1.91) &       & (11.99) &       & (5.25) &       & (2.28) &       & (12.85) &       & (5.89) &       & (2.77) &  \\
				& 5     & 57.41 & 4     & 27.01 & 0     & 10.92 & 0     & 92.09 & 1     & 49.10 & 0     & 22.17 & 0     & 113.51 & 0     & 71.49 & 0     & 36.71 & 0 \\
				&       & (12.36) &       & (5.26) &       & (2.22) &       & (17.14) &       & (6.95) &       & (3.10) &       & (17.50) &       & (8.16) &       & (3.71) &  \\
				& $d_{\text{CV}}$ & 34.18 & --    & 16.17 & --    & 6.69  & --    & 34.77 & --    & 16.47 & --    & 6.76  & --    & 32.07 & --    & 17.43 & --    & 8.60  & -- \\
				&       & (10.06) &       & (4.61) &       & (1.62) &       & (10.12) &       & (4.71) &       & (1.80) &       & (8.26) &       & (4.13) &       & (1.70) &  \\
				\bottomrule
			\end{tabular}%
		}
		\label{tab:CV_d}%
	\end{table}%

	\subsection{Simulation Study 2: Functions on Irregularly Shaped Spherical Patches}
	\label{sec:simstudy2}
	
	In this simulation study, we generate the response variable $Y_i$ from the following model:
	\begin{equation}
		Y_i = m_3(\bb{X}_i) + \sigma_2(\bb{X}_i)\epsilon_i, ~\epsilon_i \overset{i.i.d}{\sim} N(0, 1), ~ i = 1, \ldots, n, 
		\label{SIM:model2}
	\end{equation}
	with mean function $m_3: \Omega \to \mathbb{R}$ and varying standard deviation function $\sigma_2(\bb x) = 1-(1.5(x_1+1)^2 + x_2^2 + x_3^2)/30$ defined on irregularly shaped spherical patch $\Omega$. Specifically, for $\bb{x} \in \Omega$, we define $m_3(\mathbf{x}) = \widetilde m_3(\bb{x}')$, where $\widetilde m_3(\bb x) = (0.08\pi + 0.84 + x_1)I(x_1\geq -0.84, x_3 > 0)-(0.08\pi + 0.84 + x_1) I(x_1\geq -0.84, x_3 \leq 0)-0.16\arctan\{x_3/(0.84 + x_1)\}I(x_1 < -0.84)$; the spherical coordinates of $\bb{x}$, $(\theta,\phi)$, and the spherical coordinates of $\bb{x}'$, $(\theta',\phi')$, satisfy $\theta' = \theta, ~ \phi' = \phi + \theta/6$.  
	This choice of $m_3$ allows us to test the performance of the methods for functions with complex patterns on non-standard spherical domains. See \Cref{fig:eg12} for a visualization of the mean function and standard deviation function. We generate $\bb{X}_i$'s on $n = 400,~900$ and $2,500$ random locations on the complex domains for training. The SNRs of Simulation Study 2 are reported in Table \ref{tab:snr1}.
	
	For TSSS, given that $m_3$ is not wildly oscillating, we select a spline basis based on $d = 2$ and a triangulation mesh with $56$ triangles. The triangulation is constructed by mapping the spherical domain to the planar domain, triangulating the planar domain using \texttt{TriMesh} from R package ``Triangulation'' \citep{Triangulation}, and mapping the planar triangulation back to spherical coordinates. This procedure would require the user to provide boundary points and interior holes' boundary points and adjust the identical boundary points that wrap around the longitude or co-latitude. This practice is suitable and recommended for simple domains due to the nice properties of \texttt{TriMesh} function. Refer to \Cref{fig:eg12} (d) for the triangulation.
	
	Then, we calculate the PMSE by evaluating the performance on $11,986$ grid points for $m_3 | \Omega$. The estimation results are shown in \Cref{tab:sim3}. The table shows that TSSS, both with a fixed degree ($d = 2$) and a cross-validation-selected degree ($d_{\text{CV}}$),  outperforms TPSOS, Kernel and Tensor-Sphere in terms of both PMSE and TMSE across all settings. In addition to its superior estimation accuracy, TSSS demonstrates significant computational benefits over the Kernel approach. It achieves a reasonable computation efficiency that is comparable to TPSOS, especially for larger sample sizes. Although the Tensor-Sphere method frequently records minimal computational times, at times as low as zero seconds, it falls short of TSSS in terms of estimation accuracy.  The success of TSSS is attributed to its ability to address the ``leakage'' problem by adopting ``domain-aware'' splines. This simulation study highlights the advantages of TSSS in addressing complex domain issues and the ``leakage'' problem, while maintaining computational efficiency.
	
	\begin{table}[!ht]
		\centering
		\caption{Simulation Study 2 results for different estimation methods on mean function $m_{3}|\Omega$  with varying standard deviation function {$\sigma_2(\bb x)$}: the average (and standard deviations) of predicted mean squared error (PMSE), training mean squared error (TMSE), dimension of the design parameters (Dim), and computation time per iteration in seconds (Time). TSSS settings: $d = 2$ or a CV-selected $d_{\text{CV}}$, $N = 58$, $r = 1$. TPSOS and Tensor-Sphere have dimensions $k = 100$ and $k =64$, respectively. Kernel method results have only $30$ iterations instead of $100$ due to computation inefficiency. A factor of $10^3$ scales the reported average (and standard deviations) of PMSEs and TMSEs. Additional results for mean functions $m_3|\Omega$ with constant standard deviation function can be found in Table \ref{tab:sim3_const} in Supplementary Material.}
		\scalebox{0.53}{
			\begin{tabular}{c|l|cccccc|cccccc|cccccc}
				\toprule
				\multicolumn{20}{c}{$m_3|\Omega$} \\
				\midrule
				&       & \multicolumn{6}{c|}{$n=400$}                  & \multicolumn{6}{c|}{$n=900$}                  & \multicolumn{6}{c}{$n=2500$} \\
				\midrule
				$c_\sigma$ & \multicolumn{1}{c|}{Method} & \multicolumn{2}{c}{PMSE} & \multicolumn{2}{c}{TMSE} & Dim   & Time(s) & \multicolumn{2}{c}{PMSE} & \multicolumn{2}{c}{TMSE} & Dim   & Time(s) & \multicolumn{2}{c}{PMSE} & \multicolumn{2}{c}{TMSE} & Dim   & Time(s) \\
				\midrule
				\multirow{5}[2]{*}{0.5} & Kernel & 141.48 & (31.6) & 132.00 & (33.7) & 400   & 1928  & 108.22 & (44.3) & 107.12 & (45.5) & 900   & 4978  & 51.16 & (7.0) & 52.35 & (7.2) & 2500  & 21692 \\
				& TPSOS & 95.68 & (12.8) & 64.73 & (7.0) & 100   & 10    & 53.36 & (5.3) & 44.31 & (4.2) & 100   & 20    & 31.64 & (1.5) & 31.16 & (1.7) & 100   & 43 \\
				& Tensor-Sphere & 58.26 & (7.0) & 51.14 & (5.6) & 64    & 0     & 42.73 & (4.0) & 40.03 & (4.8) & 64    & 0     & 30.01 & (1.5) & 30.64 & (1.6) & 64    & 0 \\
				& TSSS, $d=2$ & 17.69 & (5.3) & 16.02 & (4.4) & 45    & 25    & 9.72  & (2.5) & 9.08  & (2.1) & 45    & 25    & 4.94  & (0.8) & 4.81  & (0.8) & 45    & 25 \\
				& TSSS, $d_{\text{CV}}$ & 18.08 & (6.0) & 16.25 & (4.3) & 100   & 263   & 9.88  & (2.2) & 9.30  & (2.1) & 85    & 325   & 4.69  & (0.9) & 4.53  & (0.9) & 132   & 270 \\
				\midrule
				\multirow{5}[2]{*}{0.75} & Kernel & 167.81 & (13.7) & 157.96 & (18.6) & 400   & 1928  & 133.84 & (35.4) & 132.22 & (35.3) & 900   & 4948  & 82.62 & (41.6) & 84.45 & (43.3) & 2500  & 21301 \\
				& TPSOS & 139.64 & (17.0) & 107.79 & (12.8) & 100   & 10    & 78.65 & (8.6) & 68.12 & (7.7) & 100   & 21    & 41.77 & (2.9) & 41.15 & (2.8) & 100   & 44 \\
				& Tensor-Sphere & 79.56 & (12.9) & 71.57 & (10.3) & 64    & 0     & 55.80 & (5.3) & 54.42 & (5.5) & 64    & 0     & 36.30 & (3.1) & 37.01 & (3.1) & 64    & 0 \\
				& TSSS, $d=2$ & 32.27 & (10.8) & 29.72 & (9.2) & 45    & 25    & 17.17 & (4.7) & 16.30 & (4.4) & 45    & 25    & 8.24  & (1.8) & 8.03  & (1.8) & 45    & 25 \\
				& TSSS, $d_{\text{CV}}$ & 30.42 & (9.3) & 28.39 & (8.0) & 54    & 341   & 17.62 & (5.4) & 16.75 & (4.9) & 63    & 333   & 8.66  & (1.9) & 8.46  & (1.8) & 68    & 359 \\
				\bottomrule
			\end{tabular}%
		}
		\label{tab:sim3}
	\end{table}

	\subsection{Simulation Study 3: Stability of TSSS Estimator}
	\label{sec:simstudy3}
	
	In this simulation study, we verify the stability of the TSSS estimator through a bootstrap procedure. 
	We conduct a total of {$200$} Monte Carlo (MC) replications. In each replication, we simulate observations $\{(\mathbf{X}_i, Y_i)\}_{i = 1}^{n}$, using the same data generation settings as those employed in Simulation Study 2 in Section \ref{sec:simstudy2}. For simplicity, we fix $\mathbf{X}_i$'s on a grid with $n = 3,065$, $d = 3$, and $c_\sigma = 0.5$.
	
\begin{figure}[!ht]
		\centering
		\scalebox{0.64}{\includegraphics[width = \textwidth]{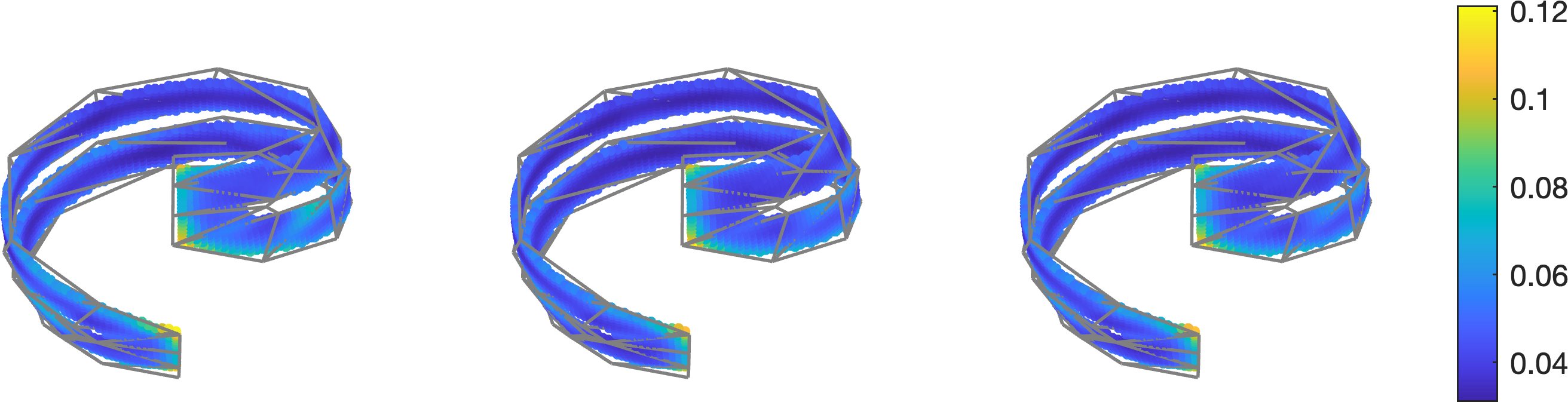}
		}
		\caption{Standard error function plots for $m_{3}|\Omega$ in Simulation Study 3: from left to right, $\text{SE}^{\text{MC}}$, $\text{SE}^{\text{Boot}}_{\text{mean}}$, and $\text{SE}^{\text{Boot}}_{\text{median}}$.}
		\label{fig:bootvar}
	\end{figure}

	Then, the bootstrap standard errors (SE) of TSSS estimator are generated through \Cref{alg:uncertain} using $B = 100$ wild bootstrap samples, denoted as $\{\text{SE}_{R}^{\text{Boot}}\}_{R = 1}^{{200}}$. We report the mean and median of $\{\text{SE}_{R}^{\text{Boot}}\}_{R= 1}^{{200}}$, denoted them as $\text{SE}^{\text{Boot}}_{\text{mean}}$ and $\text{SE}^{\text{Boot}}_{\text{median}}$, respectively. The Monte Carlo standard error  $\text{SE}^{\text{MC}}$ from ${{200}}$ TSSS estimators are calculated from the MC replications, and serves as the ground truth. 
	We then compare $\text{SE}^\text{MC}$ to $\text{SE}^{\text{Boot}}_{\text{mean}}$ and $\text{SE}^{\text{Boot}}_{\text{median}}$ in Figure \ref{fig:bootvar}. The similarity between the Bootstrap SE estimator and the Monte Carlo SE estimator demonstrates the stability of the TSSS estimator.
	
	\section{Data Applications}
	\label{sec:data-app}
	
	\subsection{Cortical Surface fMRI Data} 
	\label{sec:hcp}
	
	In our first application, we apply the proposed TSSS method to estimate the mean neural activity of the brain using the Human Connectome Project (HCP) data \citep{van2013}. The HCP is a comprehensive data project that maps macroscopic human brain circuits and their relationship to human behavior. Specifically, we use the cs-fMRI data from the Motor task study of the HCP 500-subject data release. In this study, participants are asked to tap their left or right fingers, squeeze their left or right foot, or move their tongue following visual cues. During the experiment, two runs of task fMRI scans were collected. One run was collected by scanning the brain from left to right (LR), and the other run was from right to left (RL) \citep{woolrich2001temporal}.  
	
	In this study, we use the RL scan run of a randomly sampled subject (Subject \#100307) registered to the low-resolution Conte 69 standard mesh, which contains approximately $32,000$ vertices per hemisphere \citep{glasser2013, van2012parcellations, jenkinson2012fsl, fischl2012freesurfer,van2012parcellations}. In addition, to evaluate the performance of TSSS in both the resting and task states, the first frame at the subject's resting state ($t = 0$s) and the 20th frame after left-hand tapping is onset ($t \approx 15$s) are studied. For each frame, we perform a 10-fold CV procedure to examine the prediction accuracy. For visualization, we map the predicted values from the sphere to the mid-thickness brain surface, using R package {``ciftiTools''} \citep{PHAM2022ciftitools}. 
	
\begin{figure}
		\centering
		\begin{subfigure}{0.43\textwidth}
			\centering
			\includegraphics[width = 0.35\textwidth]{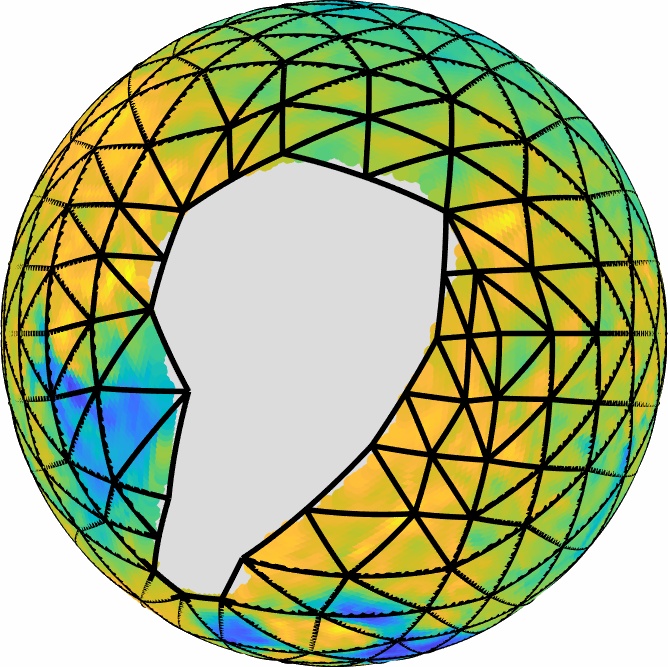}
			\includegraphics[width = 0.35\textwidth]{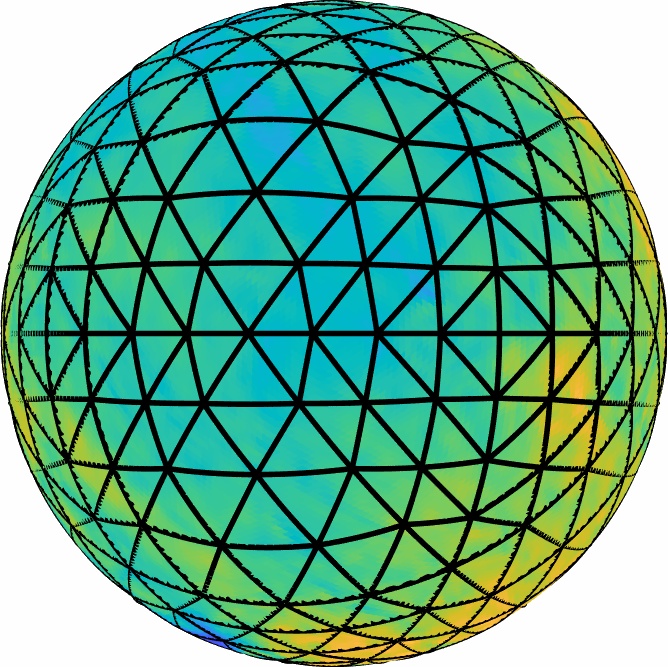}
			\caption{}
		\end{subfigure}
		\begin{subfigure}{0.43\textwidth}
			\centering
			\includegraphics[width = 0.35\textwidth]{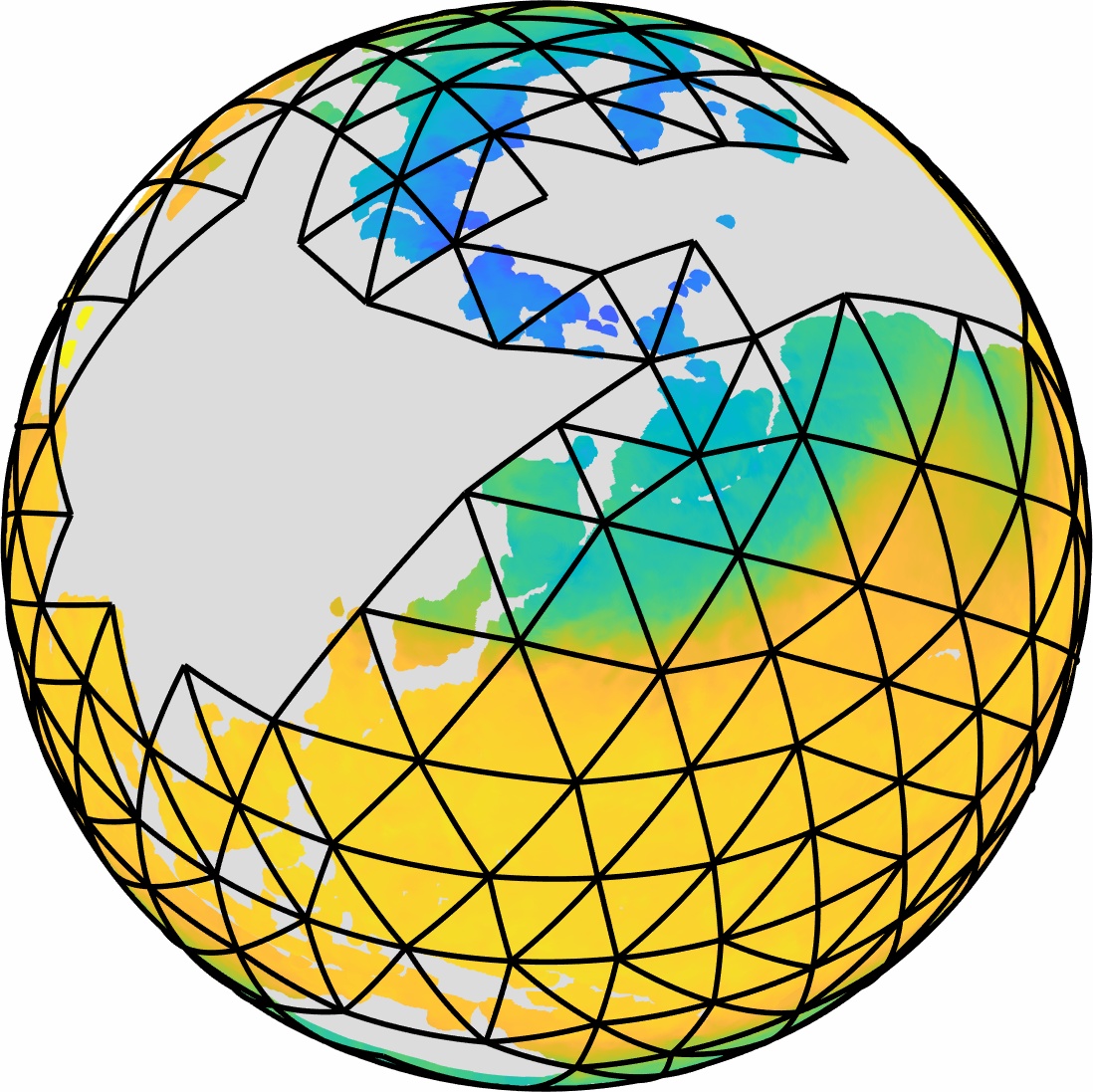}
			\includegraphics[width = 0.35\textwidth]{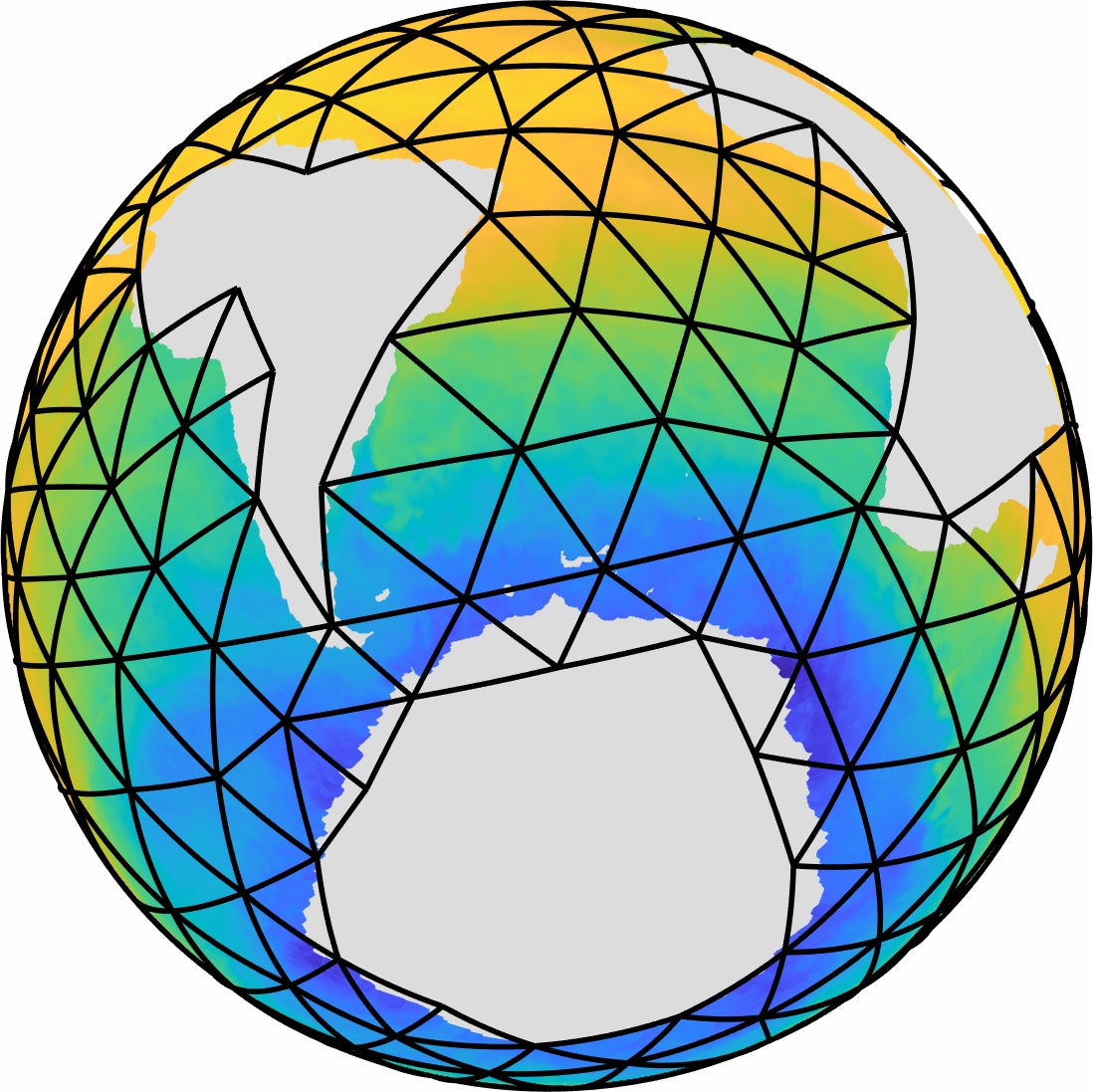}
			\caption{}
		\end{subfigure}
		\caption{Triangulation for (a) Human Right Hemisphere from two viewpoints ($N = 473$); and (b) the ocean surface from two viewpoints ($N = 380$).}
		\label{fig:HCP_tri}
	\end{figure}

	For the TSSS method, the triangulation is constructed by first obtaining a uniform triangulation of the whole sphere. Next, we select only the spherical triangles containing observations and adjust the boundary points to obtain a smoother domain shape. This step can be more complex, as it requires careful consideration of the underlying data distribution and geometry of the domain. However, in our experience, the triangulation works well even without the adjustment step as long as there are no triangles containing extremely sparse observations.
	Two settings are studied for TSSS: (i) $r = 1,~d = 5$ with a modest fine triangulation $N = 473$, as shown in Figure \ref{fig:HCP_tri}; and (ii) $r = 1,~d = 3$ with a fine triangulation $N = 1,913$ for each hemisphere.
	
	Table \ref{tab:hcp-32k} reports the mean and standard deviation of the cross-validated root mean squared prediction error (CV-RMSPE), $[N_k^{-1} \sum_{i: k[i] =k} \{Y_i - \widehat m^{-k}(\bb X_i)\}^2]^{-1/2},\, k = 1, \ldots, 10$, for TSSS, TPSOS and Tensor-Sphere, where the Kernel method is omitted due to its computation inefficiency. Here $k[i]$ is the index of the fold that contains the $i$-th observation, $N_k$ is the cardinality of the $k$-th fold, and $\widehat m^{-k}$ is the fitted mean function using data excluding the $k$-th fold. From Table \ref{tab:hcp-32k}, we observe that when the dimensions of TPSOS, TSSS and Tensor-Sphere are similar, around $2,000$, TSSS demonstrates a slight advantage over TPSOS and a more significant advantage compared to Tensor-Sphere. However, as the dimension of TPSOS increases to around $2,900$, the advantage of TSSS over both TPSOS and Tensor-Sphere becomes more pronounced. These results hold consistently for both hemispheres and different brain activity periods. Importantly, TSSS offers researchers greater flexibility in achieving their objectives. Depending on the desired outcome, researchers can adjust the parameters accordingly. Higher values of $d$ and $N$ can be employed to recover detailed signals, while finer triangulation can be used to incorporate prior domain knowledge. Moreover, by tuning the smoothness coefficient $r$, researchers can control the level of smoothness in the estimated function, aligning it with their expectations. TSSS thus provides a versatile tool for researchers with diverse needs and preferences.
	
	\begin{table}[!ht]
		\centering
		\caption{The mean (and standard deviation) of CV-RMSPE in the HCP data application based on TPSOS, Tensor-Sphere and TSSS. The number of parameters for TPSOS and Tensor-Sphere are set to be $2,000$ and $2,025$, respectively. The settings for TSSS are as follows: (i) $r = 1,~d = 5,~N = 473$, yielding a dimensionality (Dim) of $2,921$; and (ii) $r = 1,~d = 3,~N = 1,913$, yielding a dimensionality (Dim) of  approximately $2,000$.} 
		\scalebox{0.8}{
			\begin{tabular}{clcc@{\extracolsep{4pt}}clcc@{}}\toprule
				\multicolumn{4}{c}{Left Hemisphere} & \multicolumn{4}{c}{Right Hemisphere}\\\cline{1-4}\cline{5-8}
				Frame & Method & \multicolumn{1}{c}{CV-RMSPE} & Dim
				& Frame & Method & \multicolumn{1}{c}{CV-RMSPE}& Dim\\ \cline{1-4}\cline{5-8}
				1 & TPSOS & 471 (26) & 2000 & 1 & TPSOS & 421 (11)& 2000 \\
				1 & Tensor-Sphere & 528  (26) & 2025 & 1 & Tensor-Sphere & 468 (15)& 2025 \\
				1 & TSSS (i) & 412  (11) & 2921  & 1 & TSSS (i) & 370 ~~(8)& 2921\\
				1 & TSSS (ii)& 460  (15) & 2003  & 1 & TSSS (ii)& 409 (10)& 2008\\
				\cline{1-4}\cline{5-8}
				20 & TPSOS & 467 (26) & 2000 & 20 & TPSOS & 415  (12)& 2000 \\
				20 & Tensor-Sphere & 527 (29) & 2025 & 20 & Tensor-Sphere & 463 (14)& 2025 \\
				20 & TSSS (i)& 407 (12) & 2921 & 20 & TSSS (i) & 365 ~~(9)& 2921\\
				20 & TSSS (ii)& 457 (16) & 2003 & 20 & TSSS (ii) & 411  (10)& 2008\\
				\bottomrule
		\end{tabular}}
		\label{tab:hcp-32k}
	\end{table} 

\begin{figure}[!ht]
		\centering
		\begin{subfigure}{0.95\textwidth}
			\includegraphics[width = 1\textwidth]{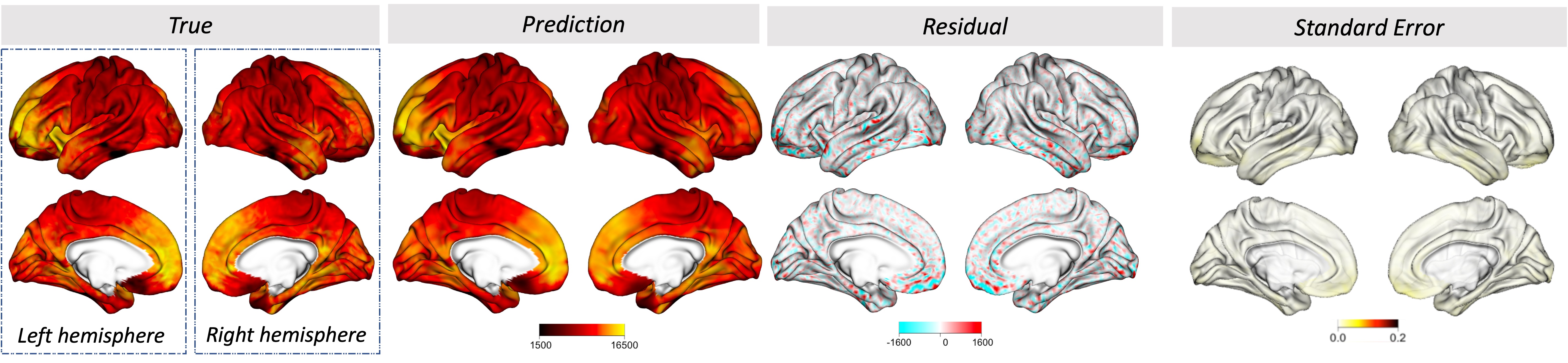}
			\caption{HCP}
		\end{subfigure}
		\begin{subfigure}{1\textwidth}
			\centering
			\includegraphics[width = 0.6\textwidth]{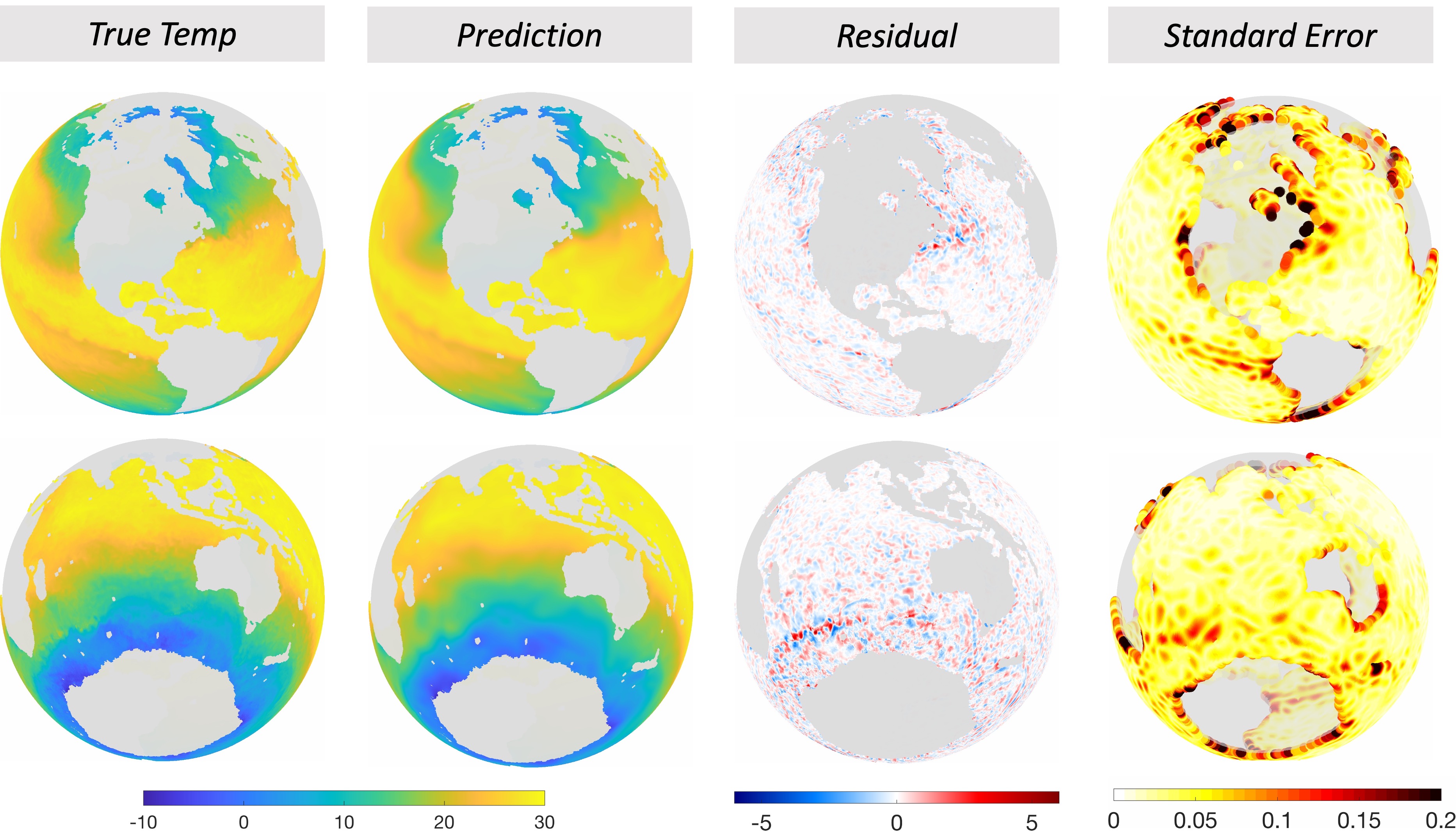}
			\caption{Temp}
		\end{subfigure}
		\begin{subfigure}{1\textwidth}
			\centering
			\includegraphics[width = 0.6\textwidth]{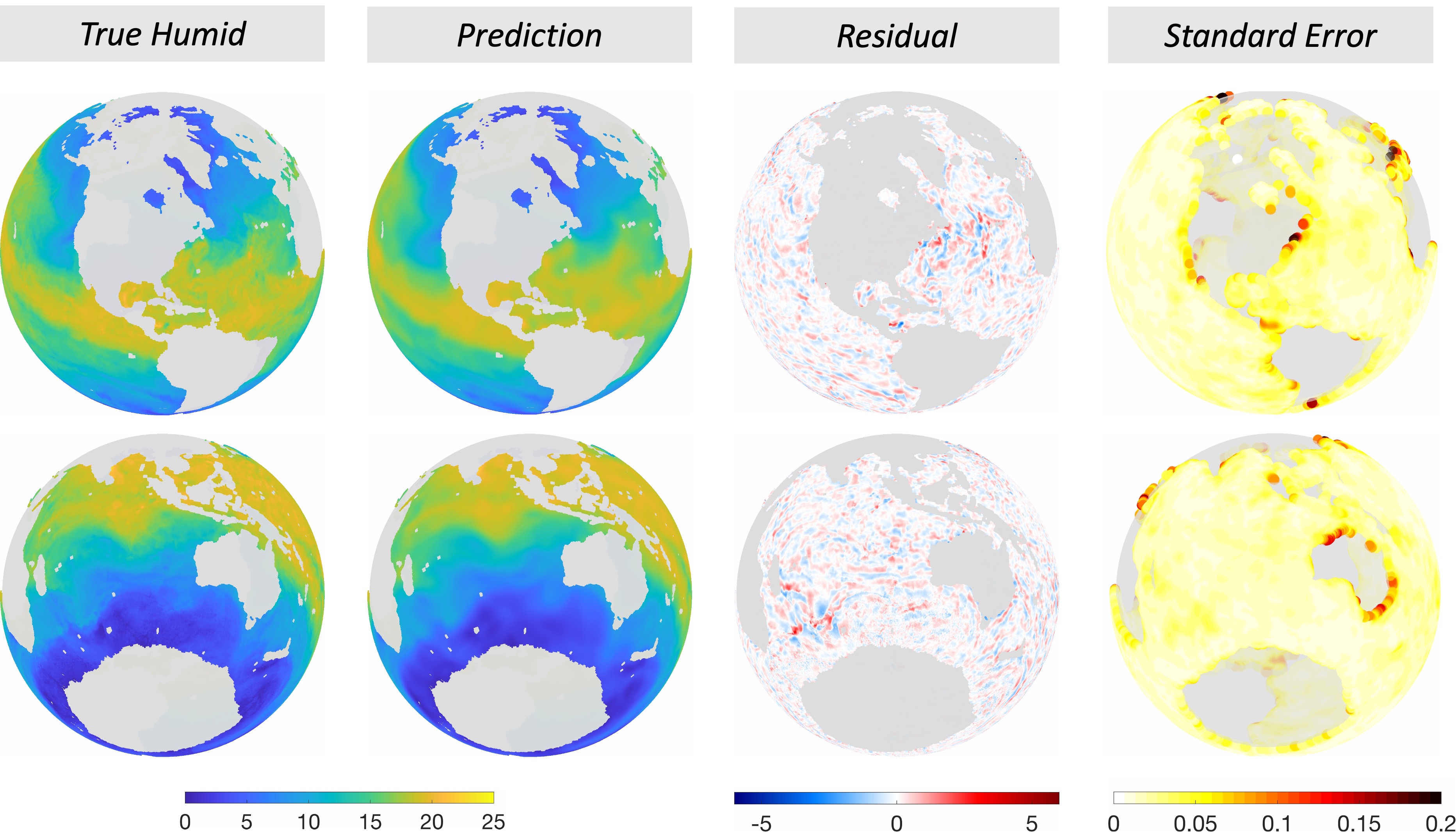}
			\caption{Humid}
		\end{subfigure}
		
		\caption{(a) The observed, predicted, residual values along with the estimated standard error of the TSSS estimator for the first frame in HCP data application. Rows 1 and 2 depict two different viewpoints for each hemisphere. The TSSS settings used are: $r = 1,~d = 3$, and $N = 1,913$;
			(b) \& (c): The observed, predicted, and residual values along with the estimated standard error of the TSSS estimator for Temp and Humid in oceanic atmospheric data application. The TSSS settings used are: $d=3$, $r = 1$, and $N = 1,513$.}
		\label{fig:res-hcp}
	\end{figure}

	Furthermore, we employ the bootstrap method outlined in \Cref{alg:uncertain} to estimate the standard error of the TSSS estimator. To facilitate better visualization, we present the standard error normalized by the maximum observed value, denoted as $\text{SE}^{\text{Boot}}/\max_i \{Y_i\}$. In Figure \ref{fig:res-hcp}, we provide visual representations of the observed, predicted, and residual values for the first frame of both hemispheres using TSSS with setting (ii). The estimated standard errors are consistently small across the whole domain.

	\subsection{Oceanic Atmospheric Data}
	\label{sec:ncei}
	
	In this application, we apply our method to the NOAA Ocean Surface Bundle Climate Data Record (CDR) from the National Centers for Environmental Information (NCEI)  \citep{clayson2016}. In particular, we consider the dataset \url{SEAFLUX-OSB-CDR_V02R00_ATMOS_D20210831_C20211223.nc} and focus on the specific humidity (Humid) and air temperature (Temp) on August 31, 2021. There are approximately one million lattice locations on the surface of Earth, among which approximately 485,000 locations are observed.
	
	Again, we apply a 10-fold CV procedure to evaluate the performance of TSSS, TPSOS, and Tensor-Sphere. For TSSS, we use settings (i) $d = 5$, $r = 1$ with a modest fine triangulation $N = 380$, as shown in Figure \ref{fig:HCP_tri}; and (ii) $d = 3$, $r = 1$ with a fine triangulation $N = 1,513$ for the ocean surface. The triangulation is constructed in the same way as described in \Cref{sec:hcp}. The results in Table \ref{tab:ncei} indicate that TSSS outperforms Tensor-Sphere under both settings. At the same time, TSSS with setting (i) outperforms TPSOS but is inferior to TPSOS under setting (ii). Therefore, the results show that TSSS outperforms TPSOS in terms of prediction accuracy when appropriate triangulation and tuning parameters are selected, while the advantage of TSSS over Tensor-Sphere remains consistent. Figure \ref{fig:res-hcp} displays the observed, predicted values, residuals of humidity and temperature, along with the estimated standard error of TSSS with setting (ii). The residuals do not exhibit patterns, and the estimated standard errors are satisfyingly small, with slightly larger values along the boundary of domains. This is reasonable since fewer observations are available near the domain boundary.

	\begin{table}[!ht]
		\centering
		\caption{The mean (and standard deviation) of the CV-RMSPEs in NCEI data application based on TSSS and TPSOS. The settings for TPSOS and Tensor-Sphere are $2,000$ and $2,025$, respectively. The TSSS settings used are: (i) $r = 1,~d = 5,~N = 380$, yielding a dimensionality (Dim) of $2,600$; and (ii) $r = 1,~d = 3,~N = 1,513$, yielding a dimensionality (Dim) of $1,947$.}
		\scalebox{0.8}{
			\begin{tabular}{@{\extracolsep{4pt}}clccclcc}
				\toprule
				Item & Methods & CV-RMSPE & Dim & Item & Methods & CV-RMSPE&Dim \\
				\cline{1-4} \cline{5-8}
				\multirow{4}{*}{Temp}  & TPSOS & 0.4913 (0.004) &2000&
				\multirow{4}{*}{Humid} & TPSOS & 0.3808 (0.004)&2000\\
				& Tensor-Sphere  & 0.5667 (0.008) & 2025&& Tensor-Sphere  &0.4992 (0.011)&2025\\
				& TSSS (i)  & 0.4684 (0.003) & 2600&& TSSS (i) & 0.3522 (0.001)&2600\\
				& TSSS (ii) & 0.5171  (0.003) & 1947&& TSSS (ii) & 0.3925  (0.001)&1947\\
				\bottomrule
		\end{tabular}}
		\label{tab:ncei}
	\end{table}

	
	\section{Summary and Discussion}
	\label{sec:conclusion}
	
	Motivated by ubiquitous surface-based data in various fields, we introduce a novel TSSS estimator for data that resides on complex domains on the sphere or any surface that can be properly mapped to and back from the sphere. The combination of penalized spline methods and triangulation techniques on spherical surfaces or patches presents a novel methodology makes TSSS advantageous in handling complex spherical domains while enjoying enhanced computation efficiency.
	
	Firstly, TSSS provides enhanced flexibility in capturing complex spatial patterns, allowing for accurate modeling of intricate variations in the data. By employing the ``domain aware'' splines, TSSS enables estimation to enjoy accuracy without suffering from ``leakage'' or boundary effects. This distinguishes TSSS from its competitors (e.g., Tensor-Sphere, TPSOS, and kernel-based methods). Secondly, TSSS offers improved computational efficiency compared to the kernel-based approach, making it more scalable for large-scale datasets. Kernel-based approaches use observed locations as the centers of kernels. However, such kernel functions do not necessarily constitute the most efficient basis for approximating highly localized or rapidly changing functions. On the contrary, TSSS constructs basis functions based on triangulated domains and avoids the need for excessive basis functions. By adaptively refining the triangulation and utilizing spline basis functions with a flexible degree and smoothness, TSSS can better capture the complexity of the underlying signal.
	Thirdly, we provide rigorous theoretical guarantees and an implementation algorithm for the uncertainty quantification of TSSS. Specifically, Theorem \ref{THE:normal} demonstrates the asymptotic normality of the TSSS estimator, and Algorithm \ref{alg:uncertain} introduces a wild Bootstrap algorithm to obtain the standard error map of the TSSS estimator. These advantages have all been demonstrated through extensive numerical experiments. 
	
	The studies in this paper also open up opportunities for further research. Note that TSSS usually requires well-constructed triangulations that accurately approximate the shape of domains. When the domain contains isolated sub-domains, determining whether to connect these sub-domains in the triangulation can be challenging and may require human input. To overcome this challenge, researchers could develop automated triangulation procedures to construct triangulations that work well with the TSSS method, where users' desired configuration could be easily tuned. Moreover, inspired by \cite{basna2022data}, there is potential to explore function-dependent triangulation methods under functional data settings. In particular, random samples of functional data can be used to customize triangulations using data-specific features. 
	
	Another interesting future research direction is to advance TSSS to accommodate datasets that possess physical constraints, such as non-negativity \citep{BL18, kim2021spatially, zhang2022estimating} and shape constraints \citep{Wang:Xue:Yang:20, Fang:Xue:MartinsFilho:Yang:22}. Incorporating non-negativity constraints into the TSSS framework can broaden the method's applicability to datasets where negative values are physically infeasible or meaningless. Moreover, integrating constraints (e.g., monotonicity, convexity, symmetry, or specific geometric patterns) into the TSSS framework can provide additional regularization to ensure the estimated function adheres to predefined physical constraints. TSSS can also be extended to handle longitudinal or time series data observed on spherical domains. Allowing researchers to model the dynamics of surface-based data over time could offer essential insights, such as estimating the temporal change of Earth's atmospheric properties. Furthermore, in functional data regression where surface-based imaging serves as functional responses or covariates \citep{lila:2020}, TSSS can be used as the first smoothing step. 
	
	It is also of immense interest to extend the proposed TSSS method to the hypersphere setting. Data that lies on the surface of hyperspheres is commonly referred to as ``directional data''. It is widely observed during the investigation of cell cycle data \citep{schafer1998cell}; the shape analysis of manifold-valued data \citep{LinEtal:2017}; compositional household expenditure \citep{ScealyWelsh2017}; and gene expression clustering \citep{DingRegev:2021}.
	Within this body of literary works, the hypersphere-valued data $\mathbf x\in \mathbb S^d$ are regarded as the response of interest, and their traits carry information. Although our framework focuses more on the estimation of mean function with $\mathbf x\in \mathbb S^d$ being the covariates, we can utilize dimension reduction methods proposed in the directional data literature to counter the high-dimension setting in our framework.
	
	In particular, it is possible to project data that lies on a hypersphere onto an optimal sphere using the method proposed by \cite{luo2022spherical}, followed by the TSSS procedure. Specifically, if $\mathbf x \in \Omega \subseteq \mathbb S^d$, the geometric distance can be utilized to find the best lower dimension representation of the given domain $\Omega$ in $\mathbb S^{d'}$. When the targeting dimension $d' = 2$, we can get the dataset's best unit sphere projection. The estimation of the mean function with hypersphere support can then be treated as one with unit sphere support. Then, the proposed TSSS method can be directly used without the triangulation of hypersphere domains, which can be easily infeasible when $d$ is very large.  
	
	In addition to the location variable $\mathbf{x}$, datasets frequently incorporate explanatory variables $\{z_k, k=1,\ldots,p\}$, which may also contribute to the model. To account for that, we can utilize the generalized additive mean structure introduced in \cite{Yu:etal:20}, $g\{\mu(\mathbf z, \mathbf x)\} = \sum_{k=1}^p f_k(z_k) + m(\mathbf x)$. Here $g$ can be any link function, and $m(\cdot), f_k(\cdot)$ can be estimated nonparametrically depending on the assumption of the function space of $m(\cdot), f_k(\cdot)$. {Another potential approach is the generalized spatially varying coefficient models \citep{Kim:Wang:21}. These modeling frameworks offer researchers the opportunity to explore the relationship between local features and responses of interest that reside on a surface.}

\section*{Acknowledgements}
Guannan Wang's research was supported in part by National Institutes of Health grant 1R01AG085616-01 and Simons Foundation collaboration grant \#963447. Ming-Jun Lai's research was partially supported by Simons Foundation collaboration grant \#864439. Li Wang's research was partially supported by National Institutes of Health grant 1R01AG085616-01 and National Science Foundation grant DMS-2203207. 
The HCP data were provided by the Human Connectome Project, WU-Minn Consortium (Principal Investigators: David Van Essen and Kamil Ugurbil; 1U54MH091657) funded by the 16 NIH Institutes and Centers or Neuroscience Research; and by the McDonnell Center for Systems Neuroscience at Washington University. The Ocean Near Surface Atmospheric Properties CDR was acquired from NOAA's National Centers for Environmental Information (http://www.ncdc.noaa.gov). 
	
\newpage
\appendix

\vspace{0.8pc} \centerline{\Large Supplemental Materials for \Large \bf ``TSSS: A Novel Triangulated Spherical} \centerline{\Large \bf  Spline Smoothing for Surface-based Data''}
\vspace{0.8pc} 

\renewcommand{\thesection}{\Alph{section}} 
\newcounter{myclemma}[section] 
\renewcommand{\thelemma}{{\thesection.\arabic{myclemma}}}
\let\olemma\lemma 
\renewenvironment{lemma}{\stepcounter{myclemma}\olemma} 

\newcounter{myctheorem}[section] 
\renewcommand{\thetheorem}{{\thesection.\arabic{myctheorem}}} 
\let\otheorem\theorem 
\renewenvironment{theorem}{\stepcounter{myctheorem}\otheorem}

\newcounter{mycremark}[section] 
\renewcommand{\theremark}{{\thesection.\arabic{mycremark}}}
\let\oremark\remark 
\renewenvironment{remark}{\stepcounter{mycremark}\oremark} 

\renewcommand{\thefigure}{\thesection.\arabic{figure}}
\setcounter{figure}{0}
\renewcommand{\thetable}{\thesection.\arabic{table}}
\setcounter{table}{0}
\renewcommand{\theequation}{\thesection.\arabic{equation}}
\setcounter{equation}{0}
\renewcommand{\theproposition}{\thesection.\arabic{proposition}} 
\renewcommand{\thedefinition}{\thesection.\arabic{definition}} 
\renewcommand{\theexample}{\thesection.\arabic{example}} 
\renewcommand{\thesubsection}{\thesection.\arabic{subsection}.}

\noindent
In this document, we provide additional simulation results, proofs of the main asymptotic results, and in-depth technical details pertaining to the proofs.
\par

\section{Numerical Results}

In this section, we provide supplementary results for simulation studies considering homogeneous random errors with a constant standard deviation.

\begin{table}[!ht]
    \caption{Signal-to-noise ratios (SNRs) for simulation studies. }
    \label{tab:snr1}
    \centering
    \scalebox{0.7}{
    \begin{tabular}{cc|cc|cc|cc|cc|cc}
    \toprule
    \multicolumn{8}{c|}{Simulation Study 1} & \multicolumn{4}{c}{Simulation Studies 2 \& 3} \\
     \midrule
\multicolumn{4}{c|}{Constant standard deviation}&\multicolumn{4}{c|}{Varying standard deviation} & \multicolumn{2}{c|}{Constant standard deviation}&\multicolumn{2}{c}{Varying standard deviation} \\\midrule 
     \multicolumn{2}{c|}{$m_1|\mathbb S^2$} & \multicolumn{2}{c|}{$m_2|\mathbb S^2$} &
     \multicolumn{2}{c|}{$m_1|\mathbb S^2$} & \multicolumn{2}{c|}{$m_2|\mathbb S^2$} &
     \multicolumn{2}{c|}{$m_3|\Omega$} & 
     \multicolumn{2}{c}{$m_3|\Omega$} \\\midrule
      $\sigma$ & \multicolumn{1}{c|}{SNR} & $\sigma$ & \multicolumn{1}{c|}{SNR} &
      $c_\sigma$ & \multicolumn{1}{c|}{SNR} &  $c_\sigma$ & \multicolumn{1}{c|}{SNR} &    $\sigma$ & \multicolumn{1}{c|}{SNR} & 
      $c_\sigma$ & \multicolumn{1}{c}{SNR} \\\midrule
      0.50 & 5.14& 0.50& 4.78& 0.50 & 6.72& 0.50 &6.25 &
       0.50& 7.83& 0.50 &7.81\\
      0.75 & 2.28& 0.75& 2.13& 0.75 &2.98& 0.75 & 2.78 & 
       0.75& 3.48 & 0.75 & 3.47 \\
      \bottomrule
    \end{tabular}}
\end{table}

\begin{table}[!ht]
\centering
\caption{Simulation Study 1 results for different estimation methods on two mean functions $m_{1}|\mathbb S^2$ and $m_{2}|\mathbb S^2$ with homogeneous errors: the average (and standard deviations) of predicted mean squared error (PMSE), training mean squared error (TMSE), dimension of the design parameters (Dim), and computation time per iteration in seconds (Time). A factor of $10^3$ scales the reported average (and standard deviations) of PMSEs and TMSEs. The results for TSSS are based on a triangulation with $N = 32$ triangles and spline basis functions with either a fixed degree $d=3$ or a CV-selected degree $d_{\text{CV}}$. The results for TPSOS and Tensor-Sphere are based on dimensions $k=100$ and $k=64$, respectively. Kernel method results have only $30$ iterations instead of $100$ due to computation inefficiency.}
\label{tab:sim1_const}
    \scalebox{0.55}{
    \begin{tabular}{c|l|cccccc|cccccc|cccccc}
    \multicolumn{20}{c}{$m_1|\mathbb S^2$} \\
    \midrule
          &       & \multicolumn{6}{c|}{$n=400$}                  & \multicolumn{6}{c|}{$n=900$}                  & \multicolumn{6}{c}{$n=2500$} \\
    \midrule
    $\sigma$ & \multicolumn{1}{c|}{Method} & \multicolumn{2}{c}{PMSE} & \multicolumn{2}{c}{TMSE} & Dim   & Time(s) & \multicolumn{2}{c}{PMSE} & \multicolumn{2}{c}{TMSE} & Dim   & Time(s) & \multicolumn{2}{c}{PMSE} & \multicolumn{2}{c}{TMSE} & Dim   & Time(s) \\
    \midrule
    \multirow{5}[2]{*}{0.5} & Kernel & 28.27 & (5.8) & 24.71 & (4.9) & 400   & 6036  & 12.52 & (2.7) & 12.15 & (2.7) & 900   & 3532  & 6.11  & (1.0) & 6.00  & (1.0) & 2500  & 11085 \\
          & TPSOS & 30.80 & (5.7) & 28.04 & (5.2) & 100   & 2     & 16.20 & (2.8) & 15.51 & (2.6) & 100   & 4     & 7.59  & (1.2) & 7.43  & (1.2) & 100   & 8 \\
          & Tensor-Sphere & 57.18 & (13.1) & 44.45 & (7.4) & 64    & 0     & 42.70 & (5.1) & 36.82 & (4.4) & 64    & 0     & 37.38 & (2.1) & 33.48 & (2.4) & 64    & 0 \\
           & TSSS, $d=3$ & 25.91 & (5.6) & 23.30 & (4.9) & 57    & 3     & 15.17 & (2.5) & 14.27 & (2.4) & 57    & 3     & 8.82  & (1.2) & 8.53  & (1.1) & 57    & 3 \\
          & TSSS, $d_{\text{CV}}$ & 27.06 & (7.0) & 24.20 & (5.8) & 101   & 20    & 13.36 & (2.9) & 12.68 & (2.7) & 126   & 27    & 5.71  & (1.0) & 5.61  & (1.0) & 150   & 48 \\
    \midrule
    \multirow{5}[2]{*}{0.75} & Kernel & 59.84 & (12.5) & 52.65 & (10.6) & 400   & 6040  & 25.04 & (6.0) & 24.43 & (5.9) & 900   & 3521  & 10.75 & (2.3) & 10.59 & (2.1) & 2500  & 11030 \\
          & TPSOS & 54.96 & (11.0) & 51.44 & (10.6) & 100   & 1     & 29.77 & (5.5) & 28.87 & (5.3) & 100   & 4     & 13.98 & (2.4) & 13.76 & (2.3) & 100   & 8 \\
          & Tensor-Sphere & 86.75 & (15.6) & 71.32 & (12.1) & 64    & 0     & 58.83 & (8.1) & 51.60 & (6.8) & 64    & 0     & 44.07 & (3.2) & 39.66 & (3.2) & 64    & 0 \\
           & TSSS, $d=3$ & 49.01 & (11.4) & 44.82 & (10.2) & 57    & 3     & 27.61 & (5.2) & 26.22 & (5.0) & 57    & 3     & 14.03 & (2.5) & 13.71 & (2.4) & 57    & 3 \\
          & TSSS, $d_{\text{CV}}$ & 48.61 & (12.9) & 44.52 & (10.8) & 110   & 20    & 26.35 & (5.8) & 25.33 & (5.7) & 108   & 26    & 12.68 & (2.4) & 12.46 & (2.4) & 109   & 46 \\
      \midrule
     \multicolumn{20}{c}{$m_2|\mathbb S^2$} \\
    \midrule
          &       & \multicolumn{6}{c|}{$n=400$}                  & \multicolumn{6}{c|}{$n=900$}                  & \multicolumn{6}{c}{$n=2500$} \\
    \midrule
    $\sigma$ & \multicolumn{1}{c|}{Method} & \multicolumn{2}{c}{PMSE} & \multicolumn{2}{c}{TMSE} & Dim   & Time(s) & \multicolumn{2}{c}{PMSE} & \multicolumn{2}{c}{TMSE} & Dim   & Time(s) & \multicolumn{2}{c}{PMSE} & \multicolumn{2}{c}{TMSE} & Dim   & Time(s) \\
    \midrule
    \multirow{5}[2]{*}{0.5} & Kernel & 25.44 & (5.9) & 22.52 & (5.1) & 400   & 1452  & 10.27 & (2.7) & 9.96  & (2.7) & 900   & 3339  & 3.74  & (1.0) & 3.70  & (1.0) & 2500  & 9820 \\
          & TPSOS & 25.25 & (5.4) & 23.86 & (4.9) & 100   & 2     & 13.25 & (2.7) & 12.77 & (2.5) & 100   & 4     & 6.19  & (1.1) & 6.06  & (1.0) & 100   & 7 \\
          & Tensor-Sphere & 28.96 & (7.6) & 22.79 & (5.1) & 64    & 0     & 14.60 & (3.9) & 11.92 & (2.6) & 64    & 0     & 6.24  & (1.2) & 5.38  & (0.9) & 64    & 0 \\
          & TSSS, $d=3$ & 19.79 & (4.9) & 18.26 & (4.3) & 57    & 3     & 10.24 & (2.3) & 9.85  & (2.2) & 57    & 3     & 4.27  & (1.1) & 4.23  & (1.1) & 57    & 4 \\
          & TSSS, $d_{\text{CV}}$ & 21.33 & (6.1) & 19.69 & (5.2) & 87    & 23    & 10.36 & (2.3) & 9.99  & (2.3) & 73    & 29    & 3.90  & (1.0) & 3.87  & (1.0) & 60    & 44 \\
    \midrule
    \multirow{5}[2]{*}{0.75} & Kernel & 55.88 & (12.4) & 49.76 & (10.9) & 400   & 1449  & 23.03 & (6.1) & 22.33 & (6.0) & 900   & 3312  & 8.40  & (2.2) & 8.32  & (2.1) & 2500  & 9820 \\
          & TPSOS & 45.92 & (10.6) & 44.20 & (10.0) & 100   & 2     & 24.45 & (5.3) & 23.77 & (5.0) & 100   & 4     & 11.37 & (2.1) & 11.19 & (2.1) & 100   & 8 \\
          & Tensor-Sphere & 53.91 & (14.0) & 44.27 & (10.3) & 64    & 0     & 27.94 & (7.7) & 23.37 & (5.6) & 64    & 0     & 11.89 & (2.4) & 10.35 & (2.0) & 64    & 0 \\
          & TSSS, $d=3$ & 43.27 & (10.9) & 40.10 & (9.6) & 57    & 3     & 22.72 & (5.2) & 21.86 & (5.0) & 57    & 3     & 9.47  & (2.4) & 9.39  & (2.3) & 57    & 4 \\
          & TSSS, $d_{\text{CV}}$ & 39.20 & (11.9) & 37.06 & (10.7) & 75    & 23    & 19.01 & (5.5) & 18.42 & (5.3) & 63    & 28    & 7.79  & (1.9) & 7.75  & (1.9) & 58    & 44 \\
    \bottomrule
    \end{tabular}
}
  \label{tab:m12_random}%
\end{table}%

\begin{table}[!htbp]
\centering
\caption{Performance comparison of TSSS Estimators for $m_{1}|\mathbb S^2$ and $m_{2}|\mathbb S^2$ in Simulation Study 1: $d_{\text{CV}}$ vs. fixed degree $d= 2,~3,~4,~5$, in the setting of homogeneous standard error with constant standard deviation $\sigma$. The frequency of selecting $d_{\text{CV}}$ among $100$ replications is denoted as $f_{d_{\text{CV}}}$. The reported average (and standard deviations) of PMSEs are scaled by a factor of $10^3$.}
\scalebox{0.58}{
    \begin{tabular}{ll|rr|rr|rr|rr|rr|rr|rr|rr|rr}
    \multicolumn{20}{c}{$m_1|\mathbb S^2$} \\
    \midrule
          &       & \multicolumn{6}{c|}{$N=8$}                    & \multicolumn{6}{c|}{$N=32$}                   & \multicolumn{6}{c}{$N=128$} \\
\cmidrule{3-20}          &       & \multicolumn{2}{c|}{$n=400$} & \multicolumn{2}{c|}{$n=900$} & \multicolumn{2}{c|}{$n=2500$} & \multicolumn{2}{c|}{$n=400$} & \multicolumn{2}{c|}{$n=900$} & \multicolumn{2}{c|}{$n=2500$} & \multicolumn{2}{c|}{$n=400$} & \multicolumn{2}{c|}{$n=900$} & \multicolumn{2}{c}{$n=2500$} \\
\cmidrule{3-20}    $\sigma$ & d     & PMSE  & $f_{d_{\text{CV}}}$ & PMSE  & $f_{d_{\text{CV}}}$ & PMSE  & $f_{d_{\text{CV}}}$ & PMSE  & $f_{d_{\text{CV}}}$ & PMSE  & $f_{d_{\text{CV}}}$ & PMSE  & $f_{d_{\text{CV}}}$ & PMSE  & $f_{d_{\text{CV}}}$ & PMSE  & $f_{d_{\text{CV}}}$ & PMSE  & $f_{d_{\text{CV}}}$ \\
    \midrule
    \multirow{10}[2]{*}{0.5} & 2     & 335.03 & 0     & 323.74 & 0     & 318.80 & 0     & 136.40 & 0     & 129.96 & 0     & 125.93 & 0     & 33.49 & 30    & 26.08 & 0     & 21.38 & 0 \\
          &       & (8.35) &       & (3.04) &       & (1.38) &       & (5.78) &       & (2.59) &       & (1.02) &       & (5.45) &       & (1.91) &       & (0.81) &  \\
          & 3     & 24.11 & 65    & 14.26 & 38    & 8.87  & 1     & 25.91 & 55    & 15.17 & 27    & 8.82  & 2     & 27.04 & 48    & 15.52 & 93    & 7.60  & 99 \\
          &       & (4.67) &       & (2.27) &       & (0.78) &       & (5.56) &       & (2.50) &       & (1.16) &       & (5.52) &       & (2.56) &       & (1.19) &  \\
          & 4     & 25.32 & 30    & 13.42 & 55    & 6.34  & 75    & 30.59 & 44    & 16.76 & 73    & 7.69  & 98    & 39.99 & 18    & 23.84 & 7     & 12.34 & 1 \\
          &       & (6.79) &       & (2.72) &       & (1.15) &       & (6.57) &       & (2.75) &       & (1.21) &       & (6.87) &       & (2.96) &       & (1.46) &  \\
          & 5     & 33.35 & 5     & 15.69 & 7     & 6.43  & 24    & 49.89 & 1     & 26.48 & 0     & 11.88 & 0     & 61.78 & 4     & 38.59 & 0     & 19.71 & 0 \\
          &       & (7.13) &       & (2.86) &       & (1.23) &       & (8.98) &       & (3.70) &       & (1.65) &       & (9.22) &       & (4.15) &       & (1.96) &  \\
          & $d_{\text{CV}}$ & 25.81 & --    & 14.43 & --    & 6.51  & --    & 27.06 & --    & 13.36 & --    & 5.71  & --    & 29.75 & --    & 13.15 & --    & 5.78  & -- \\
          &       & (5.96) &       & (2.73) &       & (1.17) &       & (7.00) &       & (2.92) &       & (1.03) &       & (7.08) &       & (2.72) &       & (1.10) &  \\
    \midrule
    \multirow{10}[2]{*}{0.75} & 2     & 342.96 & 0     & 328.20 & 0     & 320.39 & 0     & 146.63 & 0     & 135.05 & 0     & 127.77 & 0     & 47.57 & 53    & 33.40 & 21    & 24.36 & 0 \\
          &       & (11.36) &       & (4.83) &       & (1.96) &       & (10.33) &       & (4.62) &       & (1.73) &       & (10.95) &       & (3.96) &       & (1.62) &  \\
          & 3     & 41.67 & 75    & 23.52 & 62    & 12.54 & 34    & 49.01 & 44    & 27.61 & 47    & 14.03 & 45    & 58.42 & 33    & 34.05 & 79    & 16.80 & 100 \\
          &       & (10.31) &       & (4.59) &       & (1.65) &       & (11.43) &       & (5.15) &       & (2.54) &       & (12.22) &       & (5.72) &       & (2.65) &  \\
          & 4     & 50.99 & 22    & 25.46 & 36    & 11.62 & 60    & 67.30 & 56    & 37.25 & 52    & 17.18 & 55    & 88.21 & 12    & 53.07 & 0     & 27.59 & 0 \\
          &       & (14.28) &       & (5.51) &       & (2.47) &       & (14.45) &       & (6.15) &       & (2.71) &       & (15.32) &       & (6.68) &       & (3.29) &  \\
          & 5     & 70.56 & 3     & 33.59 & 2     & 13.63 & 6     & 110.15 & 0     & 58.91 & 1     & 26.56 & 0     & 136.45 & 2     & 85.91 & 0     & 44.04 & 0 \\
          &       & (15.16) &       & (6.30) &       & (2.66) &       & (19.88) &       & (8.25) &       & (3.70) &       & (20.66) &       & (9.34) &       & (4.40) &  \\
          & $d_{\text{CV}}$ & 44.03 & --    & 24.56 & --    & 12.13 & --    & 48.61 & --    & 26.35 & --    & 12.68 & --    & 47.37 & --    & 25.14 & --    & 11.21 & -- \\
          &       & (11.70) &       & (5.24) &       & (2.35) &       & (12.88) &       & (5.76) &       & (2.45) &       & (13.40) &       & (6.03) &       & (2.10) &  \\
   \midrule
    \multicolumn{20}{c}{$m_2|\mathbb S^2$} \\
    \midrule
    \multirow{10}[2]{*}{0.5} & 2     & 121.44 & 0     & 115.64 & 0     & 112.39 & 0     & 119.71 & 0     & 113.11 & 0     & 109.30 & 0     & 116.10 & 0     & 108.21 & 0     & 103.17 & 0 \\
          &       & (4.37) &       & (2.17) &       & (0.95) &       & (5.38) &       & (2.38) &       & (1.01) &       & (5.92) &       & (2.53) &       & (1.14) &  \\
          & 3     & 18.38 & 67    & 9.06  & 58    & 4.14  & 33    & 19.79 & 70    & 10.24 & 83    & 4.27  & 96    & 25.88 & 69    & 15.02 & 100   & 7.40  & 99 \\
          &       & (4.97) &       & (2.22) &       & (0.70) &       & (4.88) &       & (2.34) &       & (1.07) &       & (5.32) &       & (2.61) &       & (1.17) &  \\
          & 4     & 21.45 & 31    & 10.13 & 40    & 4.00  & 67    & 29.97 & 29    & 16.54 & 17    & 7.64  & 4     & 39.22 & 29    & 23.60 & 0     & 12.28 & 1 \\
          &       & (6.42) &       & (2.47) &       & (0.98) &       & (6.27) &       & (2.75) &       & (1.20) &       & (6.68) &       & (3.06) &       & (1.46) &  \\
          & 5     & 30.86 & 2     & 14.48 & 2     & 5.84  & 0     & 49.37 & 1     & 26.30 & 0     & 11.86 & 0     & 60.68 & 2     & 38.19 & 0     & 19.61 & 0 \\
          &       & (6.48) &       & (2.74) &       & (1.16) &       & (8.99) &       & (3.65) &       & (1.64) &       & (9.14) &       & (4.25) &       & (1.95) &  \\
          & $d_{\text{CV}}$ & 19.86 & --    & 9.70  & --    & 4.00  & --    & 21.33 & --    & 10.36 & --    & 3.90  & --    & 21.08 & --    & 10.01 & --    & 4.75  & -- \\
          &       & (5.68) &       & (2.71) &       & (0.95) &       & (6.08) &       & (2.33) &       & (0.99) &       & (5.81) &       & (2.29) &       & (0.92) &  \\
    \midrule
    \multirow{10}[2]{*}{0.75} & 2     & 130.96 & 0     & 120.07 & 0     & 113.98 & 0     & 129.74 & 0     & 118.19 & 0     & 111.22 & 0     & 128.35 & 0     & 114.90 & 0     & 105.88 & 0 \\
          &       & (9.55) &       & (4.15) &       & (1.69) &       & (9.99) &       & (4.30) &       & (1.79) &       & (11.15) &       & (4.67) &       & (1.99) &  \\
          & 3     & 35.57 & 75    & 17.61 & 68    & 7.38  & 65    & 43.27 & 82    & 22.72 & 93    & 9.47  & 98    & 57.39 & 95    & 33.58 & 100   & 16.59 & 100 \\
          &       & (10.45) &       & (4.65) &       & (1.53) &       & (10.87) &       & (5.20) &       & (2.36) &       & (11.96) &       & (5.83) &       & (2.62) &  \\
          & 4     & 47.20 & 22    & 22.39 & 32    & 8.84  & 35    & 66.73 & 17    & 37.01 & 7     & 17.13 & 2     & 87.50 & 5     & 52.85 & 0     & 27.54 & 0 \\
          &       & (14.18) &       & (5.50) &       & (2.23) &       & (14.03) &       & (6.16) &       & (2.71) &       & (15.07) &       & (6.83) &       & (3.28) &  \\
          & 5     & 68.39 & 3     & 32.28 & 0     & 13.07 & 0     & 109.72 & 1     & 58.67 & 0     & 26.54 & 0     & 135.34 & 0     & 85.50 & 0     & 43.94 & 0 \\
          &       & (14.57) &       & (6.13) &       & (2.62) &       & (19.92) &       & (8.17) &       & (3.68) &       & (20.57) &       & (9.51) &       & (4.39) &  \\
          & $d_{\text{CV}}$ & 38.50 & --    & 19.01 & --    & 7.69  & --    & 39.20 & --    & 19.01 & --    & 7.79  & --    & 37.04 & --    & 20.51 & --    & 10.21 & -- \\
          &       & (10.79) &       & (5.35) &       & (1.81) &       & (11.94) &       & (5.52) &       & (1.90) &       & (9.50) &       & (4.74) &       & (1.98) &  \\
    \bottomrule
    \end{tabular}%
}

\label{tab:CV_d_const}%
\end{table}%

\begin{figure}[!ht]
  \centering
  \subfloat[$m_{1}|{\mathbb{S}^2}$]{\includegraphics[width=0.45\columnwidth]{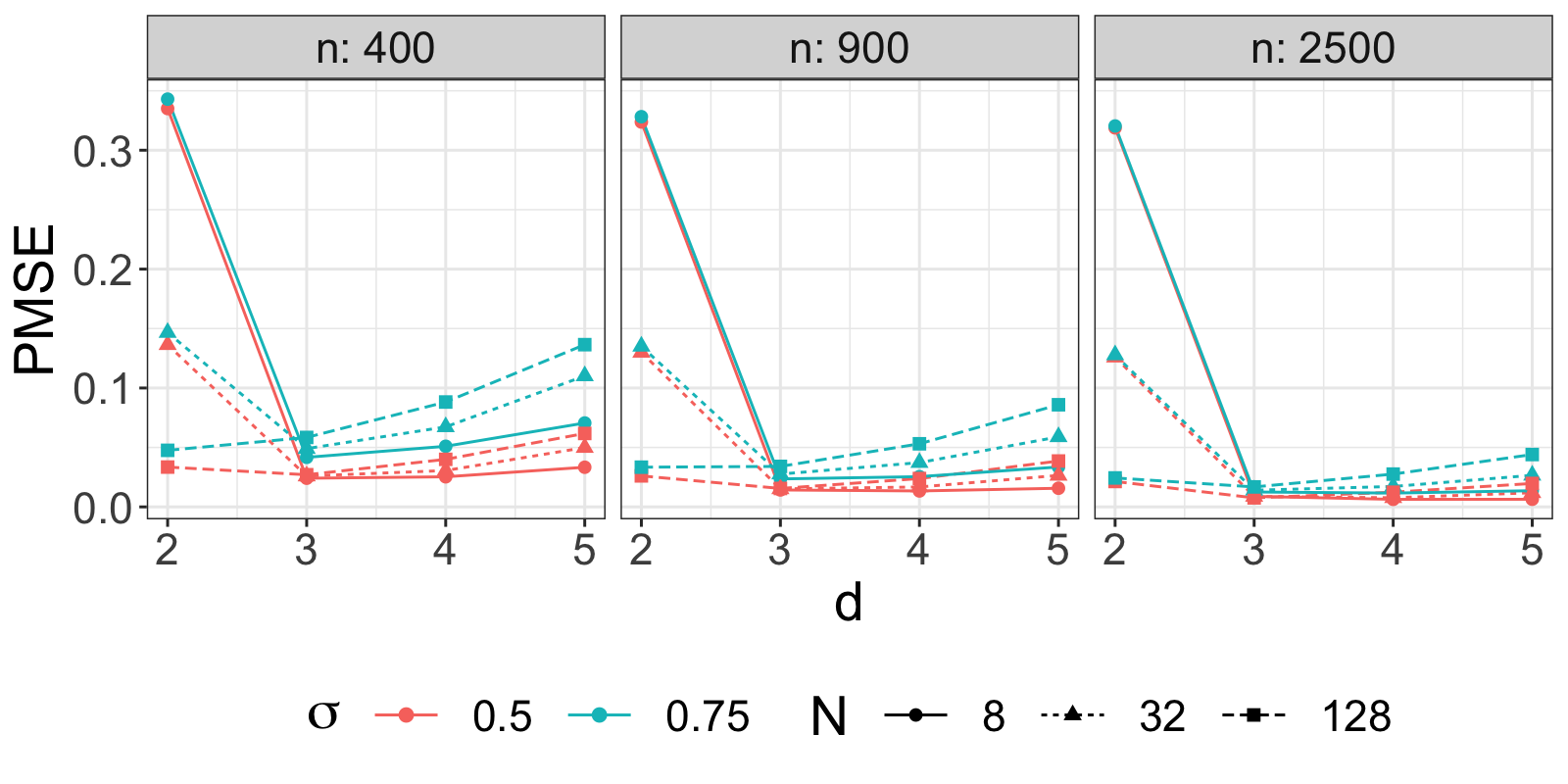}}
    \subfloat[$m_2|{\mathbb{S}^2}$]{\includegraphics[width=0.45\columnwidth]{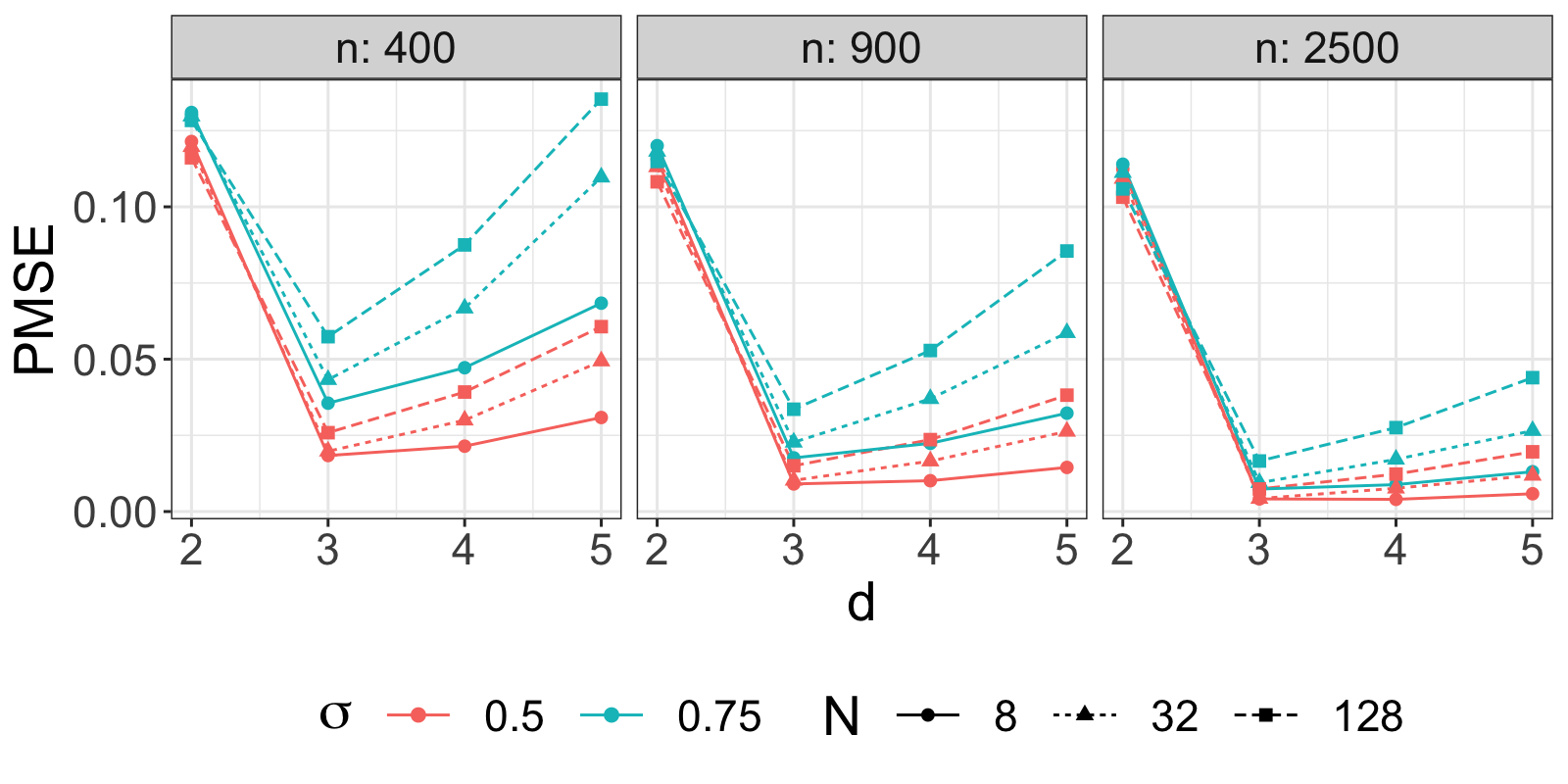}}
       \caption{Performance of TSSS for $m_1|\mathbb S^2$ and $m_2|\mathbb S^2$ in Simulation Study 1 in the setting of constant standard deviation with different $n$, $d$, $\sigma$, and $N$.}
  \label{fig:Y12TSSS-dtri_const}
\end{figure}
\begin{figure}[!ht]
\centering
 \subfloat[$m_{1}|\mathbb{S}^2$]{\includegraphics[width=0.9\columnwidth]{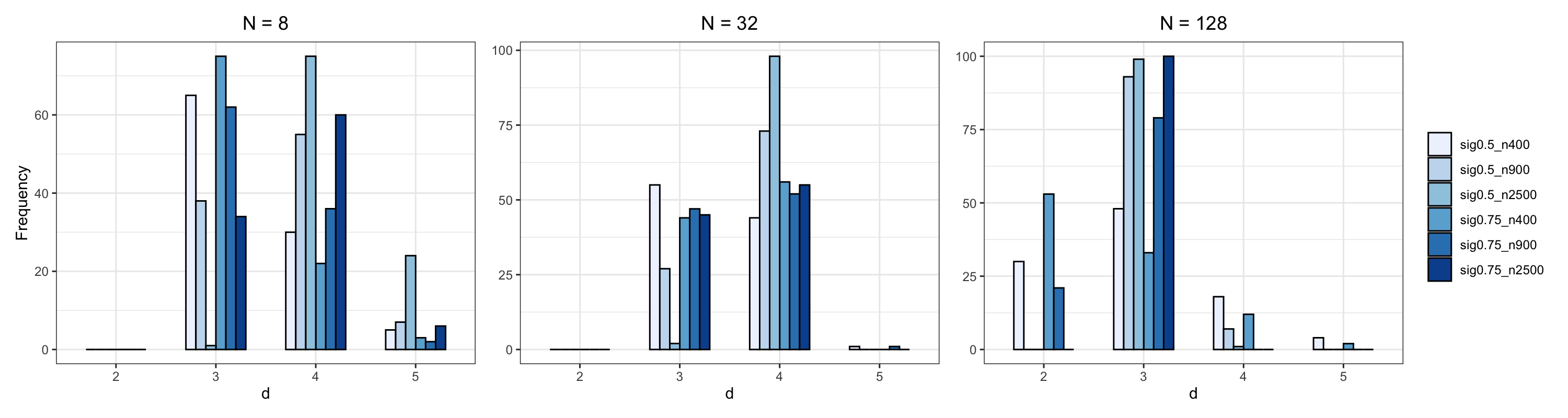}}
 \qquad
 \subfloat[$m_{2}|\mathbb{S}^2$]{\includegraphics[width=0.9\columnwidth]{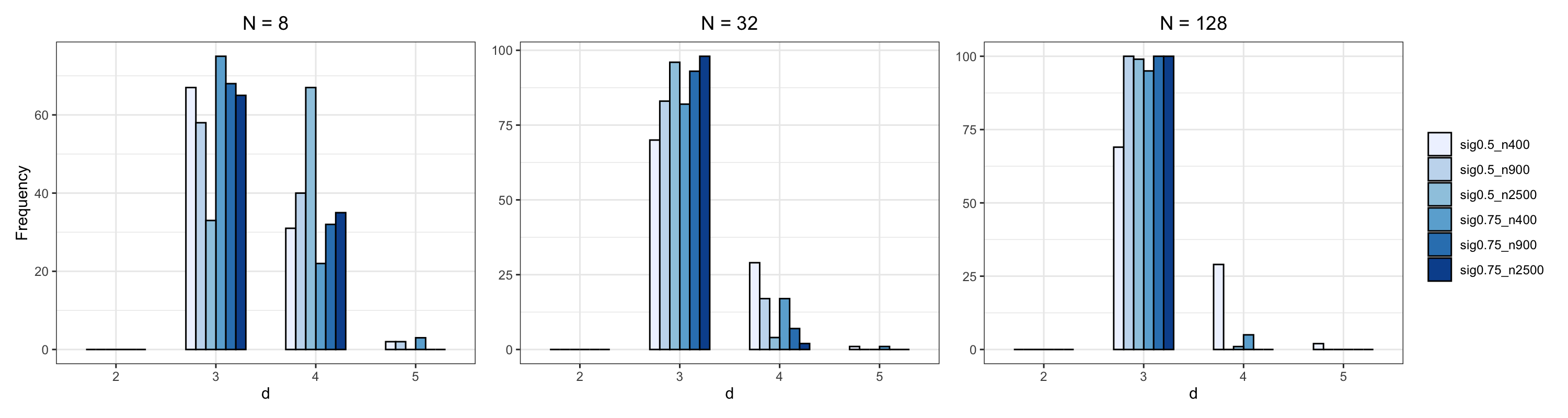}}
\caption{The bar graphs of $d_{\text{CV}}$ in Simulation Study 1 with homogeneous errors under different combinations of $\sigma = 0.50,~0.75$ (sig), $n=400, ~900, ~ 2,500$, and $N = 8, ~32, ~ 128$.}
\label{fig:d_hist_const}
\end{figure}

\begin{table}[!ht]
\centering
\caption{Simulation Study 2 results for different estimation methods of $m_{3}|\Omega$ with homogeneous errors: the average (and standard deviations) of predicted mean squared error (PMSE), training mean squared error (TMSE), dimension of the design parameters (Dim), and computation time per iteration in seconds (Time). A factor of $10^3$ scales the reported average (and standard deviations) of PMSEs and TMSEs. The results for TSSS are based on a triangulation with $N = 32$ triangles and spline basis functions with either a fixed degree $d=2$ or a CV-selected degree $d_{\text{CV}}$. The results for TPSOS and Tensor-Sphere are based on dimensions $k = 100$ and $k = 64$, respectively. Kernel method results have only $30$ iterations instead of $100$ due to computation inefficiency.}
 \scalebox{0.52}{
    \begin{tabular}{c|l|cccccc|cccccc|cccccc}
    \multicolumn{20}{c}{$m_3|\Omega$} \\
    \midrule
          &       & \multicolumn{6}{c|}{$n=400$}                  & \multicolumn{6}{c|}{$n=900$}                  & \multicolumn{6}{c}{$n=2500$} \\
    \midrule
    $\sigma$ & \multicolumn{1}{c|}{Method} & \multicolumn{2}{c}{PMSE} & \multicolumn{2}{c}{TMSE} & Dim   & Time(s) & \multicolumn{2}{c}{PMSE} & \multicolumn{2}{c}{TMSE} & Dim   & Time(s) & \multicolumn{2}{c}{PMSE} & \multicolumn{2}{c}{TMSE} & Dim   & Time(s) \\
    \midrule
    \multirow{5}[2]{*}{0.5} & Kernel & 144.14 & (30.8) & 134.83 & (33.5) & 400   & 5882  & 115.38 & (43.1) & 114.25 & (43.6) & 900   & 9273  & 54.83 & (17.9) & 56.34 & (19.3) & 2500  & 26355 \\
          & TPSOS & 103.94 & (13.7) & 72.71 & (8.1) & 100   & 11    & 57.79 & (6.0) & 48.50 & (4.7) & 100   & 19    & 33.37 & (1.7) & 32.88 & (1.8) & 100   & 44 \\
          & Tensor-Sphere & 61.78 & (8.1) & 54.50 & (6.4) & 64    & 0     & 45.89 & (3.7) & 43.59 & (4.4) & 64    & 0     & 31.14 & (1.8) & 31.74 & (1.7) & 64    & 0 \\
          & TSSS, $d=2$ & 19.78 & (5.9) & 18.12 & (5.0) & 45    & 25    & 10.71 & (2.5) & 10.16 & (2.3) & 45    & 25    & 5.44  & (0.9) & 5.31  & (0.9) & 45    & 26 \\
          & TSSS, $d_{\text{CV}}$ & 20.65 & (6.7) & 18.78 & (5.0) & 90    & 110   & 11.26 & (2.6) & 10.65 & (2.4) & 82    & 130   & 5.35  & (1.0) & 5.21  & (1.0) & 115   & 159 \\
    \midrule
    \multirow{5}[2]{*}{0.75} & Kernel & 173.34 & (17.8) & 163.18 & (24.9) & 400   & 5925  & 141.53 & (31.9) & 140.02 & (32.6) & 900   & 9271  & 94.02 & (45.1) & 96.41 & (47.3) & 2500  & 26394 \\
          & TPSOS & 152.72 & (18.3) & 121.42 & (14.9) & 100   & 11    & 87.23 & (9.8) & 76.34 & (8.8) & 100   & 21    & 45.45 & (3.4) & 44.82 & (3.3) & 100   & 42 \\
          & Tensor-Sphere & 86.48 & (15.0) & 78.23 & (12.2) & 64    & 0     & 59.44 & (6.1) & 58.18 & (6.3) & 64    & 0     & 40.05 & (3.9) & 40.84 & (4.3) & 64    & 0 \\
          & TSSS, $d=2$ & 37.37 & (12.4) & 34.57 & (10.6) & 45    & 25    & 19.77 & (5.6) & 18.85 & (5.3) & 45    & 25    & 9.27  & (2.1) & 9.06  & (2.0) & 45    & 26 \\
          & TSSS, $d_{\text{CV}}$ & 34.28 & (10.3) & 32.20 & (9.1) & 54    & 121   & 20.00 & (6.2) & 19.09 & (5.7) & 63    & 138   & 9.61  & (2.1) & 9.42  & (2.0) & 60    & 173 \\
    \bottomrule
    \end{tabular}%
   }
      \label{tab:sim3_const}
\end{table}

 \FloatBarrier
\newpage

\setcounter{equation}{0}
\setcounter{proposition}{0}
\setcounter{subsection}{0}
\setcounter{figure}{0}
\setcounter{table}{0}

\section{Technical Proofs}
\label{SEC:proofs}

\subsection{Radial Projection}
\begin{figure}[!ht]
    \centering
   \scalebox{1}{
   	 \begin{tabular}{cccc}
	\includegraphics[width = 0.18\textwidth]{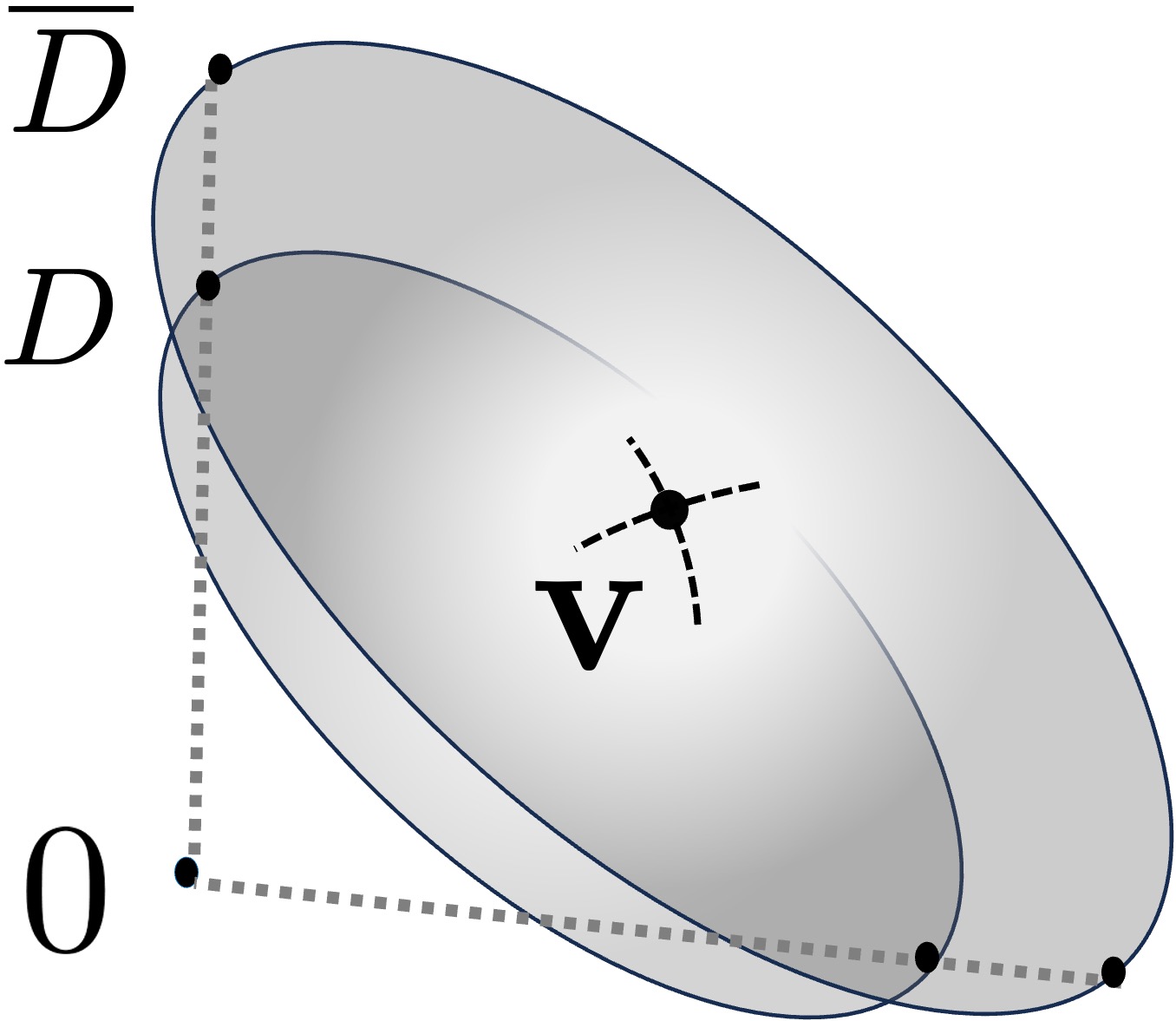}&
    \includegraphics[width = 0.15\textwidth]{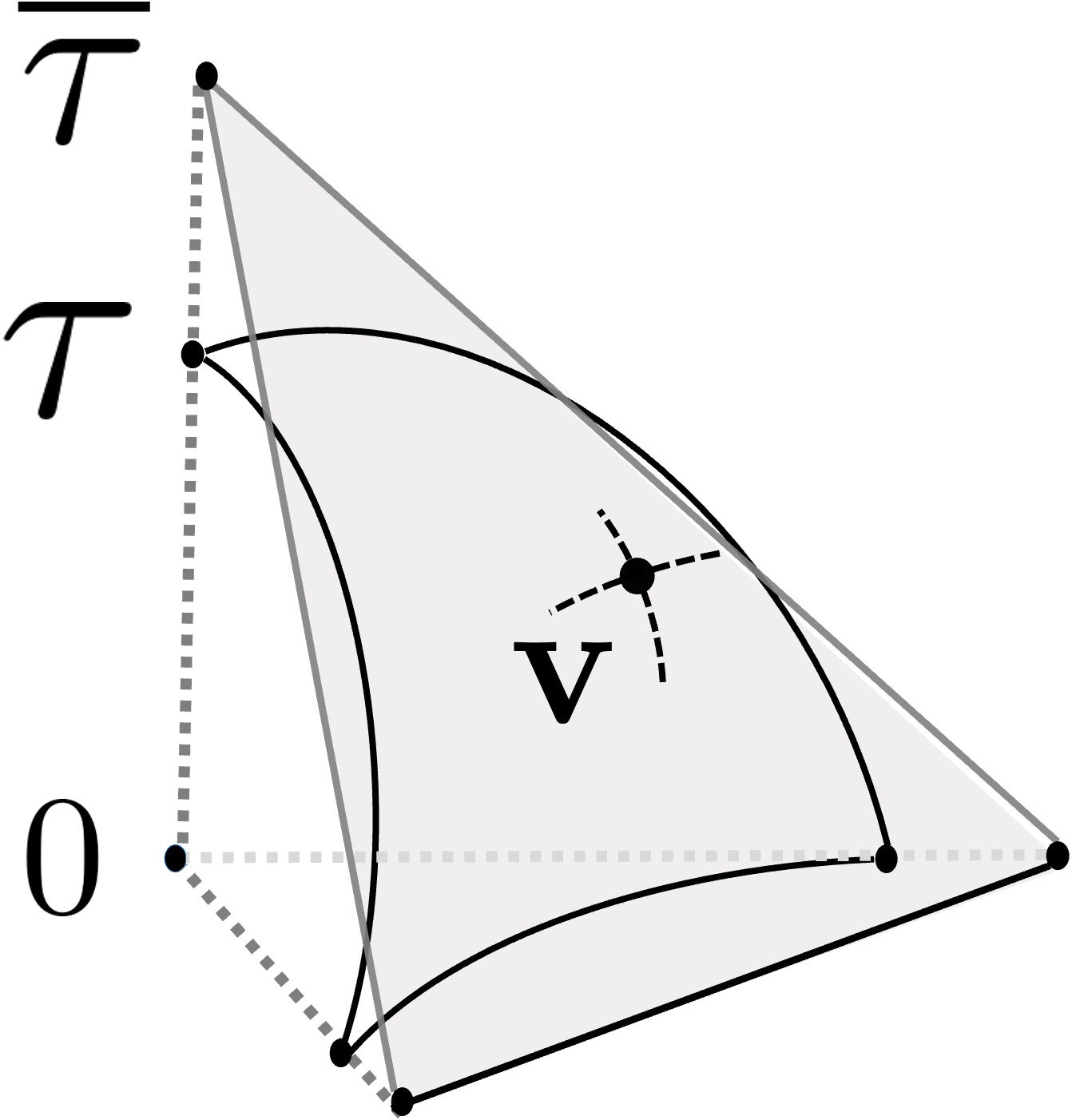}&
	 \includegraphics[width = 0.15\textwidth]{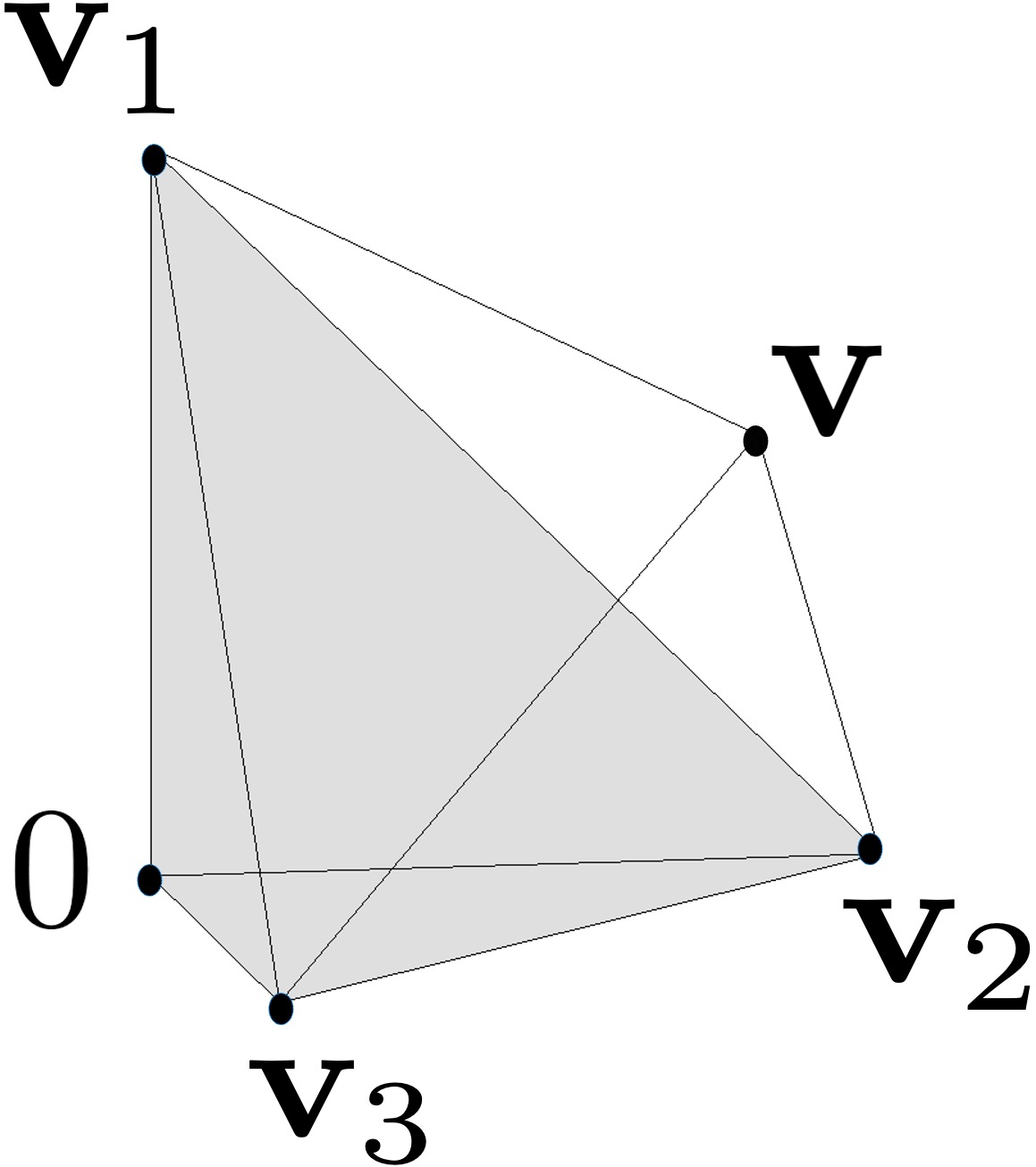} & 
	\includegraphics[width = 0.15\textwidth]{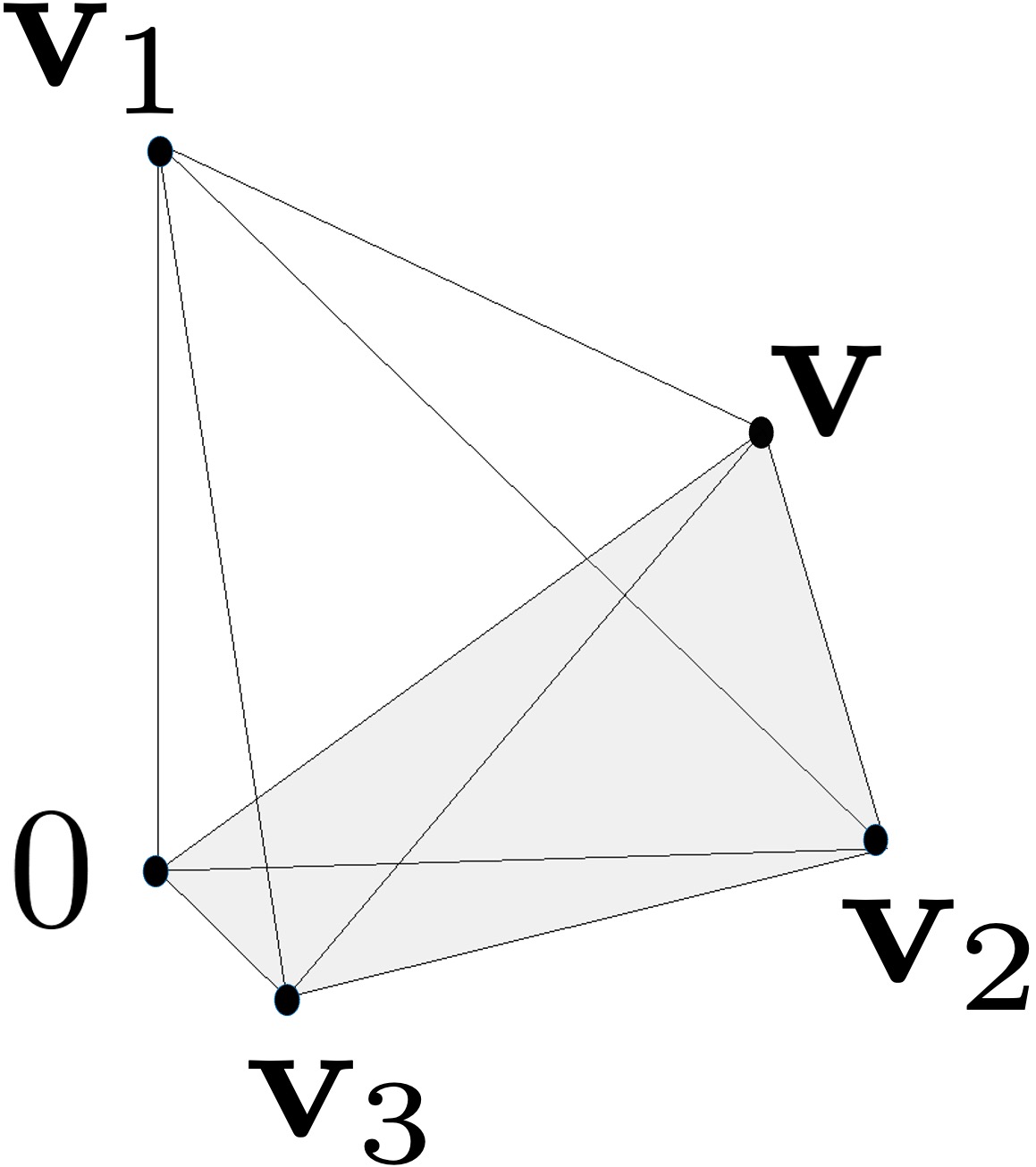}\\
	(a) & (b)& (c) & (d)
    \end{tabular}
    }
    \caption{Illustrations of (a) radial projection of spherical cap $D$, (b) radial projection of spherical triangle $\tau$, (c) tetrahedron $\uptau$, and (d) tetrahedron $\uptau_1$.}
    \label{fig:sph_proj}
\end{figure}

To prepare for the definition of norms on a sphere, we introduce some elements of radial projection, which maps a local spherical domain to a planar 2D domain. 
Define the radial projection of any point $\mathbf v$ in a spherical cap $D$ to a unique point  $\overline{\mathbf v}$ on the plane $\overline D$ tangent to the sphere at $\mathbf v_D$ through a mapping $\mathcal R_D: D\to\overline D$. Similarly, we can define the projection of a spherical triangle $\tau$ to $\overline \tau = \mathcal R_D \tau$; see illustrations of radial projection in Figure \ref{fig:sph_proj}.

\begin{remark}
If $\Omega$ is contained in a spherical cap of radius $1/2$, we have the $K_1\|f\|_{L^q(\Omega)} \leq \|\overline f_p\|_{L^q(\overline\Omega)}\leq K_2\|f\|_{L^q(\Omega)}$, where $K_1, K_2$ depends on $q$, $\overline \Omega$, and $p$, see Lemma 14.7 of \citep{Lai:Schumaker:07}. This demonstrates the $L^q$-norm of $f$'s homogeneous extension can be bounded by the $L^q$-norm of $f$. 
\end{remark}

\subsection{Properties of Spherical Sobolev Space}
\begin{figure}[!ht]
\centering
\includegraphics[width = 0.6\textwidth]{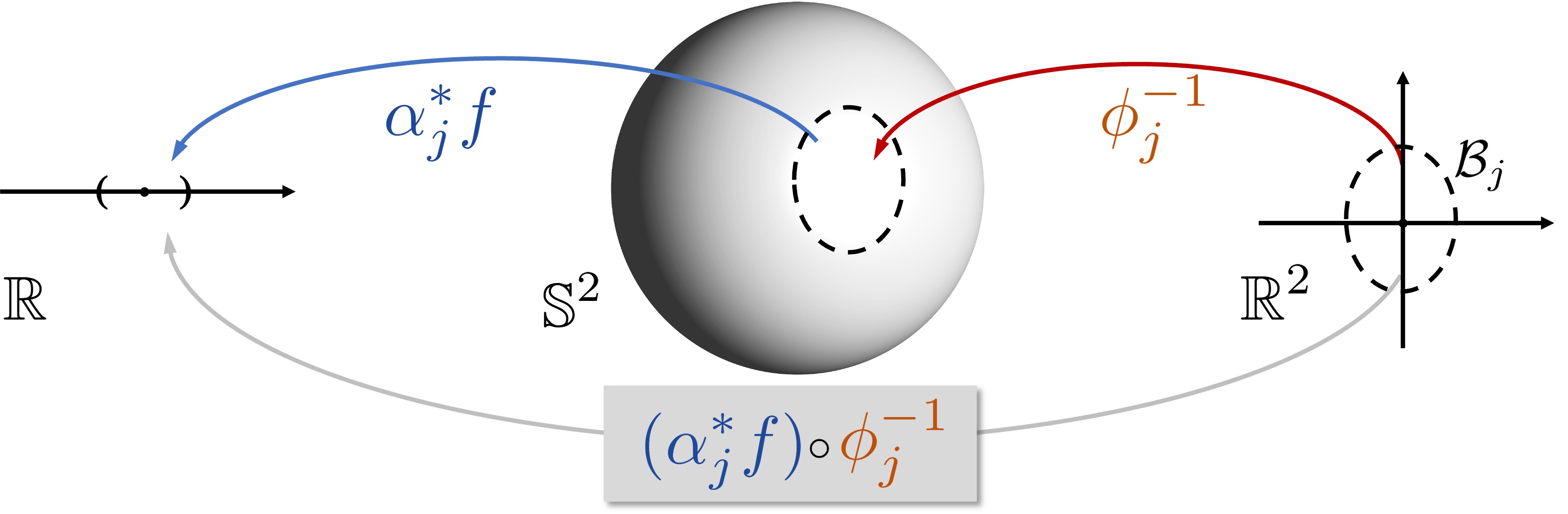}
\caption{Illustration for spherical Sobolev space construction.}
\label{fig:sobolev}
\end{figure}

In this section, we present the properties of spherical Sobolev space introduced in Section \ref{SEC:convergence}. Recall the construction of spherical Sobolev space $W^{\ell}_q(\Omega)$ from $W^{\ell}_q(\overline \Omega)$ through $$W^{\ell}_q(\Omega) := \left\{
f: (\alpha_j^* f)\circ \phi_j^{-1} \in W^{\ell}_q(\overline \Omega_j), \forall j\right\}, \, \ell >0,$$ 
where $\overline \Omega_j \subset \mathbb R^2$ is an open set and the support of $\phi_j^{-1}$; and $\alpha_j^*$ is a mapping from ${\mathbb S^2}$ to $\mathbb R$ to indicate the partition of $\mathbb S^2$, such that $\sum_j\alpha_j^* = 1$ on $\Omega$.
See Figure \ref{fig:sobolev} for an illustration of spherical Sobolev space construction.

Lemma \ref{LEM:derivatives} establishes a bound on the norm of derivatives of a spherical polynomial.
\begin{lemma}[Theorem 14.21 \cite{Lai:Schumaker:07}] 
\label{LEM:derivatives}Suppose $\tau$ is a spherical triangle with $|\tau| \leq 1$. Then, there exists a constant $K$ depending only on $d$ such that for any spherical polynomial $p$ of degree $d$ and any $1 \leq q \leq \infty$, $|p|_{k,q,\tau} \leq K \rho_{\tau}^{-k} \|p\|_{q,\tau}$,  for  $0 \leq k \leq d$,
where $\rho_{\tau}$ is the radius of the largest spherical cap contained in $\tau$. 
\end{lemma}

\begin{figure}
\centering
\includegraphics[width = 0.4\textwidth]{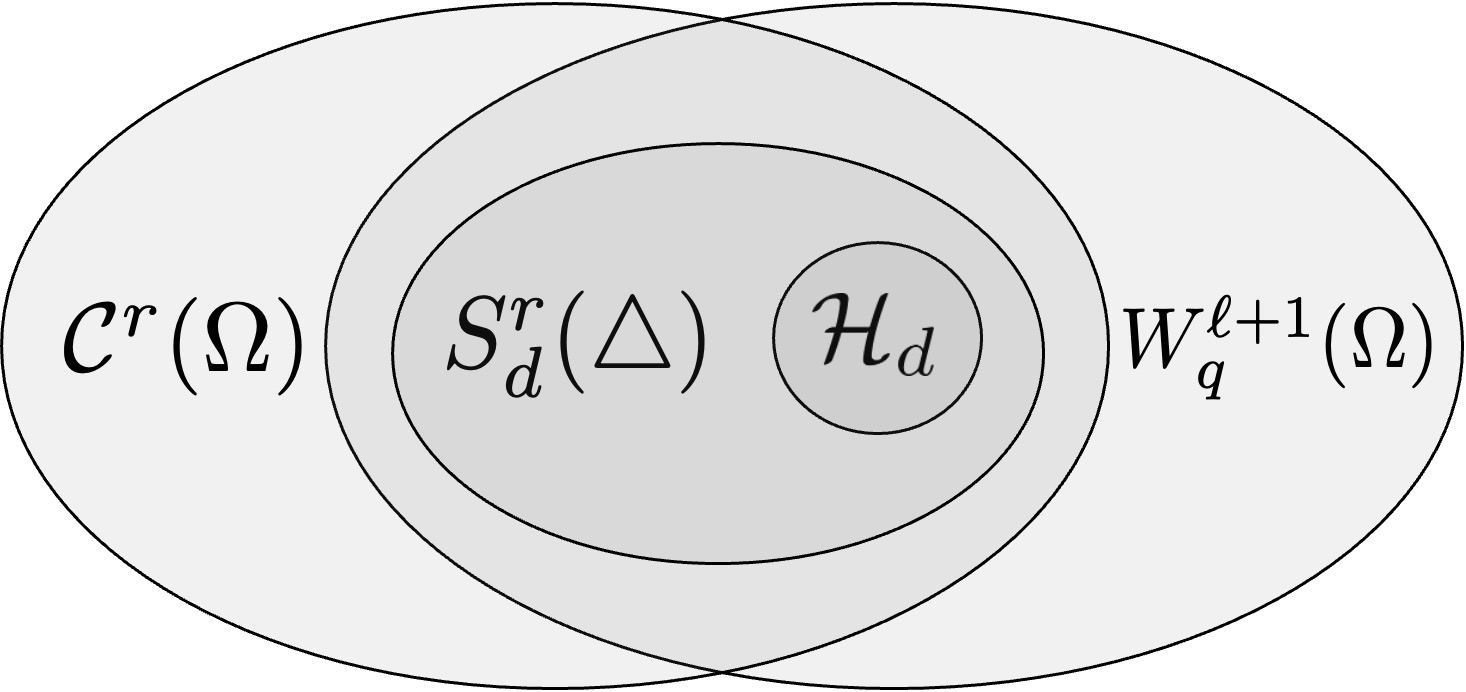}
\caption{Relationship between function spaces $\mathcal C^r(\Omega), \mathcal S_d^r(\triangle), \mathcal H_d$ and $W_q^{\ell + 1}(\Omega)$ for compact domain $\Omega$.}
\label{fig:funcspace}
\end{figure}

\begin{remark}
Note that we utilize SBB polynomials as basis to construct the homogeneous polynomial space of degree $d$, $\mathcal H_d$, over the whole domain $\Omega$. Then the smoothness constraint $\mathbf M \bs \gamma = \mathbf 0$, along all the edges and vertices of all spherical triangles $\tau \in \triangle$, is imposed to restrict the function space of interest to $\mathcal{S}_d^r(\triangle) = \{s \in C^r(\Omega), s|_\tau \in \mathcal{H}_d, \tau \in \triangle\}$. The mean function $m$ is assumed to reside in the spherical Sobolev space $W_q^{\ell + 1} (\Omega)$, which is a superset of any polynomial function space when $\Omega$ is compact. The relationship between these function spaces is illustrated in \Cref{fig:funcspace}.
\end{remark}

\subsection{Example of Spherical Bernstein Basis (SBB) Polynomials}
\label{ssec:prelimSBB}
\begin{figure}[!ht]
\centering
\includegraphics[width = 0.6 \textwidth]{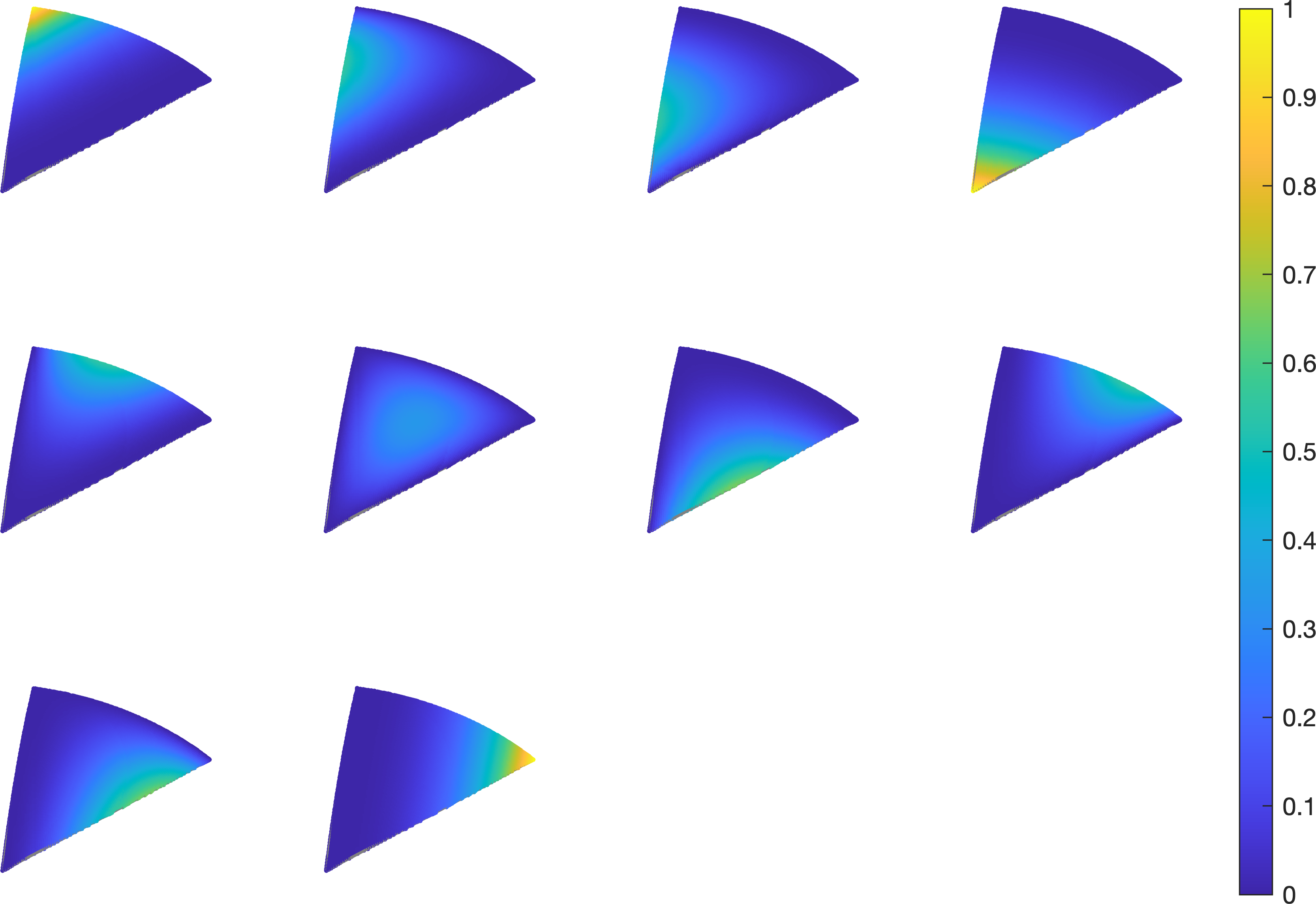}
\caption{SSB basis polynomials when $d =3$.}
\label{fig:d3SSB}
\end{figure}

The SBB polynomials are defined as $B_{ijk}^{\tau}(\bb{x}) = \frac{d!}{i!j!k!} b_1(\bb{x})^i b_2(\bb{x})^j b_3(\bb{x})^k$, $i+j+k = d$, $\tau \in \triangle$, where $b_{\kappa}(\mathbf{v}) = \mathrm{vol}(\uptau_{\kappa})/\mathrm{vol}(\uptau)$ are the spherical barycentric coordinates. Here $\mathrm{vol}$ stands for volume, $\uptau_{\kappa}$ is tetrahedron $\langle \mathbf{0}, \mathbf{v}, \mathbf{v}_{\kappa+1}, \mathbf{v}_{\kappa+2} \rangle$, and $\uptau$ is tetrahedron $\langle \mathbf{0}, \mathbf{v}_1, \mathbf{v}_2, \mathbf{v}_3 \rangle$. Illustrations of $\uptau$ and $\uptau_1$ are shown in \Cref{fig:sph_proj}.
When $d = 3$, there are ${d+2 \choose 2} = 10$ SBB polynomials of degree $3$, $\{B_{ijk}^{\tau} (\mathbf x)\}_{i+j+k = d}$, see \Cref{fig:d3SSB}.

\subsection{Proofs of Lemmas \ref{LEM:norm1} -- \ref{LEM:Vn}}
\vspace{-3pt}
In this section, we present preliminary results of SBB polynomials. 
Lemma \ref{LEM:stability} establishes the stability of SBB over a single spherical triangle.

\begin{lemma}
\label{LEM:stability}
 Let $\bs\gamma = \{\gamma_{ijk}^\tau, i+j+k = d, \tau\in\triangle\} \equiv \{\gamma_\xi,\xi\in\mathcal M\}$ be the coefficients of SBB basis $\{B_\xi, \xi\in\mathcal M\}$. Assume \ref{C3} such that all triangles in the triangulation $\triangle$ are comparable in the sense that {$c\max_{\tau\in\triangle} A_\tau \leq \min_{\tau\in\triangle} A_\tau$ for some constant $c>0$}, then
\[\frac{1}{K^2}|\triangle|^2\|\bs\gamma\|_2^2 \leq \|p\|_{\Omega}^2 \leq 3^{2d}|\triangle|^2 \|\bs\gamma\|_2^2,\]
and equivalently 
\[
\frac{1}{K^2}N^{-1}\|\bs\gamma\|_2^2 \leq \|p\|_{\Omega}^2\leq 3^{2d} N^{-1}\|\bs\gamma\|_2^2.
\]
\end{lemma}
\begin{proof}
By Assumption \ref{C3}, we have $\rho_\tau^2 \leq A_\tau \leq |\triangle|^2 \leq \rho^2 \rho_\tau^2$ for all $\tau\in\triangle$, implying $A_\tau \asymp |\triangle|^2$.
Let $\tau$ be a spherical triangle contained in a spherical cap of radius $1/2$, and let $1 \leq q \leq \infty$. Then by Theorem 14.9 of \cite{Lai:Schumaker:07}, for any SBB polynomial $p = \sum_{i+j+k = d} \gamma_{ijk}^\tau B_{ijk}^\tau$ of degree $d$, we have ${K}^{-1}A_{\tau}^{1/q}\|\bs{\gamma}^\tau\|_q \leq \|p\|_{q, \tau} \leq 3^d A_{\tau}^{1/q} \|\bs{\gamma}^\tau\|_q$, where constant $K$ only depends on $d$. Let $q=2$, we have 
$K^{-1}|\triangle|\|\bs{\gamma}^\tau\|_2 \leq \|p\|_{2,\tau} \leq 3^d |\triangle|\|\bs{\gamma}^\tau\|_2$. 
Further, we can extend the result to the closure of $\Omega\subseteq\mathbb S^2$,
$K^{-2}|\triangle|^2\|\bs{\gamma}\|_2^2 \leq \|p\|_{L^2(\Omega)}^2 \leq 3^{2d}|\triangle|^2\|\bs{\gamma}\|_2^2$.
In addition, by Assumption {\ref{C1}}, we have 
\begin{align}
c \|p\|_{L^2(\Omega)} \leq \|p\|_{\Omega}^2 \leq C \|p\|_{L^2(\Omega)},
\label{EQN:L2Omega}
\end{align}
for some positive constants $0<c, C<\infty$.
Therefore, the lemma is proved.   
\end{proof}

\begin{proof}[Proof of \Cref{LEM:norm1}]

By the definition of $\|\cdot\|_{n,\Omega}$ and $\|\cdot\|_{\Omega}$, we have 
\begin{align*}
\langle p_1, p_2\rangle_{n,\Omega} & = 
\frac{1}{n} \sum_{i = 1}^n \left\{ \sum_{\xi_1\in \mathcal M} \gamma_{\xi_1} B_{\xi_1} (\mathbf X_i)\right\}\left\{ \sum_{\xi_2 \in \mathcal M} \tilde\gamma_{\xi_2} B_{\xi_2}(\mathbf X_i)\right\} = 
\sum_{\xi_1, \xi_2 \in\mathcal M} \gamma_{\xi_1}\tilde \gamma_{\xi_2} \langle B_{\xi_1}, B_{\xi_2}\rangle_{n,\Omega},\\
\|p_1\|_{\Omega}^2 & = \sum_{\xi_1, \xi_1'} \gamma_{\xi_1}\tilde \gamma_{\xi_1'} \langle B_{\xi_1},B_{\xi_1'}\rangle_\Omega, \quad 
\|p_2\|_{\Omega}^2 = \sum_{\xi_2, \xi_2'} \gamma_{\xi_2}\tilde \gamma_{\xi_2'} \langle B_{\xi_2}, B_{\xi_2'}\rangle_\Omega.
\end{align*}
By \Cref{LEM:stability}, we have 
$$
\frac{1}{NK^2}\left\{\sum_{\xi_1\in\mathcal M}\gamma_{\xi_1}^2 \right\}^{1/2}\left\{\sum_{\xi_2\in\mathcal M}\tilde\gamma_{\xi_2}^2 \right\}^{1/2} \leq \|p_1\|_{\Omega} \|p_2\|_{\Omega} \leq \frac{3^{2d}}{N}\left\{\sum_{\xi_1\in\mathcal M}\gamma_{\xi_1}^2 \right\}^{1/2}\left\{\sum_{\xi_2\in\mathcal M}\tilde\gamma_{\xi_2}^2 \right\}^{1/2}.
$$
Therefore, by Cauchy-Schwarz inequality,
\begin{align*}
R_n &= \frac{|\langle p_1, p_2\rangle_{n,\Omega} -\langle p_1, p_2\rangle_{\Omega}|}{\|p_1\|_\Omega \|p_2\|_\Omega} \\
&\leq \frac{\sum_{\xi_1,\xi_2\in\mathcal M}|\gamma_{\xi_1}\tilde \gamma_{\xi_2}|}{N^{-1}\left\{\sum_{\xi_1\in\mathcal M}\gamma_{\xi_1}^2\sum_{\xi_2\in\mathcal M}\tilde\gamma_{\xi_2}^2 \right\}^{1/2}}  \max_{\xi_1, \xi_2 \in\mathcal M}
\left|\langle B_{\xi_1}, B_{\xi_2}\rangle_{n,\Omega} - \langle B_{\xi_1}, B_{\xi_2}\rangle_{\Omega}\right|\\
&\leq C {N} \max_{\xi_1, \xi_2 \in\mathcal M}
\left|\langle B_{\xi_1}, B_{\xi_2}\rangle_{n,\Omega} - \langle B_{\xi_1}, B_{\xi_2}\rangle_{\Omega}\right|.
\end{align*}
It only remains to prove 
$$
\max_{\xi_1, \xi_2\in\mathcal M} \left|\langle B_{\xi_1}, B_{\xi_2}\rangle_{n,\Omega} - \langle B_{\xi_1}, B_{\xi_2}\rangle_{\Omega}\right|= O_p\left\{(\log n)^{1/2} (nN)^{-1/2}\right\}.
$$
Let $Z_i = B_{\xi_1}(\mathbf X_i) B_{\xi_2}(\mathbf X_i)  -\E\{B_{\xi_1}(\mathbf X_i) B_{\xi_2}(\mathbf X_i)\}$. We have $\E Z^2_i  \asymp |\triangle|^2$. In addition, $\E |Z_i|^k  \asymp \E|B_{\xi_1}(\mathbf X_i) B_{\xi_2}(\mathbf X_i) |^k \asymp |\triangle|^2 \leq C2^{k-1} k! \E Z_i^2$, for some constant $0 <C<\infty$.
Thus, by union bound and Bernstein's inequality, we have for large $t>0$, 
\begin{align*}
& \sum_{n = 1}^\infty \mathrm P\left(\max_{\xi_1, \xi_2 \in\mathcal M} \left|\frac{1}{n} \sum_{i=1}^n  \langle B_{\xi_1}, B_{\xi_2}\rangle_{n,\Omega} - \langle B_{\xi_1}, B_{\xi_2}\rangle_{\Omega}\right|\geq t\sqrt{\frac{\log n}{nN}}\right) \\
\leq & |\mathcal M|^2 \sum_{n = 1}^\infty \mathrm P\left( \left|\frac{1}{n} \sum_{i=1}^n  \langle B_{\xi_1}, B_{\xi_2}\rangle_{n,\Omega} - \langle B_{\xi_1}, B_{\xi_2}\rangle_{\Omega}\right|\geq t\sqrt{\frac{\log n}{nN}}\right)\\
\leq & \sum_{n = 1}^\infty |\mathcal M|^2 \exp\left( \frac{-t^2}{4 + 2ct\sqrt{N\log n/n}}\right) 
\leq \sum_{n = 1}^\infty |\mathcal M|^2 n^{-4} < \sum_{n = 1}^\infty  n^{-2} <\infty.
\end{align*}
The second last inequality holds due to  $N\asymp cn^\omega, \omega <1, c>0$ by Assumption {\ref{C3}}.
With Borel-Cantelli lemma, we have 
$$\max_{\xi_1, \xi_2 \in\mathcal M} \left|\frac{1}{n} \sum_{i=1}^n  \langle B_{\xi_1}, B_{\xi_2}\rangle_{n,\Omega} - \langle B_{\xi_1}, B_{\xi_2}\rangle_{\Omega}\right| = O_p(\sqrt{\log n/nN}),$$ 
and the proof is completed.
\end{proof}

As a result of \Cref{LEM:norm1}, we have
\begin{align}
 \left|\frac{\|p\|_{n,\Omega}^2 - \|p\|_{\Omega}^2}{\|p\|_\Omega^2} \right| = 
\left|\frac{\|p\|_{n,\Omega}^2}{ \|p\|_{\Omega}^2}-1\right|
= O_p \left\{ (N\log n)^{1/2} n^{-1/2} \right\},\nonumber
\text{ and } \\
 c\|p\|_{\Omega}^2 \leq \|p\|_{n,\Omega}^2 \leq C\|p\|_{\Omega}^2, ~\text{ for some constants }  0<c,C<\infty. \label{EQN:Rn}
\end{align}

\begin{proof}[Proof of \Cref{LEM:Vn}]
Firstly, let $1\leq q\leq \infty$. By Theorem 14.8 of \cite{Lai:Schumaker:07}, for any $\tau$ contained in a spherical cap of radius $1/2$ and $p\in\mathcal S_d^r(\tau)$, there exists a constant $K$ depending only on $d, q, $ and the smallest angle of $\tau$, such that $A_\tau^{-1/q}\|p\|_{q,\tau} \leq \|p\|_{\infty, \tau}\leq K A_\tau^{-1/q} \|p\|_{q,\tau}$.
Here $A_\tau$ is the area of $\tau$, and of order $|\triangle|^2$. Let $q = 2$, we have
\begin{equation}
\label{EQN:markov_inf}
\|p\|_{\infty,\Omega} \leq \sum_{\tau\in\triangle} \|p\|_{\infty, \tau} \leq \sum_{\tau\in\triangle} \frac{K}{|\triangle|}\|p\|_{2,\tau} \asymp \frac{K}{|\triangle|}\|p\|_{L^2(\Omega)}\asymp KN^{1/2}\|p\|_{L^2(\Omega)}.
\end{equation}
Equations (\ref{EQN:L2Omega}), (\ref{EQN:Rn}) and (\ref{EQN:markov_inf}) together imply
$\|p\|_{\infty, \Omega} \leq N ^{1/2}\|p\|_{\Omega} \leq  N ^{1/2}\|p\|_{n,\Omega}$, completing the first part of \Cref{LEM:Vn}.

Secondly, by Theorem 14.21 of \cite{Lai:Schumaker:07}, for any $\tau$ with $|\tau|<1$, we have $|p|_{\ell,q,\tau}\leq K \rho_\tau^{-\ell}\|p\|_{q,\tau}$. Let $q = 2, \ell = 2<d$, {by Assumption \ref{C2}}, we have $\|p\|_{\mathcal E, \tau} \leq K\rho_{\tau}^{-2}\|p\|_{2,\tau}\leq K|\triangle|^{-2}\|p\|_{2,\tau}$. Take summation over $\tau\in\triangle$, we have 
\begin{align} \label{EQN:markov}
\|p\|_{\mathcal E, \Omega} \leq \frac{K}{|\triangle|^{2}} \|p\|_{L^2(\Omega)}
\asymp {K N\|p\|_{L^2(\Omega)}},
\end{align}
which completes the second part of the lemma by applying Equations (\ref{EQN:L2Omega}) -- (\ref{EQN:Rn}).
\end{proof}

\subsection{Minimal determining set and approximation power of spherical splines}
\label{SSEC:MDS}

To extend our results on a single spherical triangle $\tau$ (or a local spherical domain) to the whole triangulation $\triangle$ (or a large spherical domain), we make use of Minimal Determining Set (MDS). A local MDS, $\mathcal M$, is the determining set for $\mathcal S$ with the smallest cardinality, such that if $s\in \mathcal S$, and the B-coefficients $c_{\xi}(s) = 0$ for all $\xi\in \mathcal M$, then $s\equiv 0$. Here $\mathcal S\subseteq \mathcal S_d^0(\triangle)$ is the set of functions defined over $\triangle$. Such construction enables us to consider a local set of points, other than which, the B-coefficients are fully dependent on those in $\mathcal M$ through smoothness conditions. We also define $\text{star}(v):= \text{star}^1(v)$ to be the set of all triangles sharing the vertex $v$, and $\text{star}^j(v):= \cup\{\text{all triangles sharing vertices of those in }\text{star}^{j-1}(v)\}$.
Similarly, for a triangle, $\tau$, we define $\text{star}^0(\tau) := \tau$, and $\text{star}^j(\tau):= \cup \{\text{star}(v): v\in \text{star}^{j-1}(\tau)\}$.

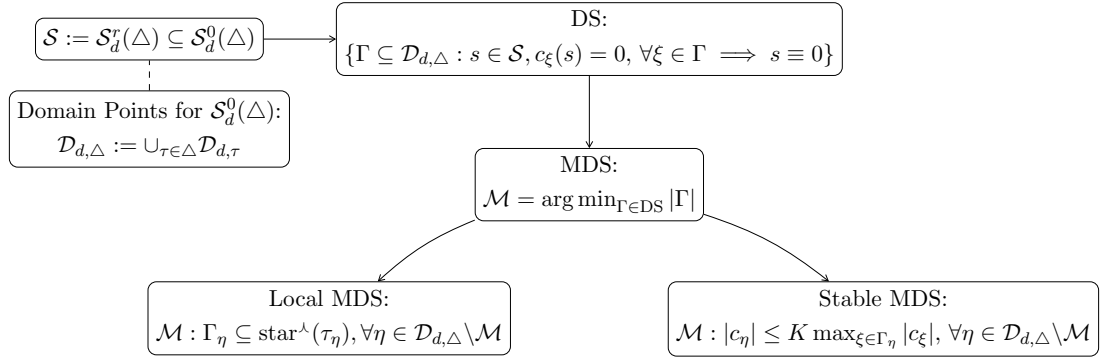
\begin{figure}[!ht]
\centering
\scalebox{0.8}{
\begin{tikzpicture}[
node distance = 12mm and 12mm, 
box/.style = {draw, rounded corners, 
minimum width=22mm, minimum height=5mm, align=center},
> = {Straight Barb[angle=60:2pt 3]},
bend angle = 15,above,sloped, auto = left]
\node (n1)  [box]{$\mathcal S:= \mathcal S_d^r(\triangle)\subseteq\mathcal S_d^0(\triangle)$};
\node (n2)  [box, right=of n1]{DS: \\$\{\Gamma\subseteq \mathcal D_{d,\triangle}: s\in\mathcal S, c_\xi(s) = 0,\, \forall \xi \in \Gamma \implies s \equiv 0\}$};
\node (n3)  [box, below=of n2] {MDS: \\$\mathcal M = \argmin_{\Gamma\in\mathrm{DS}} |\Gamma|$};
\node (n4a)  [box, below left=1 and -0.6 of n3] {Local MDS:  \\$\mathcal M: \Gamma_\eta \subseteq \text{star}^\curlywedge (\tau_\eta),\forall \eta\in \mathcal D_{d,\triangle}\backslash \mathcal M$};
\node (n4b)  [box, below right=1 and -0.6 of n3] {Stable MDS: \\ $\mathcal M: |c_\eta|\leq K\max_{\xi\in\Gamma_\eta}|c_\xi|$, $\forall \eta\in \mathcal D_{d,\triangle}\backslash \mathcal M$};
\node (n5)  [box, below=0.5 of n1] {Domain Points  for $\mathcal S_d^0(\triangle)$: \\$\mathcal D_{d,\triangle} := \cup_{\tau\in \triangle} \mathcal D_{d,\tau}$};
\draw[->] (n1) to []  (n2);
\draw[->] (n2) to []  (n3);
\draw[dashed] (n5) to (n1);
\draw[->] (n3) to [bend left=-10] (n4a);
\draw[->] (n3) to [bend right=-10] (n4b);
\end{tikzpicture}
}
\caption{Illustration of local stable MDS construction. Note $\mathcal M$ is not necessarily unique, and $|\mathcal M| = \text{dim} \mathcal S$; $\Gamma_\eta=\{\xi\in\mathcal M: c_\eta(s)$ depends on $c_\xi(s)\}$ is the set of domain points in $\mathcal M$ whose B-coefficients affects $c_\eta$;  $\tau_\eta$ is the triangle that contains $\eta$. }
\label{fig:MDS}
\end{figure}

In the main paper and the rest of this supplementary material, we focus on local and stable MDS.
We say an MDS, $\mathcal M$, is local if there exists a constant $\curlywedge$ not depending on $\triangle$, such that $\Gamma_\eta$ is contained in $\text{star}^\curlywedge (\tau_\eta)$, for all $\eta\in \mathcal D_{d,\triangle}\backslash \mathcal M$, where $\Gamma_\eta$ consists of all the domain points whose B-coefficients affect the B-coefficient $\eta$, and $\tau_\eta$ is the triangle containing $\eta$. The lower $\curlywedge$ is, the more locally defined a function is relative to $\triangle$. In addition, we say an MDS, $\mathcal M$, is stable if there exists a constant $K$ depending only on $\curlywedge$ and the smallest angle in $\triangle$, such that $|c_\eta|\leq K\max_{\xi\in\Gamma_\eta}|c_\xi|$, for all $\eta\in \mathcal D_{d,\triangle}\backslash \mathcal M$.
The construction of MDS, local MDS, and stable MDS are illustrated in Figure \ref{fig:MDS}, also see in Definition 13.38 of \cite{Lai:Schumaker:07}.

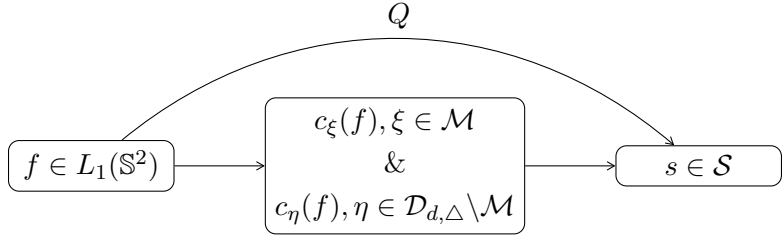
\begin{figure}[!ht]
\centering
\begin{tikzpicture}[
node distance = 12mm and 12mm, box/.style = {draw, rounded corners, minimum width=22mm, minimum height=5mm, align=center},
> = {Straight Barb[angle=60:2pt 3]},bend angle = 15, auto = left]
\node (n1)  [box]{$f\in L_1(\mathbb S^2)$};
\node (n2)  [box, right=of n1]{$c_\xi(f), \xi\in \mathcal M$\\ \& \\ $c_\eta(f), \eta\in \mathcal D_{d,\triangle}\backslash \mathcal M$};
\node (n4)  [box, right=of n2] {$s \in\mathcal S$};
\draw[->] (n1) to [""] (n2);\draw[->] (n2) to [""] (n4);
\draw[->] (n1) to [bend right=-40, "$Q$"] (n4);
\end{tikzpicture}
\caption{Illustration of quasi-interpolation operator $Q$. Here $c_\xi(f) = \gamma_\xi(F_{d,\tau_\xi} f)$ is the B-coefficients corresponding to $\xi$, $\tau_\xi$ is the triangle that contains  $\xi$, $F_{d,\tau_\xi} f$ is the averaged Taylor polynomial of degree $d$ associated with $\tau_\xi$;  $c_\eta(f)$ is a linear combination of $\{c_\xi(f)\}_{\xi\in\mathcal M_\eta}$, $\mathcal M_\eta \subseteq \text{star}^\curlywedge (\tau_\eta)$, $\tau_\eta$ is the triangle that contains $\eta$.}
\label{fig:Q}
\end{figure}

Lemma \ref{LEM:localApprox} establishes the local approximation power of the spherical splines for functions defined over a triangulation $\triangle \subseteq \mathbb S^2$. 
\begin{lemma}[Theorem 14.22 of \cite{Lai:Schumaker:07}] 
\label{LEM:localApprox}
Suppose $\mathcal M$ is a stable local MDS for a spherical spline space $\mathcal S \subseteq \mathcal S^0_d(\triangle)$, and $\Omega\subseteq \mathbb S^2$ is the domain of interest. 
The triangulation $\triangle$ covers all or part of $\mathbb S^2$. 
We consider a quasi-interpolation operator $Q: L_1(\Omega) \to \mathcal S$ which yields a spline $s = Qf$ for any $f\in L_1(\mathbb S^2)$. We can construct $s$ by assigning a coefficient functional to $f$ calculated by solving the B-coefficients corresponding to $\xi \in\mathcal M$, see Figure \ref{fig:Q}.
Then $Q: L_1(\Omega) \to \mathcal S$  is a {linear projector} such that for any triangle $\tau\in\triangle$ and $1\leq q\leq\infty$, 
$
\|Qf\|_{q,\tau} \leq K\|f\|_{q,\Omega_\tau}, \quad \forall f\in L_1(\Omega_\tau),
$
where $\Omega_\tau = \text{star}^\curlywedge (\tau)$. The constant $K$ depends on $d, \curlywedge,$ and the smallest angle in $\Omega_\tau$.
\end{lemma}

 Now we are ready to derive the global approximation power of SBB polynomials in Lemma \ref{LEM:globalApprox} and Lemma \ref{LEM:globalApprox2} using Lemma \ref{LEM:localApprox}.
\begin{lemma}[Theorem 14.24 \cite{Lai:Schumaker:07}]\label{LEM:globalApprox}
Assume the constant $\curlywedge$ and the mesh size $|\triangle|$ satisfies $\curlywedge|\triangle|\leq 1$.
Let $0 \leq \ell \leq d$ with $\ell=d(\bmod 2)$. Then for $f \in W_q^{\ell+1}(\Omega)$ with $1 \leq q \leq \infty$, $|f-Qf|_{k, q, \Omega} \leq K|\triangle|^{\ell+1-k}|f|_{\ell+1, q, \Omega}$,
for all $0 \leq k \leq \ell$. If $\Omega$ covers $\mathbb S^2$, the constant $K$ depends only on $d$ and the smallest angle in the triangles of $\triangle$. Otherwise, $K$ may also depend on the Lipschitz constant of $\partial \Omega$.
\end{lemma}
\begin{remark}
When $d$ and $\ell$ are both even or both odd, $\ell = d(\mathrm{mod}2)$. Otherwise, $\ell \neq d(\mathrm{mod}2)$, where the approximation power in the above lemma cannot be gauranteed. This issue can be circumvented by considering the approximation function space $\{f = c_1f_1 + c_2f_2: f_1\in S_d^r(\triangle), f_2\in \mathcal S_{d-1}^r(\triangle)\}$, see \cite{Baramidze:Etal:06}. 
\end{remark}
Next, we extend the results above to $\mathcal S_d^r(\triangle)$.
In particular, we can construct a triangulation such that $\curlywedge = 3, |\triangle|\leq 1/6$, which satisfies the condition $\curlywedge |\triangle| \leq 1$, and leads to Lemma \ref{LEM:globalApprox2}.
\begin{lemma} [Theorem 14.26 \cite{Lai:Schumaker:07}]
\label{LEM:globalApprox2}
Let $0 \leq \ell \leq d$ with $\ell=d(\bmod 2)$. Suppose $\triangle$ is a spherical triangulation of a set $\Omega$ on $\mathbb S^2$, and that $|\triangle| \leq 1 / 6$. Let $d \geq 3 r+2$. Suppose $f \in W_q^{\ell+1}(\Omega)$ with $1 \leq q \leq \infty$. Then there exists a spline $s \in \mathcal{S}_d^{r}(\triangle)$ such that $|f-s|_{k, q, \Omega} \leq K|\triangle|^{\ell+1-k}|f|_{\ell+1, q, \Omega}$,
for all $0 \leq k \leq \ell$. If $\Omega$ covers $\mathbb S^2$, the constant $K$ depends only on $d$ and the smallest angle in the triangles of $\triangle$. Otherwise, $K$ may also depend on the Lipschitz constant of $\partial \Omega$.
\end{lemma}
This lemma gives the global approximation power of spherical splines. By assuming $|f|_{\ell+1, 2,\Omega}$ being a finite constant, the residual of spline approximation can be bounded with order $|\triangle|^{\ell+1}$.
Let $k = 2, q = \infty$, Lemma \ref{LEM:globalApprox2} implies
\begin{equation}
    |f-s|_{2,\infty, \Omega} \leq K|\triangle|^{\ell - 1}|f|_{\ell+1, \infty ,\Omega}.
    \label{EQN:S5col1}
\end{equation}
Let $k = 0, q = \infty$, Lemma \ref{LEM:globalApprox2} implies
\begin{equation}
    \|f-s\|_{\infty, \Omega} \leq K|\triangle|^{\ell+1}|f|_{\ell+1, \infty ,\Omega}.
        \label{EQN:S5col2}
\end{equation}

\subsection{Size of bias and noise terms}
\label{SSEC:convergence}

In this section, we study the asymptotics of the bias and noise terms, and provide proof for \Cref{prop1} and \Cref{prop2}.

To prepare for the proof, we consider $\mathcal B(\Omega)$ the space of all bounded real-valued functions over $\Omega = \cup_{\tau\in\triangle}\tau$.
By definitions (\ref{EQN:tsss}), (\ref{DEF:slamm}) and (\ref{DEF:slame}),
\begin{align}
s_\lambda &= \argmin_{s\in\mathcal S} n\|m(\mathbf x) + \sigma(\mathbf x)\epsilon - s(\mathbf x)\|_{n,\Omega}^2 + \lambda \|s(\mathbf x)\|_{\mathcal E}^2,\\
s_{\lambda,m} &= \argmin_{s\in\mathcal S} n\|m(\mathbf x) - s(\mathbf x)\|_{n,\Omega}^2 + \lambda \|s(\mathbf x)\|_{\mathcal E}^2,\label{EQN:slamm}\\
s_{\lambda,\epsilon} & = \argmin_{s\in\mathcal S} n\|\sigma(\mathbf x)\epsilon - s(\mathbf x)\|_{n,\Omega}^2 + \lambda \|s(\mathbf x)\|_{\mathcal E}^2. \label{EQN:slame}
\end{align}

\begin{proof}[Proof of \Cref{prop1}]
By triangle inequality and definition of $V_n$, 
\[
\|s_{\lambda, m}-m\|_{\infty , \Omega} \leq
\|m-s_{0,m}\|_{\infty,\Omega} + 
\|s_{\lambda,m}-s_{0,m}\|_{\infty,\Omega} \leq 
\|m-s_{0,m}\|_{\infty,\Omega} + 
V_n \|s_{\lambda,m}-s_{0,m}\|_{n,\Omega}.
\]
By (\ref{EQN:slamm}), for any $u\in\mathcal S_d^r(\triangle)$ and $t\in\mathbb R$, 
\begin{align*}
n\|m - (s_{\lambda, m} + tu)\|_{n,\Omega}^2 + \lambda \|s_{\lambda, m}+tu\|_{\mathcal E}^2 & \geq  n\|m - s_{\lambda, m}\|_{n,\Omega}^2 + \lambda \|s_{\lambda, m}\|_{\mathcal E}^2,\\
n\|m - (s_{0, m} + tu)\|_{n,\Omega}^2 & \geq  n\|m - s_{0, m}\|_{n,\Omega}^2.
\end{align*}
It follows that  $s_{\lambda,m}$ is characterized by the orthogonality relationship
\begin{align}\label{EQN:s5}
    n\langle m - s_{\lambda, m}, u\rangle _{n, \Omega} = \lambda \langle s_{\lambda, m},u\rangle _{\mathcal E}, \quad \text{ for all } u\in\mathcal S_d^r(\triangle),
\end{align}
while $s_{0,m}$ is characterized by
\begin{align}\label{EQN:s6}
n\langle m- s_{0,m}, u\rangle _{n,\Omega} = 0 , \quad \text{ for all } u \in \mathcal S.
\end{align}
By taking the difference of (\ref{EQN:s5}) and(\ref{EQN:s6}), we have
$n\langle s_{0,m} - s_{\lambda, m}, u\rangle_{n,\Omega} = \lambda \langle s_{\lambda,m} , u\rangle_{\mathcal E}$. 
Plug in $u = s_{0,m} - s_{\lambda, m}$, we have
\begin{align*}
0 & \leq n \|s_{0,m} - s_{\lambda, m}\|_{n,\Omega}^2 
= \lambda \langle s_{\lambda, m}, s_{0, m} -s_{\lambda,m} \rangle_{\mathcal E}
 \leq \lambda (\langle s_{0, m} , s_{\lambda,m}\rangle_{\mathcal E}- \|s_{\lambda,m}\|_{\mathcal E}^2),
\end{align*}
and  $\|s_{\lambda, m}\|^2_{\mathcal E} \leq \langle s_{0, m}, s_{\lambda,m}\rangle_{\mathcal E} \leq \|s_{\lambda, m}\|_{\mathcal E}\|s_{0, m}\|_{\mathcal E}$.
By Cauchy-Schwarz inequality,
\begin{align*}
n \|s_{0,m} - s_{\lambda, m}\|_{n,\Omega}^2 
& = \lambda \langle s_{\lambda, m}, s_{0, m} -s_{\lambda,m} \rangle_{\mathcal E}
 \leq \lambda \|s_{\lambda,m}\|_{\mathcal E}\cdot \|s_{0, m} - s_{\lambda,m}\|_{\mathcal E}\\
& \leq \lambda \|s_{\lambda,m}\|_{\mathcal E} \cdot \overline V_n \|s_{0, m} - s_{\lambda,m}\|_{n,\Omega},
\end{align*}
which implies that $\|s_{0,m} - s_{\lambda, m}\|_{n,\Omega} \leq n^{-1}\lambda  \overline V_n \|s_{\lambda,m}\|_{\mathcal E}.$
With (\ref{EQN:S5col1}),
\begin{align*}
\|s_{0,m}-s_{\lambda, m}\|_{\infty , \Omega} & \leq n^{-1}\lambda \overline V_n  V_n \|s_{\lambda,m}\|_{\mathcal E} \leq n^{-1}\lambda \overline V_n  V_n  \|s_{0,m}\|_{\mathcal E} \\
&  \leq n^{-1}\lambda \overline V_n  V_n  (\|m-s_{0,m}\|_{\mathcal E} + \|m\|_{\mathcal E}) \\
& \leq  n^{-1}\lambda  {N^{3/2}}\left(K|\triangle|^{\ell -1}|m|_{\ell+1, \infty, \Omega} + |m|_{2, \infty, \Omega}\right).
\end{align*}
And (\ref{EQN:S5col2}) entails
$\|m-s_{0, m}\|_{\infty , \Omega} \leq
K|\triangle|^{\ell + 1}|m|_{\ell + 1, \infty, \Omega}.$
Therefore,
\begin{align*}
\|s_{\lambda, m}-m\|_{\infty , \Omega} 
& \leq 
\|m-s_{0,m}\|_{\infty,\Omega} + 
\|s_{0,m}-s_{\lambda,m}\|_{\infty,\Omega}\\
& =
O_p\left\{
|\triangle|^{\ell + 1}|m|_{\ell + 1, \infty, \Omega} + 
\frac{\lambda {N^{3/2}}}{n}  \left(|\triangle|^{\ell -1}|m|_{\ell+1, \infty, \Omega} + |m|_{2, \infty, \Omega}\right)
\right\}\\
& =
O_p\left\{
N^{-(\ell + 1)/2}|m|_{\ell + 1, \infty, \Omega} + 
\frac{\lambda {N^{3/2}}}{n}  \left(N^{-(\ell -1)/2}|m|_{\ell+1, \infty, \Omega} + |m|_{2, \infty, \Omega}\right)
\right\}\\
& =
O_p\left\{
\frac{\lambda {N^{3/2}}}{n}  |m|_{2, \infty, \Omega} + 
\left( 1+ \frac{\lambda{N^{5/2}}}{n} \right) N^{-(\ell + 1)/2} |m|_{\ell + 1, \infty, \Omega} 
\right\}.    
\end{align*}
The proof is completed.
\end{proof}

\begin{proof}[Proof of \Cref{prop2} (Part 1)]
The key is to recognize the orthogonal relation that characterizes $m_{\lambda, \epsilon}$.
By (\ref{EQN:slame}), for any $u\in\mathcal S_d^r(\triangle)$ and $t\in\mathbb R$, 
\begin{align*}
n\|\sigma(\mathbf x)\epsilon- (s_{\lambda, \epsilon} + tu)\|_{n,\Omega}^2 + \lambda \|s_{\lambda, \epsilon}+tu\|_{\mathcal E}^2 & \geq  n\|\sigma(\mathbf x)\epsilon - s_{\lambda, \epsilon}\|_{n,\Omega}^2 + \lambda \|s_{\lambda, \epsilon}\|_{\mathcal E}^2.
\end{align*}
It follows that $s_{\lambda,\epsilon}$ is characterized by
$$
n\langle s_{\lambda, \epsilon}-\sigma(\mathbf x) \epsilon, u\rangle _{n,\Omega}  + \lambda \langle s_{\lambda, \epsilon}, u\rangle_{\mathcal E} =0, \quad \forall u \in\mathcal S_d^r(\triangle).
$$
Note that we can write $s_{\lambda, \epsilon}$ in terms of SBB basis $\{B_\xi, \xi\in\mathcal M\}$, $s_{\lambda, \epsilon}(\mathbf x) = \sum_{\xi\in\mathcal M} c_{s_{\lambda,\epsilon}, \xi} B_\xi(\mathbf x)$, where $c_{s_{\lambda,\epsilon}, \xi}$ are the B-coefficients. 
Let $u = B_\xi(\mathbf X_i)$, we have
$
\sum_{i = 1}^n s_{\lambda, \epsilon}(\mathbf X_i) B_\xi(\mathbf X_i)  + \lambda \langle s_{\lambda, \epsilon}, B_\xi\rangle_{\mathcal E} =  \sum_{i = 1}^n B_\xi(\mathbf X_i) \sigma(\mathbf X_i) \epsilon_i
$
for all $\xi\in\mathcal M$. Multiply both sides by $c_{s_{\lambda, \epsilon}, \xi}/n$, and perform summation over $\xi\in\mathcal M$. We have, by Cauchy-Schwarz inequality,
\begin{align*}
\|s_{\lambda, \epsilon}\|_{n, \Omega}^2
& \leq  \|s_{\lambda, \epsilon}\|_{n, \Omega}^2 + \frac{\lambda}{n} \mathcal E (s_{\lambda, \epsilon}) 
= \frac{1}{n} \sum_{\xi\in\mathcal M} c_{s_{\lambda, \epsilon}, \xi} \sum_{i = 1}^n B_\xi(\mathbf X_i) \sigma(\mathbf X_i) \epsilon_i\\
& \leq \left( \sum_{\xi\in\mathcal M} |c_{s_{\lambda, m}, \xi}|^2\right)^{1/2}  \left[\sum_{\xi\in\mathcal M} \sum_{i =1}^n\left\{\frac{1}{n}  B_\xi (\mathbf X_i) \sigma(\mathbf X_i) \epsilon_i\right\}^2\right]^{1/2}.
\end{align*}
Apply \Cref{LEM:stability}, we have 
\begin{align*}
\|s_{\lambda, \epsilon}\|_{n, \Omega}^2
& \leq K {N^{1/2}} \|s_{\lambda, \epsilon}\|_{\Omega} \left[\sum_{\xi\in\mathcal M} \sum_{i =1}^n \left\{\frac{1}{n} B_\xi (\mathbf X_i) \sigma(\mathbf X_i) \epsilon_i\right\}^2\right]^{1/2}.
\end{align*}
In addition, by \Cref{LEM:norm1}, we have $\sup_{p\in \mathcal S_d^r(\triangle)}|\|p\|_{n,\Omega}^2/\|p\|_{\Omega}^2-1| = O_p\{(N\log n)^{1/2}n^{-1/2}\}$, which implies
\begin{align}
&  \|s_{\lambda,\epsilon}\|_{\Omega}^2
\leq \|s_{\lambda,\epsilon}\|_{n,\Omega}^2 + \left(\frac{N\log n}{n}\right)^{1/2}\|s_{\lambda,\epsilon}\|_{\Omega}^2\nonumber\\
\leq &K {N^{1/2}} \|s_{\lambda, \epsilon}\|_{\Omega} \left[\sum_{\xi\in\mathcal M}\sum_{i =1}^n \left\{ \frac{1}{n} B_\xi (\mathbf X_i) \sigma(\mathbf X_i) \epsilon_i\right\}^2\right]^{1/2}+ \left(\frac{N\log n}{n}\right)^{1/2}\|s_{\lambda,\epsilon}\|_{\Omega}^2.
\label{EQN:EB2sig2ep2a},
\end{align}
By Lemma 14.8 of \cite{Lai:Schumaker:07} and \ref{C4}, {$\|p\|_{\tau} \asymp \|p\|_{L^2(\tau)} \leq N^{-1/2}\|p\|_{\infty, \tau}, \forall p\in\mathcal S_d^r(\triangle)$.} Then, by the law of large number, 
\begin{align}
\frac{1}{n} \sum_{i =1}^n \left\{B_\xi (\mathbf X_i) \sigma(\mathbf X_i) \epsilon_i\right\}^2 
& = \E(B_\xi(\mathbf X)^2\sigma(\mathbf X)^2\epsilon(\mathbf X)^2) + o_p(1) \nonumber\\
& = \E(B_\xi(\mathbf X)^2\sigma(\mathbf X)^2)+o_{p}(1) \nonumber\\
&\leq K \|B_\xi(\mathbf X)\sigma(\mathbf X)\|_{L^2(\Omega)}^2 \nonumber\\
&\leq K N^{-1}\|B_\xi(\mathbf X)\sigma(\mathbf X)\|_{\infty,\tau^*}^2 = O_p(N^{-1}) \label{EQN:EB2sig2ep2},
\end{align}
where $\tau^*$ is the triangle where $B_\xi(\mathbf X)$ has positive values.
By Assumption \ref{C5}, we have $N = Cn^\omega, \omega <1$. Note $|\mathcal M| = N(d+1)(d+2)/2 \asymp N$. By (\ref{EQN:L2Omega}), (\ref{EQN:EB2sig2ep2a}) and (\ref{EQN:EB2sig2ep2}),
$$
\left\{1-\left(\frac{N\log n}{n}\right)^{1/2}\right\}\|s_{\lambda, \epsilon}\|_{L^2(\Omega)} \leq 
K N^{1/2}{({|\mathcal M|}/{nN})^{1/2}}
= O_p\left(n^{-1/2}N^{1/2}\right).
$$
The proof is completed by noticing $\left(\frac{N\log n}{n}\right)^{1/2} \to 0$ as $n\to\infty$.
\end{proof}

To prepare for Part 2 of \Cref{prop2}, we first introduce the following two lemmas.
Let $\bs\Gamma_\lambda$ be the symmetric positive definite matrix 
\begin{equation}
\left[ \frac{1}{n}\sum_{i=1 }^n B_{\xi_1}(\mathbf X_i) B_{\xi_2}(\mathbf X_i) + \frac{\lambda}{n}\langle B_{\xi_1}, B_{\xi_2}\rangle_{\mathcal E} \right]_{\xi_1, \xi_2\in\mathcal M}.
\label{EQN:Glambda}
\end{equation}
In particular, when $\lambda =0$, we have $\bs\Gamma_0 = \left[ \frac{1}{n}\sum_{i=1 }^n B_{\xi_1}(\mathbf X_i) B_{\xi_2}(\mathbf X_i)\right]_{\xi_1, \xi_2\in\mathcal M}$.

\begin{lemma}\label{LEM:Gamma}
Under Assumptions \ref{C3}, \ref{C4}, and \ref{C5}, for $\bs\Gamma_\lambda$, we have the following asymptotic results.
\begin{itemize}
\item [(i)] As $n\to \infty$, with probability one, 
\[
cN^{-1}\leq \rho_{\min}(\bs\Gamma_\lambda) \leq \rho_{\max} (\bs\Gamma_{\lambda}) \leq C\left(N^{-1} + \lambda N/n\right),
\]
holds for some constants $0<c<C<\infty$ depending on $\lambda$.
\item [(ii)] There exists a constant $C>0$ depending on $d$ such that $\|\bs\Gamma_\lambda^{{-1}}\|_{\infty} \leq CN$.

\item [(iii)] For any vector $\bs\sigma = (\sigma_1, \ldots, \sigma_n)^\top$, there exists a constant $C>0$ depending on $d$, such that 
$
\left\|\mathbf B^\top(\mathbf x) \bs\Gamma_{\lambda}^{-1} \frac{1}{n}\sum_{n =1 }^n \mathbf B(\mathbf X_i) \sigma_i\right\|_{\infty} \leq C\|\bs\sigma\|_{\infty}.
$
\end{itemize}
\end{lemma}
\begin{proof}
(i) For any $\bs\gamma\in\R^{|\mathcal M|}$, let $p(\mathbf x) = \mathbf B^\top(\mathbf x) \bs\gamma \in\mathcal S_d^r(\triangle)$.
Note that
$$
\bs\gamma^\top \bs\Gamma_\lambda \bs\gamma = \bs\gamma^\top \frac{1}{n}\sum_{i = 1}^n \mathbf B(\mathbf X_i) \mathbf B^\top(\mathbf X_i) \bs\gamma + \bs\gamma^\top \frac{\lambda}{n}|\langle B_{\xi_1}, B_{\xi_2}\rangle_{\mathcal E} |_{\xi_1, \xi_2\in\mathcal M}\bs\gamma = \|p\|_{n, \Omega}^2 + \frac{\lambda}{n}\|p\|_{\mathcal E}^2.
$$
By \Cref{LEM:norm1}, we have $|\|p\|_{n,\Omega}^2 / \|p\|_{2, \Omega}^2 -1|\leq R_n$.  And by \Cref{LEM:stability}, 
$$
c(1-R_n)N^{-1}\|\bs\gamma\|_2^2\leq 
(1-R_n) \|p\|_{L^2(\Omega)}^2 \leq \|p\|_{n,\Omega}^2 \leq 
(1+ R_n) \|p\|_{L^2(\Omega)}^2 \leq C(1+R_n) N^{-1} \|\bs\gamma\|_2^2.
$$
It follows that $\rho_{\min}(\bs\Gamma_\lambda)\geq c N^{-1}$ for some $c>0$.
By \Cref{LEM:Vn}, $\|p\|_{\mathcal E}^2\leq CN^2 \|p\|_{\Omega}^2$. By \Cref{LEM:stability}, we have 
$\|p\|_{\Omega}^2 \leq C N^{-1}\|\bs\gamma\|_2^2$.
Combining the bounds for $\|p\|_{n,\Omega}^2$ and $\|p\|_{\mathcal E}^2$ gives
$$
\rho_{\max} (\bs\Gamma_\lambda) \leq C\left\{ (1+ R_n) N^{-1} + \frac{\lambda N}{n}\right\} \leq C \left(N^{-1} +  \lambda N/n \right).
$$

(ii) We know $\bs\Gamma_{\lambda}$ is an invertible symmetric matrix from (i), with condition number $1< \kappa = \rho_{\max}(\bs\Gamma_\lambda)/\rho_{\min}(\bs\Gamma_\lambda)< C$, for some constant $C>1$. 
We then make use of properties of banded matrix. A matrix $\mathbf M = (M_{ij})$ is said to be banded with bandwidth $b$ if $M_{ij} =0$ for $|i-j|\geq b$, where $b$ is the smallest integer satisfying the condition. 
By construction of $B_\xi$, the bandwidth of $\bs\Gamma_{\lambda}$ is $\kappa = (d+1)(d+2)/2$. By Theorem 13.4.3 of \cite{Devore:1993}, $\|\bs\Gamma_{\lambda}^{-1}\|_{\infty} \leq 2\tau^{-2b} \|\bs\Gamma_{\lambda}^{-1}\|_2(1-\tau)^{-1}$, where $\tau = (\kappa^2-\kappa^{-2} + 1)^{1/4b} <1$. Therefore, $\|\bs\Gamma_\lambda^{-1}\|_{\infty} \leq C \|\bs\Gamma_\lambda^{-1}\|_{2} \leq C \text{dim}(\bs\Gamma_\lambda) \leq C N$, for some $C>0$.

(iii) Combining (i) and (ii), we have
$$
\left\|\mathbf B^\top(\mathbf x) \bs\Gamma_{\lambda}^{-1} \frac{1}{n}\sum_{i = 1}^n \mathbf B(\mathbf X_i) \sigma_i\right\|_{\infty}  \leq C\|\bs\sigma\|_{\infty}.
$$
The proof is completed.
\end{proof}

\begin{lemma}\label{LEM:e0inf}
Under Assumptions \ref{C2} and \ref{C5}, 
$$
\|s_{0,\epsilon}\|_{\infty, \Omega} = 
O_p\left\{ \sqrt{\frac{N\log n }{n}}\right\}.
$$
\end{lemma}
\begin{proof}
Note that $s_{0,\epsilon} = \sum_{\xi\in\mathcal M}\widehat c_{0,\xi} B_\xi(\mathbf x) $. Since $B_\xi$ are bounded, by definition (\ref{DEF:slame}),  
$$
\|s_{0, \epsilon}\|_{\infty, \Omega} \leq c\|\widehat{\mathbf c}_0\|_{\infty} 
\asymp \left\|\bs\Gamma_0^{-1} \left[n^{-1} \sum_{i = 1}^n B_\xi(\mathbf X_i) \sigma(\mathbf X_i) \epsilon_i \right]_{\xi\in\mathcal M}\right\|_{\infty, \Omega},
\quad
\widehat {\mathbf c}_0  = \left(\widehat c_{0, \xi}\right)_{\xi\in\mathcal M}.
$$
By \Cref{LEM:Gamma} (ii), 
$$
\|s_{0,\epsilon}\|_{\infty, \Omega} \leq C\|\bs\Gamma_{0}^{-1}\|_{\infty} \max_{\xi\in\mathcal M}\left| n^{-1}\sum_{i = 1}^n B_\xi(\mathbf X_i) \sigma(\mathbf X_i) \epsilon_i\right|\leq CN \max_{\xi\in\mathcal M}\left| n^{-1}\sum_{i = 1}^n B_\xi(\mathbf X_i) \sigma(\mathbf X_i) \epsilon_i\right|
$$
holds almost surely. Therefore, it suffices to show 
$$
\max_{\xi\in\mathcal M}\left| n^{-1}\sum_{i = 1}^n B_\xi(\mathbf X_i) \sigma(\mathbf X_i) \epsilon_i\right| = 
O_p\left\{ (\log n)^{1/2} (nN)^{-1/2}\right\}.
$$
Consider decomposing $\epsilon_i$ into a tail part $\epsilon_{i,1}^{D_n} $ and a truncated part $\epsilon_{i,2}^{D_n}$, 
\[
\epsilon_i = \epsilon_{i,1}^{D_n} + \epsilon_{i,2}^{D_n}, \quad
\epsilon_{i,1}^{D_n} = \epsilon_i \mathbbm{1}\{|\epsilon_i|> D_n\}, \quad 
\epsilon_{i,2}^{D_n} = \epsilon_i \mathbbm{1}\{|\epsilon_i|\leq D_n\}, 
\]
where $D_n = n^\alpha$ with $\max\{\frac{1}{2+\eta}, \frac{1+\omega}{2(1+\eta)}\} < \alpha<{\frac{1-\omega}{2}}$. Note that $\eta$ defined in Assumption {\ref{C5}} and $\omega$ defined in Assumption {\ref{C3}} need to satisfy ${\omega <\frac{\eta}{\eta+2}}$ for the condition to be feasible.

Since $\E(\epsilon_i) = 0$, $|\E\{\epsilon_i \mathbbm{1}(|\epsilon_i| \leq D_n)\}| = |\E\{\epsilon_i \mathbbm{1}(|\epsilon_i| > D_n)\} |$. We then show $\mu^{D_n}:= \E(\epsilon_{i,2}^{D_n}) = - \E(\epsilon_{i,1}^{D_n})$ is bounded. 
By Assumption {\ref{C5}}, 
$$
|\E\{\epsilon_i \mathbbm{1}(|\epsilon_i|>D_n)\}| 
\leq 
\left|\E\left\{\epsilon_i \frac{|\epsilon_i|^{1+\eta}}{D_n^{1+\eta}} \mathbbm{1}\left(|\epsilon_i|>D_n\right)\right\}\right|  
\leq 
\frac{1}{D_n^{1+\eta}}\E|\epsilon_i^{2+\eta}| \leq CD_n^{-(1+\eta)},
$$
for $D_n$ large enough.  
By the boundedness of SBB and $\sigma$,
\begin{equation}
|B_\xi(\mathbf X_i) \sigma(\mathbf X_i) \mu^{D_n}| = O_p\{D_n^{-(1+\eta)}\}. \label{EQN:S7a}
\end{equation}
Here we use bound $\frac{1+\omega}{2(1+\eta)} < \alpha$ to ensure $D_n^{-(1+\eta)} \leq {\log n}^{1/2} (nN)^{-1/2}$.

Then, we prove the tail part $\epsilon_{i,1}^{D_n}$ vanishes almost surely. By Assumption {\ref{C5}}, $\E|\epsilon_i^{2+\eta}| \leq v_{\eta}, \forall i$ for some $\eta>0$ and constant $v_{\eta}$. Therefore, by Markov's inequality,
$$
\sum_{n = 1}^\infty \mathrm{P}\{|\epsilon_n|>D_n\} \leq \sum_{i = 1}^\infty \frac{\E |\epsilon_n|^{2+\eta}}{D_n^{2+\eta}} \leq v_{\eta} \sum_{n = 1}^\infty D_n^{-(2+\eta)} <\infty,
$$
where the last inequality holds due to $\frac{1}{2+\eta}< \alpha$.
By the Borel-Cantelli's lemma, $\{|\epsilon_n|\leq D_n\}$ holds almost surely, for some $n$ large enough. Thus
$$
\mathrm{P}\{w: \exists N(w), \text{ such that } |\epsilon_n(w)|\leq D_n \text{ for } n>N(w)\} = 1.
$$ 
Let $\epsilon^* = \max\{|\epsilon_1|, \ldots, |\epsilon_n|\}$, there exists $N^*(w) > N(w)$, such that $D_{N^*(w)} > \epsilon^*$. Since $D_n$ is an increasing function of $n$, we have $D_n> D_{N^*(w)} > \epsilon^*$, for $n> N^*(w)$. Then, 
$$
\mathrm{P}\{w: \exists N(w), \text{ such that } |\epsilon_i(w)|\leq D_n, 1\leq i\leq n, \text{ for } n>N(w)\} = 1, 
$$
implying $\mathrm{P}\{w: \exists N(w), \text{ such that } |\epsilon_{i,1}^{D_n}|=0, 1\leq i\leq n, \text{ for } n>N(w)\} = 1.$
The boundedness of SBB and $\sigma$ leads to 
\begin{equation}
|B_\xi(\mathbf X_i) \sigma(\mathbf X_i) \epsilon_{i,1}^{D_n}| = O_{a.s.}(n^{-k}), \forall k>0.
    \label{EQN:S7b}
\end{equation}
Finally, let $Z_i = \frac{1}{n}(\epsilon_{i,2}^{D_n} - \mu^{D_n}) \sigma(\mathbf X_i) B_\xi(\mathbf X_i)$. Note that $\E(Z_i) = 0$, and 
\begin{align*}
\mathrm{Var}(\epsilon_{i,2}^{D_n} - \mu^{D_n})  & = \E(\epsilon_{i,2}^{D_n} - \mu^{D_n})^2 = \E\{\epsilon_{i}^2\mathbbm{1}(|\epsilon_i|\leq D_n)\} - (\mu^{D_n})^2 \\
&= \E(\epsilon_i^2) - \E\{\epsilon_{i}^2\mathbbm{1}(|\epsilon_i|>D_n)\}- (\mu^{D_n})^2 \\
&= 1 + O_p\left\{D_n^{-\eta} + D_n^{-2(1+\eta)}\right\}.  
\end{align*}
Note here the last inequality holds because $\E\{\epsilon_i^2 \mathbbm 1(|\epsilon_i|>D_n) \}\leq \E|\epsilon_i^{2 + \eta}|/D_n^{\eta} \leq v_\eta / D_n^{\eta}$ by Assumption \ref{C5}.
By (\ref{EQN:EB2sig2ep2}), $\E\{B_\xi^2(\mathbf X)\sigma^2(\mathbf X)\} = O_p(N^{-1})$. Since $\epsilon_{i,2}^{D_n} - \mu^{D_n}$ is independent from $B_\xi(\mathbf X_i)$ given $\mathbf X_i$, and that $\{\epsilon_{i,2}^{D_n} - \mu^{D_n}, 1\leq i\leq n\}$ are independent from each other, we have 
$ \var(\sum_{i = 1}^n Z_i) = C{(nN)^{-1}}$ for some $C>0$. 
Note that $|\epsilon_{i,2}^{D_n} - \mu^{D_n}| < 2D_n$, therefore 
$$
\E|\epsilon_{i,2}^{D_n} - \mu^{D_n}|^k \leq 2^{k-2} D_n^{k-2} \E|\epsilon_{i,2}^{D_n} - \mu^{D_n}|^2, ~~~ k\geq 2,
$$
and 
\begin{align*}
\E |Z_i|^k  &= n^{-k} \E|\epsilon_{i,2}^{D_n} - \mu^{D_n}|^k \E|\sigma(\mathbf X_i)B_\xi(\mathbf X_i)|^k\\
&\leq n^{-(k-2)}\E|\epsilon_{i,2}^{D_n} - \mu^{D_n}|^{k}\|B_\xi\|_{\infty, \Omega}^{k-2} \E|\sigma(\mathbf X_i)B_\xi(\mathbf X_i)|^2 \\&\leq (2D_n n^{-1})^{k-2} k!\E Z_i^2,
\end{align*}
which implies $\{Z_i\}_{i = 1}^n$ satisfies the Cram\'{e}r's condition with constant $c^* = 2D_n n^{-1}$.
By Bernstein's inequality, for $\delta>0$ large enough, we have
\begin{align*}
\mathrm{P}\left(\left|\sum_{i = 1}^n Z_i\right|  \geq \delta \sqrt{\frac{\log n}{nN}} \right) & \leq 
2\exp\left\{\frac{-\delta^2\log n /nN}{4 \var(\sum_{i = 1}^n Z_i) + 2c^*\delta \sqrt{\log n/nN}}\right\} \\
&= 
2\exp\left(\frac{-\delta^2\log n/nN }{4c/nN+ 2c^*\delta \sqrt{\log n /nN}}\right) \leq 2n^{-3}.
\end{align*}
For the last inequality to hold, we need $\alpha < \frac{1-\omega}{2}$.
Therefore, 
$$
\sum_{n = 1}^\infty \mathrm{P}\left(\max_{\xi\in\mathcal M}\left|\sum_{i = 1}^n Z_i\right|\geq \delta \sqrt{\frac{\log n}{nN}} \right)
\leq
\frac{N}{2} (d+1)(d+2) \sum_{n = 1}^\infty n^{-3}<\infty.
$$
By applying Borel-Cantelli's lemma, 
\begin{equation}
\max_{\xi\in\mathcal M}\left|\sum_{i = 1}^n Z_i\right| 
=\max_{\xi\in\mathcal M}\left|\sum_{i = 1}^n \frac{1}{n}(\epsilon_{i,2}^{D_n} - \mu^{D_n}) \sigma(\mathbf X_i) B_\xi(\mathbf X_i)\right|
= O_{a.s.}\left(\sqrt{\frac{\log n}{nN}}\right)
\label{EQN:S7c}
\end{equation}
The proof is completed by combining (\ref{EQN:S7a}), (\ref{EQN:S7b}) and (\ref{EQN:S7c}).
\end{proof}
\begin{proof}[Proof of \Cref{prop2}  (Part 2)]
By (\ref{EQN:slame}), for any $u\in\mathcal S_d^r(\triangle)$ and $t\in\mathbb R$, 
\begin{align*}
n\|\sigma(\mathbf x)\epsilon- (s_{\lambda, \epsilon} + tu)\|_{n,\Omega}^2 + \lambda \|s_{\lambda, \epsilon}+tu\|_{\mathcal E}^2 & \geq  n\|\sigma(\mathbf x)\epsilon - s_{\lambda, \epsilon}\|_{n,\Omega}^2 + \lambda \|s_{\lambda, \epsilon}\|_{\mathcal E}^2,\\
n\|\sigma(\mathbf x)\epsilon - (s_{0, \epsilon} + tu)\|_{n,\Omega}^2 & \geq  n\|\sigma(\mathbf x)\epsilon - s_{0, \epsilon}\|_{n,\Omega}^2.
\end{align*}
It follows that $s_{\lambda,\epsilon}$ is characterized by
$$
n\langle \sigma(\mathbf x) \epsilon -s_{\lambda, \epsilon}, u\rangle _{n,\Omega}  = \lambda \langle s_{\lambda, \epsilon}, u\rangle_{\mathcal E} , \quad \forall u \in\mathcal S_d^r(\triangle),
$$
and $s_{0,\epsilon}$ is characterized by 
$$
n\langle \sigma(\mathbf x) \epsilon - s_{0, \epsilon}, u\rangle _{n,\Omega}  =0, \quad \forall u \in\mathcal S_d^r(\triangle).
$$
Take the difference of the above two equations, and plug in $u = s_{0, \epsilon} - s_{\lambda, \epsilon}$, we have 
$$
n\|s_{0,\epsilon} - s_{\lambda, \epsilon}\|_{n, \Omega}^2 
= \lambda (\langle s_{\lambda, \epsilon}, s_{0,\epsilon}\rangle_{\mathcal E} - \|s_{\lambda, \epsilon}\|_{\mathcal E}^2 )= \lambda \langle s_{\lambda, \epsilon}, s_{0,\epsilon} - s_{\lambda, \epsilon}\rangle_{\mathcal E} \geq 0.
$$
Therefore $\|s_{\lambda,\epsilon}\|_{\mathcal E}^2 \leq \langle s_{\lambda, \epsilon}, s_{0, \epsilon}\rangle_{\mathcal E} \leq \|s_{\lambda, \epsilon}\|_{\mathcal E} \|s_{0, \epsilon}\|_{\mathcal E}$, and $\|s_{\lambda,\epsilon}\|_{\mathcal E} \leq \|s_{0,\epsilon}\|_{\mathcal E}$.
By the definition of $\overline V_n$ and Cauchy-Schwarz inequality,
$$
n\|s_{0,\epsilon}- s_{\lambda, \epsilon}\|_{n,\Omega}^2 \leq \lambda \|s_{\lambda, \epsilon}\|_{\mathcal E} \|s_{0,\epsilon} - s_{\lambda, \epsilon}\|_{\mathcal E} \leq \lambda \|s_{\lambda, \epsilon} \|_{\mathcal E}  \cdot \overline V_n\|s_{0,\epsilon}- s_{\lambda, \epsilon}\|_{n,\Omega},
$$
which implies 
$\|s_{0,\epsilon} - s_{\lambda, \epsilon}\|_{n,\Omega} \leq \frac{\lambda \overline V_n}{n} \|s_{0,\epsilon}\|_{\mathcal E}$.
Further, by the definition of $V_n$, we have 
$$
\|s_{0,\epsilon} - s_{\lambda, \epsilon}\|_{\infty,\Omega} \leq V_n\|s_{0,\epsilon} - s_{\lambda, \epsilon}\|_{\mathcal E}\leq \frac{\lambda V_n\overline V_n}{n} \|s_{0,\epsilon}\|_{\mathcal E}.
$$
Then by (\ref{EQN:markov}), $\|s_{0,\epsilon} - s_{\lambda, \epsilon}\|_{\infty, \Omega} \leq\frac{\lambda V_n\overline V_n}{n} N\|s_{0,\epsilon}\|_{L^2(\Omega)}$. Therefore,
\begin{align*}
\|s_{\lambda, \epsilon}\|_{\infty, \Omega}&\leq \|s_{0,\epsilon}\|_{\infty,\Omega} + \|s_{0,\epsilon}-s_{\lambda, \epsilon}\|_{\infty, \Omega} \\
&\leq \|s_{0,\epsilon}\|_{\infty,\Omega} + \frac{\lambda V_n\overline V_n}{n} N\|s_{0,\epsilon}\|_{L^2(\Omega)}
\end{align*}
With \Cref{LEM:e0inf}, \Cref{LEM:Vn}, and  Part 1 of \Cref{prop2},
\begin{align*}
\|s_{\lambda, \epsilon}\|_{\infty, \Omega}&= O_p\left(\sqrt{\frac{{N}\log n }{n}} +  \frac{\lambda V_n\overline V_n}{n} N n^{-1/2}N^{1/2}\right)\\
& = O_p\left(\sqrt{\frac{{N}\log n }{n}} + \lambda N^{3} n^{-3/2}\right).
\end{align*}
Thus the proof is completed.
\end{proof}

\subsection{Variance of noise term}

Theorem \ref{THE:variance} below states the asymptotic property of the variance of the TSSS estimator. 
\begin{theorem}
\label{THE:variance}
Under Assumptions \ref{C1}, \ref{C2}, \ref{C3}, \ref{C4}, 
as $n\to \infty$, we have 
\[
\frac{C_1 c_\sigma^2 N}{n}\left(1+ \frac{\lambda N^2}{n}\right)^{-2}  \leq \var\{s_{\lambda,\epsilon}(\mathbf x)|\mathbb X\} \leq  \frac{C_2 C_\sigma^2 N}{n},
\]
for some positive constants $c$ and $C$, where $c_\sigma$ and $C_\sigma$ are defined in \ref{C2}.
\end{theorem}

\begin{proof}
Similar to the proof of \Cref{LEM:e0inf}, we can write $s_{\lambda, \epsilon}(\mathbf x) = \mathbf B(\mathbf x)^\top \widehat {\bs\gamma}_{\lambda, \epsilon}$, where $\widehat{\bs\gamma}_{\lambda, \epsilon}$ is the coefficient vector for $s_{\lambda, \epsilon}$ using basis functions $B_\xi, ~\xi\in\mathcal M$. 
Note that 
$$
\widehat{\bs\gamma}_{\lambda, \epsilon} = 
\left[\sum_{i = 1}^n B_{\xi_1}(\mathbf X_i) B_{\xi_2} (\mathbf X_i) + \lambda \langle B_{\xi_1} , B_{\xi_2}\rangle_{\mathcal E}\right]_{\xi_1, \xi_2\in\mathcal M}^{-1}
\left[\sum_{i = 1}^n B_{\xi}(\mathbf X_i) \sigma(\mathbf X_i) \epsilon_i \right]_{\xi\in\mathcal M},
$$
and $\var\{s_{\lambda, \epsilon}(\mathbf x)|\mathbb X\} = \mathbf B(\mathbf x)^\top \E(\widehat{\bs\gamma}_{\lambda,\epsilon}\widehat{\bs\gamma}_{\lambda,\epsilon}^\top |\mathbb X)\mathbf B(\mathbf x)$. 
In addition, by the definition of $\bs\Gamma_{\lambda}$, we can write 
\begin{align*}
\E(\widehat{\bs\gamma}_{\lambda,\epsilon}\widehat{\bs\gamma}_{\lambda,\epsilon}^\top |\mathbb X) = &\bs\Gamma_\lambda^{-1} \E\left(\left[\frac{1}{n}\sum_{i = 1}^n B_{\xi_1}(\mathbf X_i) \sigma(\mathbf X_i)\epsilon_i \right]_{\xi_1\in\mathcal M} \left[\frac{1}{n}\sum_{i = 1}^n B_{\xi_2}(\mathbf X_i) \sigma(\mathbf X_i)\epsilon_i \right]_{\xi_2\in\mathcal M}^\top \bigg | \mathbb X\right)\bs\Gamma_\lambda^{-1}\\
= &\bs\Gamma_\lambda^{-1}\frac{1}{n^2} \E\left\{\sum_{i = 1}^n \left[B_{\xi_1}(\mathbf X_i) B_{\xi_2}(\mathbf X_i) \right]_{\xi_i,\xi_2\in\mathcal M} \sigma^2(\mathbf X_i)\epsilon_i^2 \bigg | \mathbb X\right\}\bs\Gamma_\lambda^{-1}\\
 = &\bs\Gamma_\lambda^{-1}\frac{1}{n^2} \sum_{i = 1}^n \left[B_{\xi_1}(\mathbf X_i) B_{\xi_2}(\mathbf X_i) \right]_{\xi_i,\xi_2\in\mathcal M} \sigma^2(\mathbf X_i)\bs\Gamma_\lambda^{-1},
\end{align*}
which implies 
$$
\frac{c_\sigma^2}{n}\bs\Gamma_\lambda^{-1} \bs\Gamma_0\bs\Gamma_\lambda^{-1} \leq \E(\widehat{\bs\gamma}_{\lambda,\epsilon}\widehat{\bs\gamma}_{\lambda,\epsilon}^\top |\mathbb X)\leq \frac{C_\sigma^2}{n}\bs\Gamma_\lambda^{-1} \bs\Gamma_0\bs\Gamma_\lambda^{-1},
$$
and 
$$
\mathbf B(\mathbf X_i)^\top \frac{c_\sigma^2}{n}\bs\Gamma_\lambda^{-1} \bs\Gamma_0\bs\Gamma_\lambda^{-1}\mathbf B(\mathbf X_i)\leq \var (s_{\lambda, \epsilon}|\mathbb X) \leq \mathbf B(\mathbf X_i)^\top \frac{C_\sigma^2}{n}\bs\Gamma_\lambda^{-1} \bs\Gamma_0\bs\Gamma_\lambda^{-1}\mathbf B(\mathbf X_i).
$$
Let $\rho_{\max}(\bs\Gamma_{\lambda})$ and $\rho_{\min}(\bs\Gamma_{\lambda})$ be the largest and smallest eigenvalues of $\bs\Gamma_{\lambda}$, we have
$$
\frac{c_\sigma^2}{n}\rho_{\max}(\bs\Gamma_{\lambda})^{-2} \rho_{\min}(\bs\Gamma_0)\|\mathbf B(\mathbf x)\|^2 \leq \var \{ s_{\lambda, \epsilon} (\mathbf x)|\mathbb X\} \leq \frac{C_\sigma^2}{n}\rho_{\min}(\bs\Gamma_{\lambda})^{-2} \rho_{\max}(\bs\Gamma_0)\|\mathbf B(\mathbf x)\|^2.
$$
It is important to note $\|\mathbf B(\mathbf x)\|^2$ is bounded away from $0$ and infinity. This can be proved by recognizing: (i) $B_\xi(\mathbf x)$ cannot be zero for all $\xi\in\mathcal M$; and (ii) $B_\xi$ is bounded and that for any $\mathbf x$, there are finite nonzero terms in $B_\xi(\mathbf x), \xi\in\mathcal M$. 
Therefore, with probability approaching one,
$$
\frac{c_\sigma^2}{n}\rho_{\max}(\bs\Gamma_{\lambda})^{-2} \rho_{\min}(\bs\Gamma_0)\leq \var \{ s_{\lambda, \epsilon} (\mathbf x)|\mathbb X\} \leq \frac{C_\sigma^2}{n}\rho_{\min}(\bs\Gamma_{\lambda})^{-2} \rho_{\max}(\bs\Gamma_0).
$$
By \Cref{LEM:Gamma} (i), we have 
$cN^{-1} \leq \rho_{\min}(\bs\Gamma_{\lambda})\leq \rho_{\max}(\bs\Gamma_{\lambda})\leq C\left(N^{-1}+ \lambda N/n \right)$ and 
$cN^{-1} \leq \rho_{\min}(\bs\Gamma_{0})\leq \rho_{\max}(\bs\Gamma_{0})\leq CN^{-1}$.
Therefore, 
\begin{align*}
\frac{C_1 c_\sigma^2}{n}N^{-1}\left(N^{-1}+ \lambda N/n\right)^{-2}  \leq \var\{s_{\lambda,\epsilon}(\mathbf x)|\mathbb X\} 
\leq \frac{C_2 C_\sigma^2}{n}(N^{-1})^{-2}N^{-1} = \frac{C_2 C_\sigma^2 }{n}N.    
\end{align*}
The proof is completed.
\end{proof}

\subsection{Proof of Theorem 2}

\begin{proof}[Proof of \Cref{THE:normal}]
Firstly, we show under Assumptions \ref{C1}, \ref{C2}, \ref{C3}, \ref{C4}, \ref{C5}, 
\begin{align}\label{EQN:them2prop}
\frac{s_{\lambda, \epsilon}(\mathbf x)}{\sqrt{\var\{s_{\lambda, \epsilon}|\mathbb X\}}}\stackrel{d}{\to} N(0,1), \quad n\to\infty.
\end{align}
Let $\{B_\xi\}_{\xi\in\mathcal M}$ be the SBB for $\mathcal S_d^r$, where $\mathcal M$ is the index sets for SBB. Then we can write $s_{\lambda, \epsilon}(\mathbf x) = \mathbf B(\mathbf x)^\top \widehat {\bs\gamma}_{\lambda, \epsilon}$, where $\widehat{\bs\gamma}_{\lambda, \epsilon}$ is the coefficient vector for $s_{\lambda, \epsilon}$ using basis functions $B_\xi, \xi\in\mathcal M$. 
By the definition of $\bs\Gamma_\lambda$ in (\ref{EQN:Glambda}), 
$$
s_{\lambda, \epsilon}(\mathbf x) = \mathbf B(\mathbf x)^\top \bs\Gamma_\lambda^{-1} \left[\frac{1}{n}\sum_{i = 1}^n B_\xi(\mathbf X_i) \sigma(\mathbf X_i)\epsilon_i \right]_{\xi\in\mathcal M} = \frac{1}{n}\sum_{i = 1}^n \mathbf B(\mathbf x)^\top \bs\Gamma_{\lambda}^{-1}\mathbf B(\mathbf X_i) \sigma(\mathbf X_i) \epsilon_i.
$$
Let $a_i = \frac{1}{n}\mathbf B(\mathbf x)^\top \bs\Gamma_{\lambda}^{-1}\mathbf B(\mathbf X_i)\sigma(\mathbf X_i)$, we have $s_{\lambda, \epsilon}(\mathbf x) = \sum_{i = 1}^n a_i\epsilon_i$, and 
$$
a_i^2 = \frac{1}{n^2}\mathbf B(\mathbf x)^\top \bs\Gamma_{\lambda}^{-1}\mathbf B(\mathbf X_i) \mathbf B(\mathbf X_i)^\top \bs\Gamma_{\lambda}^{-1}\mathbf B(\mathbf x) \sigma^2(\mathbf X_i).
$$
Note that for any $\bs\gamma = (\gamma_\xi, \xi\in\mathcal M)^\top$, we have 
$
\bs\gamma^\top \mathbf B(\mathbf X_i) \mathbf B(\mathbf X_i) ^\top \bs\gamma \leq \|\bs\gamma\|_2^2 \sum_{\xi\in\mathcal M} B_\xi^2(\mathbf X_i).
$
By \Cref{LEM:Gamma} (i) and the boundedness of $\sigma$, with probability approaching one,  
\begin{align*}
a_i^2 & \leq \frac{1}{n^2}\sum_{\xi\in\mathcal M} B_\xi^2(\mathbf x)\mathbf B(\mathbf x)^\top \bs\Gamma_\lambda^{-1}\bs\Gamma_\lambda^{-1} \mathbf B(\mathbf x) \sigma^2(\mathbf X_i)\\
& \leq \frac{C_\sigma^2}{n^2}
{N^2}\sum_{\xi\in\mathcal M} B_\xi^2(\mathbf x)\mathbf B(\mathbf x)^\top \mathbf B(\mathbf x) \sigma^2(\mathbf X_i) = \frac{C_\sigma^2}{n^2}{N^2}\sum_{\xi\in\mathcal M}B_\xi^2(\mathbf x) \sum_{\xi\in\mathcal M}B_\xi^2(\mathbf X_i). 
\end{align*}
And by \Cref{THE:variance}, we have $\sum_{i = 1}^n a_i^2 = \var \{s_{\lambda, \epsilon}|\mathbb X\} \geq \frac{c_\sigma^2 N}{n}$ when $\lambda = 0$.
By \Cref{LEM:Gamma} (iii), we have $\|\bs a\|_\infty \leq c_1\|\bs\sigma\|_\infty \leq c_1 C_\sigma$, where $\bs\sigma = (\sigma(\mathbf x_1),\ldots, \sigma(\mathbf x_n))^\top$.
Therefore, there exists a positive constant $c_2$, such that
$$
\frac{\max_{1\leq i\leq n}a_i^2}{\sum_{i= 1}^n a_i^2} \leq \frac{ c_2 C_\sigma^2 N}{
c_\sigma^2 n}\sum_{\xi\in\mathcal M} B_\xi^2(\mathbf X_i)  = O_p(N/n) = o_p(1).
$$
Therefore, $\sum_{i = 1}^n a_i\epsilon_i/\sqrt{\sum_{i = 1}^n a_i\epsilon_i^2} \stackrel{d}{\to} N(0,1)$ by Linderberg-Feller Central Limit Theorem. 
Similarly, when $\lambda >0$, and $\lambda = o(nN^{-2})$ by Assumption \ref{C6}, the following holds
$$
\frac{\max_{1\leq i\leq n}a_i^2}{\sum_{i= 1}^n a_i^2} 
\leq 
\frac{c_2 C_\sigma^2 N^2 n \left(1+ \frac{\lambda N^2}{n}\right)^{2}}{c_\sigma^2 n^2  N}\sum_{\xi\in\mathcal M} B_\xi^2(\mathbf X_i) = O_p(N/n) = o_p(1).
$$
Therefore, $\sum_{i = 1}^n a_i\epsilon_i/\sqrt{\sum_{i = 1}^n a_i\epsilon_i^2} \stackrel{d}{\to} N(0,1)$ by Linderberg-Feller Central Limit Theorem, which completes the proof for (\ref{EQN:them2prop}).
Under the Assumptions \ref{C5'} and \ref{C6}, $\|s_{\lambda, m} -m\|_{L^2(\Omega)} \ll \sqrt{\var(s_{\lambda,\epsilon}|\mathbb X)} \asymp \sqrt{N/n}$, and \Cref{THE:normal} is proved.
\end{proof}


\bibliographystyle{asa}
\bibliography{references}

\end{document}